\documentclass{aa}

\usepackage{graphicx}
\usepackage[varg]{txfonts}

\usepackage{textcomp}
\usepackage{lineno}
\newcommand*\patchAmsMathEnvironmentForLineno[1]{%
  \expandafter\let\csname old#1\expandafter\endcsname\csname #1\endcsname
  \expandafter\let\csname oldend#1\expandafter\endcsname\csname end#1\endcsname
  \renewenvironment{#1}%
     {\linenomath\csname old#1\endcsname}%
     {\csname oldend#1\endcsname\endlinenomath}}%
\newcommand*\patchBothAmsMathEnvironmentsForLineno[1]{%
  \patchAmsMathEnvironmentForLineno{#1}%
  \patchAmsMathEnvironmentForLineno{#1*}}%
\AtBeginDocument{%
\patchBothAmsMathEnvironmentsForLineno{equation}%
\patchBothAmsMathEnvironmentsForLineno{align}%
\patchBothAmsMathEnvironmentsForLineno{flalign}%
\patchBothAmsMathEnvironmentsForLineno{alignat}%
\patchBothAmsMathEnvironmentsForLineno{gather}%
\patchBothAmsMathEnvironmentsForLineno{multline}%
}

\usepackage{color}
\usepackage{ulem}
\usepackage{hyperref}
\hypersetup{colorlinks=true, urlcolor=blue, citecolor=blue}

\makeatletter
\renewcommand*\aa@pageof{, page \thepage{} of \pageref*{LastPage}}
\makeatother

\begin{document}

\title{Ceres' opposition effect observed by the Dawn framing camera}

\author{Stefan~E.~Schr\"oder \inst{1}\fnmsep\thanks{Corresponding author, \email{stefanus.schroeder@dlr.de}} \and Jian-Yang~Li \inst{2} \and Marc~D.~Rayman \inst{3} \and Steven~P.~Joy \inst{4} \and Carol~A.~Polanskey \inst{3} \and Uri~Carsenty \inst{1} \and Julie~C.~Castillo-Rogez \inst{3} \and Mauro~Ciarniello \inst{5} \and Ralf~Jaumann \inst{1} \and Andrea~Longobardo \inst{5} \and Lucy~A.~McFadden \inst{6} \and Stefano~Mottola \inst{1} \and Mark~Sykes \inst{2} \and Carol~A.~Raymond \inst{3} \and Christopher~T.~Russell \inst{4}}

\institute{Deutsches Zentrum f\"ur Luft- und Raumfahrt (DLR), 12489 Berlin, Germany \and Planetary Science Institute (PSI), Tucson, AZ 85719, U.S.A. \and Jet Propulsion Laboratory (JPL), California Institute of Technology, Pasadena, CA 91109, U.S.A. \and Institute of Geophysics and Planetary Physics (IGPP), University of California, Los Angeles, CA 90095-1567, U.S.A. \and Istituto di Astrofisica e Planetologia Spaziali, Istituto Nazionale di Astrofisica (INAF), 00133 Rome, Italy \and Goddard Space Flight Center (GSFC), Greenbelt, MD 20771, U.S.A.}

%\date{Received September 15, 1996; accepted March 16, 1997}

% \abstract{}{}{}{}{}
% 5 {} token are mandatory

\abstract
{The surface reflectance of planetary regoliths may increase dramatically towards zero phase angle, a phenomenon known as the opposition effect (OE). Two physical processes that are thought to be the dominant contributors to the brightness surge are shadow hiding (SH) and coherent backscatter (CB). The occurrence of shadow hiding in planetary regoliths is self-evident, but it has proved difficult to unambiguously demonstrate CB from remote sensing observations. One prediction of CB theory is the wavelength dependence of the OE angular width.}
{The Dawn spacecraft observed the OE on the surface of dwarf planet Ceres. We aim to characterize the OE over the resolved surface, including the bright Cerealia Facula, and to find evidence for SH and/or CB. It is presently not clear if the latter can contribute substantially to the OE for surfaces as dark as that of Ceres.}
{We analyze images of the Dawn framing camera by means of photometric modeling of the phase curve.}
{We find that the OE of most of the investigated surface has very similar characteristics, with an enhancement factor of 1.4 and a full width at half maximum{\sc } of $3^\circ$ (``broad OE''). A notable exception are the fresh ejecta of the Azacca crater, which display a very narrow brightness enhancement that is restricted to phase angles $< 0.5^\circ$ (``narrow OE''); suggestively, this is in the range in which CB is thought to dominate. We do not find a wavelength dependence for the width of the broad OE, and lack the data to investigate the dependence for the narrow OE. The prediction of a wavelength-dependent CB width is rather ambiguous, and we suggest that dedicated modeling of the Dawn observations with a physically based theory is necessary to better understand the Ceres OE. The zero-phase observations allow us to determine Ceres' visible geometric albedo as $p_{\rm V} = 0.094 \pm 0.005$. A comparison with other asteroids suggests that Ceres' broad OE is typical for an asteroid of its spectral type, with characteristics that are primarily linked to surface albedo.}
{Our analysis suggests that CB may occur on the dark surface of Ceres  in a highly localized fashion. While the results are inconclusive, they provide a piece to the puzzle that is the OE of planetary surfaces.}


\maketitle
%
%-------------------------------------------------------------------

\section{Introduction}

The reflectance of planetary regoliths can increase dramatically towards zero solar phase angle, a phenomenon known as the \textit{opposition effect} (OE). On April 29, 2017, the framing cameras of the Dawn spacecraft captured the OE on the surface of dwarf planet Ceres in images with a spatial resolution of 1.8~km per pixel. The OE is usually observed by Earth-based telescopes, and resolved observations by spacecraft are still rare. Zero-phase images of asteroids are often acquired on approach during a flyby, as for 2867~\v{S}teins \citep{S10,Sp12} and 21~Lutetia \citep{M15,S15,H16}. Sometimes, the small mass of an asteroid allows the spacecraft to slowly hover into the opposition geometry, as for 25143~Itokawa \citep{LI18}. In the case of Ceres, Dawn reached the opposition geometry through an ingenious scheme of orbital navigation to avoid eclipse at the other side of the asteroid \citep{R17}. Ground-based observations of Ceres that extend to very low phase angle have been reported by \citet{T83} (minimum phase angle $1.1^\circ$) and \citet{R15} (minimum $0.85^\circ$). Ceres is also the largest asteroid in the main belt and is classified as C-type in the SMASS taxonomy \citep{BB02}. This paper studies the Ceres OE as observed by the Dawn cameras in the context of observations for other rocky solar system bodies, especially asteroids and the Moon, with the aim of linking the OE characteristics to the physical properties of the regolith.

In practice, the OE as observed is not always consistently defined \citep{BS00}. In the first description of the OE for an asteroid, \citet{G56} wrote: ``...close to opposition there is a pronounced nonlinear increase in brightness''. As he was referring to telescopic observations of 20~Massalia as an integrated point source, the OE represents a departure of the phase curve from linearity on the visual magnitude scale, which is logarithmic in intensity. The asteroid OE range typically extends from $0^\circ$ to $\sim 5^\circ$ phase angle, whereas the ``linear'' part of the phase curve extends to at least $25^\circ$, but not necessarily beyond \citep{BS00,R02}. Two physical processes that are thought to be the dominant contributors to the OE brightness surge are shadow hiding (SH) and coherent backscatter (CB). {Shadow hiding}  refers to the shadows cast by regolith particles and larger objects. Near phase angle zero, when the Sun is directly behind the observer, shadows are hidden, increasing the reflectance of the surface \citep{H84,H86,S94,SH01}. While the SH opposition effect (SHOE) refers only to the contribution of SH to the OE, SH affects the phase curve over the full range of phase angles. Coherent backscatter  is a form of constructive interference of light at very small phase angles, the physics of which is well understood \citep{TM10,PT11}. While originally described for media of widely separated particles in suspension \citep{KI84,AL85}, CB was predicted to also play an important role in surfaces of high packing density, such as planetary regoliths \citep{MD93,Mi09,DM13}. For CB to produce an OE peak that is wide enough to be observed, the presence of grains with sizes on the order of the wavelength is required; much larger or smaller grains would produce a coherent peak that is too narrow \citep{MD93}. The scattering behavior of individual regolith particles (often described by the single particle phase function) could be considered as a third process contributing to the OE \citep{K13}, but the balance between back- and forward scattering by a single particle can also be interpreted in terms of its ability to cast shadows \citep{S12}. It is generally held that the CBOE peak is narrow, at most a few degrees wide, whereas the SHOE peak can be much broader \citep{MD93,H97,H98}, although modeling suggests that CB may contribute to backscattering up to at least $10^\circ$ phase angle \citep{S99}.

Given its reliance on multiple scattering, it is likely that CB is experienced by bright, atmosphereless solar system bodies like the Saturnian satellites \citep{He98,Mi09} and E-type asteroids \citep{MD93,R09}, and not by the darkest objects \citep{SB10}. Whether CB is important for bodies of intermediate albedo is not clear. The optical properties of the lunar regolith have been studied extensively, and this particular question has been subject of a lively debate \citep{B96,H97,H98,S99,H12,S12}. Recent modeling suggests that CB contributes substantially to the OE of lunar maria \citep{M11}, which are almost as dark as the surface of Ceres at visible wavelengths \citep{V11}. Therefore, we cannot \textit{a priori} dismiss all thought of a CB contribution to the Ceres OE.

In practice, it is difficult to unambiguously demonstrate CB from remote sensing data. When both photometric and polarimetric observations are available, the latter may be more easily inferred and modeled \citep{M93,S02,R06,M10}, although quantitative interpretations of the observations in terms of sizes of particles, their refractive index, and packing density remain challenging \citep{PT11}. Ceres has a very broad polarization minimum that is almost $20^\circ$ wide \citep{M10}. Polarization data are not available for very small phase angles, so it is not known if Ceres displays the so-called polarization OE, a narrow negative polarization peak, thought to be related to CB \citep{R06,R09,DM13}. Dawn only acquired photometric observations of Ceres. Therefore, the challenge is to distinguish SH from CB in the absence of polarimetric data. One particular prediction of CB theory that we hope to test is that the width of the CB peak should depend on wavelength \citep{M92}.

\section{Data and methods}

\begin{figure*}
\centering
\includegraphics[angle=90,width=16cm]{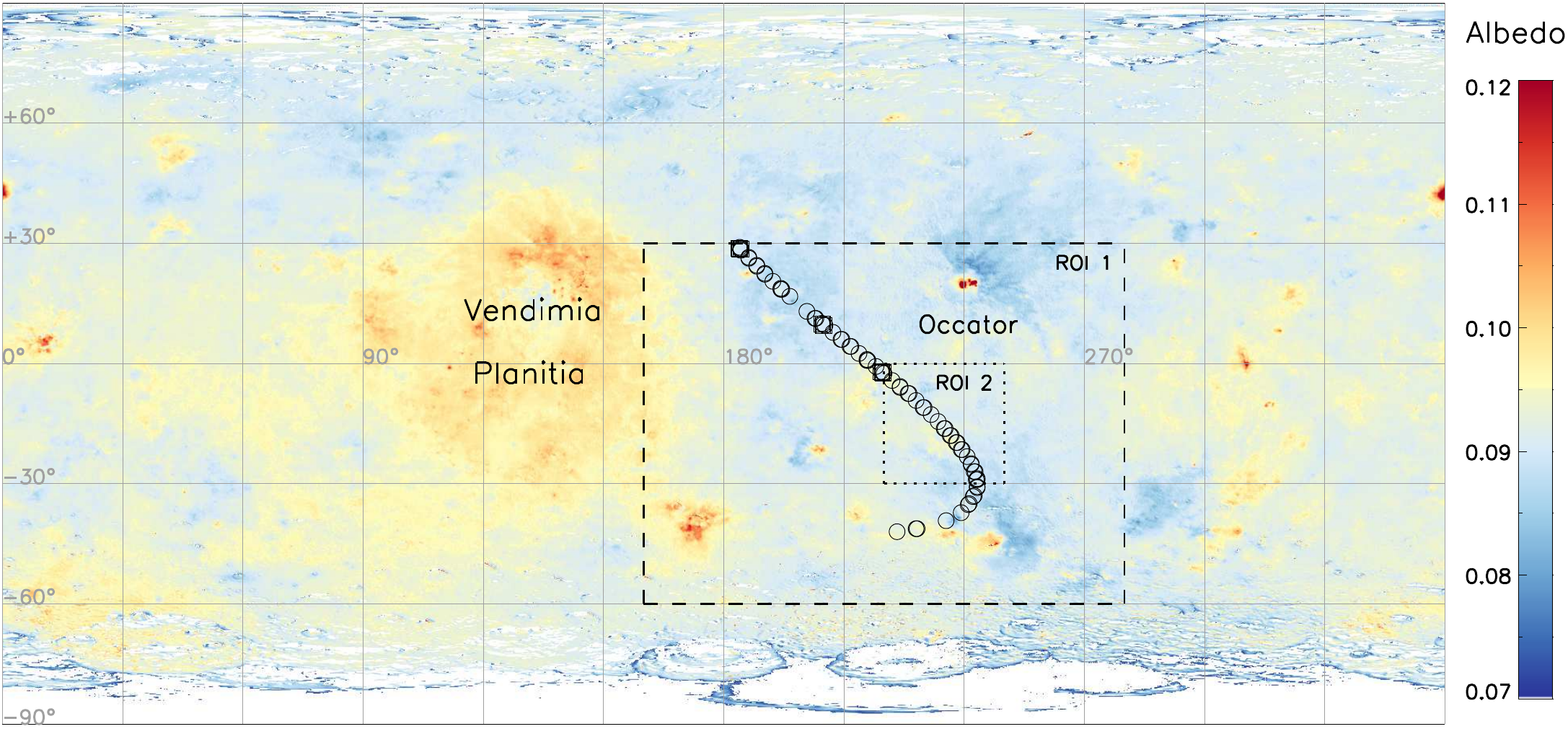}
\caption{The locations of zero phase in clear ($\bigcirc$) and narrow-band ($\Box$) framing camera images displayed on a map of the normal albedo in the clear filter \citep{S17}. The regions delineated by the dotted and dashed lines are two regions of interest (ROI~1 and ROI~2).}
\label{fig:albedo_map}
\end{figure*}

\begin{figure}
\resizebox{\hsize}{!}{\includegraphics{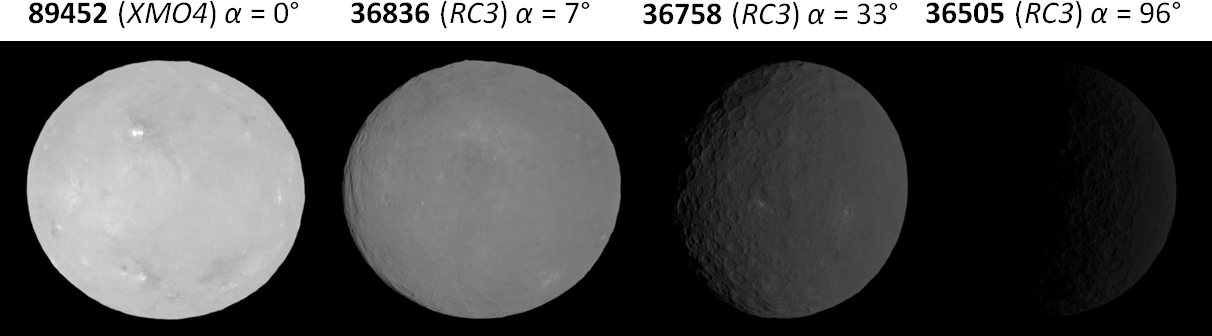}}
\caption{Ceres images representing different parts of the phase curve, shown at their correct relative brightness on a linear scale. The image number, mission phase, and phase angle at the disk center are indicated. Only the Cerealia and Vinalia Faculae (the bright spots in the opposition image at far left) are not displayed at their correct brightness, as they are much brighter than the average Ceres surface \citep{L16,S17}.}
\label{fig:phot_seq}
\end{figure}

Images that captured Ceres at zero phase were acquired by the Dawn FC1 and FC2 framing cameras \citep{Si11} as part of extended mission orbit~4 (\textit{XMO4}\footnote{Mission phase names are printed in \textit{italics}.}; see \citealt{R17} for an overview of the Dawn mission phases at Ceres). The path of zero phase over the surface is indicated in Fig.~\ref{fig:albedo_map}, superposed on an albedo map. The path is situated between big, and relatively bright, Vendimia Planitia and Occator crater with the bright, enigmatic Cerealia Facula \citep{SB16,Ru17}. We focus our analysis on an area immediately surrounding the path: region-of-interest \#1 (ROI~1, with $160^\circ <$ lon $< 280^\circ$ and $-60^\circ <$ lat $< +30^\circ$). In our study of the OE we use only FC2 images; the FC1 images are affected by residual charge \citep{S13a} and will not be considered here\footnote{A simple fix for residual charge exists in the form of acquisition of a zero-second exposure immediately prior to the regular exposure \citep{S13a}, which was, unfortunately, not implemented for \textit{XMO4}.}. The \textit{XMO4} images (\textbf{89387}-\textbf{89580}\footnote{Image numbers are printed \textbf{bold}.}) were acquired in the extended mission at Ceres at a relatively low spatial resolution of 1.9~km per pixel (Table~\ref{tab:image_data}). A total of 170 clear filter images were taken continuously in rapid succession, 155 of which captured zero phase somewhere on the surface. At four instances, a set of six narrow-band images was acquired, each set consisting of two cycles of the filter sequence F3 (effective wavelength 749~nm), F5 (964~nm), and F8 (437~nm). The footprints of only three sets are shown in Fig.~\ref{fig:albedo_map}, as the fourth one did not capture zero phase. The full width of the Ceres OE may exceed the phase angle range of the \textit{XMO4} data set ($< 7^\circ$), and therefore we include \textit{RC3} images \citep{R17}. These were acquired early in the mission at a comparable spatial resolution. The \textit{RC3} images cover a very wide phase angle range. We can distinguish three groups of \textit{RC3} images: a, b, and c (Table~\ref{tab:image_data}). Two of these (a and b) were acquired at very large phase angles, but cover ROI~1 in only a restricted latitude range \citep{L15}. The third group (c) is most relevant to our purposes, as it provides complete coverage of ROI~1 at moderate phase angles ($<50^\circ$). Figure~\ref{fig:phot_seq} shows a few examples of images acquired at different phase angles in the two mission phases, displayed at their correct relative brightness.

\begin{table*}
\centering
\caption{Available FC2 image data covering ROI~1 for the \textit{RC3} and \textit{XMO4} mission phases with no restrictions on the photometric angles. The date refers to the day of acquisition of the first image of the series. The latitude range is a rough indication of the coverage inside ROI~1. Resolution is in km per pixel.}
\begin{tabular}{lllllll}
\hline\hline
Phase & & Date & Image \# & Phase angles & Latitudes & Resolution \\
\hline
\textit{RC3} & a & 2015-04-25 & \textbf{34720}-\textbf{35420} & ($110^\circ$, $155^\circ$) & $(-60^\circ,  -10^\circ)$ & 1.26 \\
    & b & 2015-04-30 & \textbf{36357}-\textbf{36517} & ($95^\circ$, $120^\circ$) & $(0^\circ, +30^\circ)$ & 1.26 \\
    & c & 2015-05-04 & \textbf{36529}-\textbf{37264} & ($6^\circ$, $48^\circ$) & $(-60^\circ, +30^\circ)$ & 1.26 \\
\textit{XMO4} & & 2017-04-29 & \textbf{89387}-\textbf{89580} & ($0^\circ$, $7^\circ$) & $(-60^\circ, +30^\circ)$ & 1.87 \\
\hline
\end{tabular}
\label{tab:image_data}
\end{table*}

\begin{figure}
\resizebox{\hsize}{!}{\includegraphics{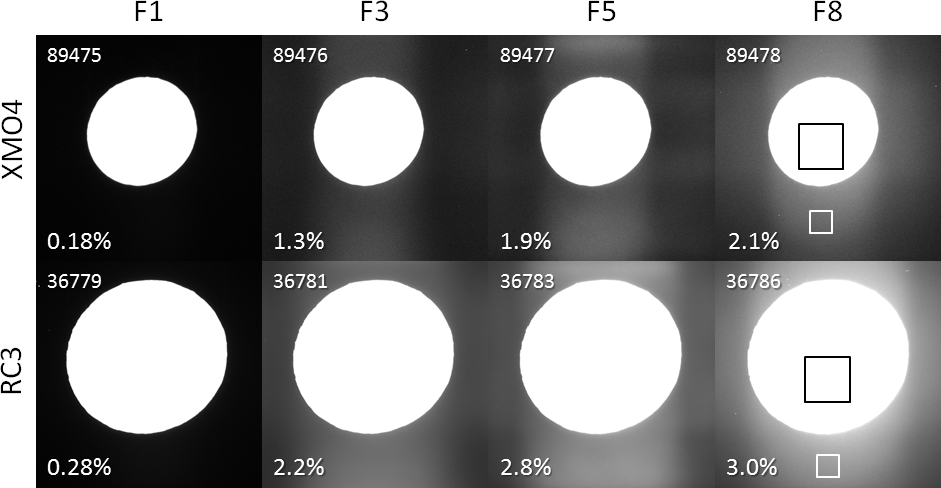}}
\caption{In-field stray light in full-frame FC2 images acquired with different filters (F1...F8) during the \textit{RC3} and \textit{XMO4} mission phases (image numbers indicated). The percentages refer to how the average signal in the white square compares to that in the black square in each image (for clarity, the squares are only shown for F8). Cerealia Facula is outside the black square. The images are displayed with black being zero signal and white being 5\% of the average signal in the black square. The stray light in the \textit{RC3} images is about 50\% stronger than in the \textit{XMO4} images.}
\label{fig:stray_light}
\end{figure}

All images were calibrated to reflectance as described by \citet{S13a,S14a}. Images through several of the narrow-band filters are affected by substantial in-field stray light \citep{Si11}. For images in which the surface of Ceres fills the field-of-view (FOV) and is evenly illuminated, a satisfactory stray light correction method exists \citep{S14a}. However, Ceres fills only a small fraction of the FOV in the images in this study, which invalidates the method. Nevertheless, while we did not attempt to subtract stray light from the images, we did apply a first-order correction to the data as follows. In Fig.~\ref{fig:stray_light}, the stray light is visible as an interference pattern around the illuminated disk of Ceres. We assume that a similar amount of stray light is also present on the disk. The stray light is stronger in \textit{RC3} images than in \textit{XMO4} images because of the larger size of Ceres' disk. For the \textit{XMO4} mission phase and one particular filter, let us consider $I_{\rm S,XMO4}$, the radiance emanating as stray light from empty space. We assume that the radiance emanating from the disk, averaged over a certain area, is $I_{\rm C,XMO4} = I_{\rm C} + I_{\rm S,XMO4}$, with $I_{\rm C}$ the true radiance of Ceres. The percentages in the top row of Fig.~\ref{fig:stray_light} refer to the ratio of the two: $I_{\rm S,XMO4} / (I_{\rm C} + I_{\rm S,XMO4}) \approx I_{\rm S,XMO4} / I_{\rm C}$, because $I_{\rm S,XMO4} \ll I_{\rm C}$. The ratio of the observed radiance of Ceres in a \textit{XMO4} and \textit{RC3} image acquired at the same phase angle is therefore
\begin{equation}
\frac{I_{\rm C,XMO4}}{I_{\rm C,RC3}} = \frac{1 + I_{\rm S,XMO4} / I_{\rm C}}{1 + I_{\rm S,RC3} / I_{\rm C}}
\label{eq:stray_light_cor}
.\end{equation}
Using the numbers in Fig.~\ref{fig:stray_light}, we find this ratio to be 0.991 for all narrow-band filters and 0.999 for the clear filter. We therefore apply a correction factor of 0.99 to all narrow-band \textit{RC3} reflectances to better match the low phase angle \textit{RC3} reflectances to the \textit{XMO4} reflectances (a 1\% correction). We did not attempt to retrieve the true Ceres reflectance ($I_{\rm C}$) from the observed reflectance ($I_{\rm C,XMO4}$), as this is not necessary in light of the purpose of this paper and the uncertainties associated with the stray light. At least the two data sets will now match up better. We note that although this type of interference stray light does not affect the clear filter, the stray light in \textit{RC3} F1 images is still higher than that in \textit{XMO4} F1 images.

All images were mapped to the same equirectangular projection with a spatial resolution of 4 pixels per degree by means of the USGS Integrated Software for Imagers and Spectrometers ISIS3 \citep{A04,B12} using bi-linear interpolation. Thereby, the projected pixels on the equator have approximately the same spatial resolution as the pixels at the center of Ceres' disk in the \textit{XMO4} images. We also used ISIS for calculating the photometric angles (incidence, emission, phase) for each image pixel. We employed the \textit{HAMO} shape model described by \citet{Pr16}, which covers about 98\% of Ceres' surface and has a vertical accuracy of about 10~m. The model is oriented in the Ceres reference frame, which is defined by crater Kait \citep{Ro16} and the Ceres rotation state derived by \citet{Pr16}. We did not attempt to improve the projection by registering the images to the shape model \citep{S17}, as the opposition images are devoid of shading. Consequently, projected pixels associated with more extreme geometries suffer from obvious projection errors, and it was necessary to limit the incidence and emission angles of the pixels included in our analysis, typically to $(\iota, \epsilon) < 50^\circ$.

In this paper we fit model curves to data several times. Fitting was performed using the Levenberg-Marquardt algorithm with constrained search spaces for the model parameters\footnote{Using the MPFITFUN library retrieved from \url{http://cow.physics.wisc.edu/~craigm/idl/idl.html}} \citep{M78,Ma09}. In the fitting process we adopted the photon noise for the reflectance error, which does not account for potential projection errors. The formal error for the model parameters as reported by the fitting algorithm is often a very small fraction of the best-fit value. However, in light of the unavoidable projection errors, the true uncertainty is almost certainly larger. We estimate that the true uncertainty of the best-fit parameter values given in this paper is on the order of unity in the last digit provided.

\section{Model description}

The surface reflectance of a planetary body depends on the wavelength $\lambda$ and the angles of observation: the local angle of incidence $\iota$ of sunlight, the local angle of emergence $\epsilon$, and the phase angle $\alpha$. For convenience we define $\mu_0 = \cos\iota$ and $\mu = \cos\epsilon$. The radiance factor is defined as
\begin{equation}
r_{\rm F}({\mu_0, \mu}, \alpha, \lambda) = \pi I({\mu_0, \mu}, \alpha, \lambda) / J(\lambda),
\label{eq:reflectance}
\end{equation}
where $I$ is the radiance in W~m$^{-2}$~\textmu m$^{-1}$~sr$^{-1}$ and $J$ is the normal solar irradiance in W~m$^{-2}$~\textmu m$^{-1}$, which depends on the distance of the planet to the Sun \citep{H81}. The radiance factor is also known as $I/F$, with $F \equiv J / \pi$. We use the term ``reflectance'' for the radiance factor in the remainder of this paper.

In this paper we apply essentially two different photometric models to the surface of Ceres. In the first, the reflectance can be separated into two parts, the equigonal albedo and the disk function \citep{K01,S11}:
\begin{equation}
r_{\rm F} = A_{\rm eq}(\alpha) D(\mu_0, \mu, \alpha).
\label{eq:type_I}
\end{equation}
We often refer to the equigonal albedo as the ``phase function''. Plotting the phase function as a function of phase angle produces the phase curve. It describes the phase dependence of the brightness \citep{S11}:
\begin{equation}
A_{\rm eq}(\alpha) = A_0 f(\alpha),
\label{eq:equigonal_albedo}
\end{equation}
where $A_0$ is the normal albedo\footnote{We note that $A_0$ is not equal to the normal albedo of \citet{H81}, which is defined at zero phase but arbitrary incidence angle ($\iota = \epsilon$), and therefore depends on the local topography.}, and $f(\alpha)$ is the phase function normalized to unity at $\alpha = 0^\circ$. The latter depends on the choice of disk function $D$, which describes how the reflectance varies over the planetary disk at constant phase angle. An equigonal albedo image has no brightness trend from limb to terminator for a surface of constant albedo. The normal albedo in Eq.~\ref{eq:equigonal_albedo} is independent of local topography, as such brightness variations are contained in the disk function in Eq.~\ref{eq:type_I}. If the disk function is uniform at $\alpha = 0^\circ$, that is,\ $D(\mu_0, \mu, 0^\circ) \approx 1$, then $A_{\rm eq}(0^\circ) = A_0$ is equal to the geometric albedo of the body. This condition is met for Ceres \citep{S17}. Our strategy is to adopt the most appropriate disk function and choose a convenient form for the phase function, such that we may link spatial variations of the phase function parameters to the physical properties of the regolith.

The Akimov phase function was developed for the lunar surface \citep{A88b,K07,S11}. It also approximates the observed phase curves of asteroids with a good accuracy and is simple and convenient to use \citep{BS00}. It has the form:
\begin{equation}
f(\alpha) = \frac{e^{-\nu_1 \alpha} + m e^{-\nu_2 \alpha}}{1 + m},
\label{eq:Akimov_phase_fie}
\end{equation}
in which $\alpha$ is in degrees. It has three parameters: the slope $\nu_1$ of the phase curve, the OE amplitude $m$, and slope $\nu_2$. Occasionally, we fit Eq.~\ref{eq:Akimov_phase_fie} to data restricted to only the smallest phase angles ($< 2^\circ$). In such cases we set $m = 0$, and $\nu_1$ plays the role of $\nu_2$ in the sense that it really models the slope of the OE instead of that of the full phase curve. Two quantities that are often used to characterize the OE are the enhancement factor and the half-width at half-maximum (\textsc{hwhm}) \citep{R02}. We use the enhancement factor $\zeta$ in the definition of \citet{MD93}, as the equigonal albedo divided by the ``linear'' (on a logarithmic scale) part of the phase curve:
\begin{equation}
\zeta(\alpha) = 1 + m e^{(\nu_1 - \nu_2) \alpha}.
\label{eq:Akimov_zeta}
\end{equation}
The term ``enhancement factor'' is also often used to denote only $\zeta(0^\circ) = 1 + m$. The \textsc{hwhm} of the enhancement factor is then:
\begin{equation}
{\rm HWHM} = \frac{\ln 2}{\nu_2 - \nu_1}.
\label{eq:Akimov_HWHM}
\end{equation}

We employ two disk functions, both normalized at $\iota = \epsilon = \alpha = 0^\circ$. For the general Ceres surface we use the Akimov disk function \citep{A76,A88a,S94,S11}, which has no parameters:
\begin{equation}
D_{\rm A}({\alpha, \beta, \gamma}) = \cos \frac{\alpha}{2} \cos \left[ \frac{\pi}{\pi - \alpha} \left( \gamma - \frac{\alpha}{2} \right) \right] \frac{(\cos \beta)^{\alpha / (\pi - \alpha)}}{\cos \gamma},
\label{eq:Akimov_disk_fie}
\end{equation}
in which $\alpha$ is in radians. The photometric latitude $\beta$ and longitude $\gamma$ depend on the incidence, emergence, and phase angles as follows:
\begin{equation}
\begin{split}
\mu_0 & = \cos \beta \cos (\alpha - \gamma) \\
\mu & = \cos \beta \cos \gamma.
\end{split}
\end{equation}
The Akimov disk function was analytically derived for a dark, extremely rough surface with a hierarchical, self-similar structure \citep{S03}. It was found to provide a good match for the typical Ceres surface, although it does not perform well for Cerealia Facula, the bright central area in the Occator crater \citep{S17,L18}. For the latter we use the combined Lambert / Lommel-Seeliger law \citep{BV83,McE96}:
\begin{equation}
D_{\rm L}({\mu_0, \mu}, \alpha) = c_{\rm L}(\alpha) \frac{2 \mu_0}{\mu_0 + \mu} + [1 - c_{\rm L}(\alpha)] \mu_0,
\label{eq:L-S+Lam}
\end{equation}
with free parameter $c_{\rm L}$, which governs the relative contribution of the Lambert and Lommel-Seeliger terms.

We also employ the \citet{H81,H84,H86} model, which is not explicitly separated into a disk and phase function:
\begin{equation}
r_{\rm F}(\mu_0^\prime, \mu^\prime, \alpha) = \frac{w}{4} \frac{\mu_0^\prime}{\mu_0^\prime + \mu^\prime} [B_{\rm SH}(\alpha) P(\alpha) + H(\mu_0^\prime,w)H(\mu^\prime,w) - 1] f(\bar{\theta}),
\label{eq:Hapke}
\end{equation}
with $w$ the so-called single scattering albedo. The $H$-function is given by
\begin{equation}
H(\mu,w) = \frac{1 + 2\mu}{1 + 2 \mu \gamma(w)},
\end{equation}
with $\gamma = \sqrt{1-w}$. The SHOE is described by
\begin{equation}
B_{\rm SH}(\alpha) = 1 + B_{\rm S0} B_{\rm S}(\alpha) = 1 + \frac{B_{\rm S0}}{1 + \tan(\alpha/2) / h_{\rm S}},
\label{eq:SHOE}
\end{equation}
with $B_{\rm S0}$ and $h_{\rm S}$ the OE amplitude and width, respectively. \citet{H02} introduced two separate parameters for the CBOE, but we decided not to include those here for reasons of simplicity; we use Eq.~\ref{eq:Hapke} to merely describe the OE, which is better done with two rather than four parameters. Furthermore, \citet{S17} found that the \citet{H02} version of the model did not perform well for the Ceres surface. The model in Eq.~\ref{eq:Hapke} includes $f(\bar{\theta})$, which describes the effects of ``macroscopic roughness'', a measure of the roughness of the surface on a scale up to the image resolution limit with $\bar{\theta}$ being the mean slope angle of the surface facets. The inclusion of macroscopic roughness changes $\mu_0$ and $\mu$ to $\mu_0^\prime$ and $\mu^\prime$ in a way that we do not reproduce here; details can be found in \citet{H84}. The factor $P(\alpha)$ is the phase function of a single particle. We employ two versions of the Henyey-Greenstein function. One is the single-term version \citep{HG41}:
\begin{equation}
P(\alpha) = \frac{1-b^2}{(1 + 2b \cos \alpha + b^2)^{3/2}},
\label{eq:sHG}
\end{equation}
with parameter $b$ (often written as $g$), with $|b|<1$. This function has a single lobe, which leads to either backscattering ($b < 0$) or forward scattering ($b > 0$) behavior. Isotropic scattering corresponds to $b = 0$. The other version of the Henyey-Greenstein function has two terms \citep{SH07}:
\begin{equation}
P(\alpha) = \frac{1+c}{2} \frac{1-b^2}{(1 + 2b \cos \alpha + b^2)^{3/2}} + \frac{1-c}{2} \frac{1-b^2}{(1 - 2b \cos \alpha + b^2)^{3/2}},
\label{eq:dHG}
\end{equation}
with two parameters $b$ and $c$. We note that Eq.~\ref{eq:dHG} reduces to Eq.~\ref{eq:sHG} for $c = 1$. This function has two lobes, one for forward scattering, and the other for backscattering. The width of both lobes is governed by $b$, with $0 \leq b < 1$, and $b = 0$ representing isotropic scattering. The relative amplitude of the two lobes is determined by $c$, with $|c|<1$. When $b \neq 0$, forward scattering dominates\footnote{\citet{SH07} erroneously stated the opposite.} for $c > 0$ and backscattering dominates for $c < 0$, with $c = 0$ meaning back- and forward scattering in equal strength. Of course, abundant data in the $(90^\circ, 180^\circ)$ phase angle range are required to reliably derive values for $c$. However, as the $b$ parameter offers flexibility to shape the backscattering lobe, the double term Henyey-Greenstein function is always expected to improve the fit quality of the Hapke model over the single-term version.

\section{Resolved photometry}

\subsection{Regions of interest}

\begin{figure*}
\centering
\includegraphics[width=8cm]{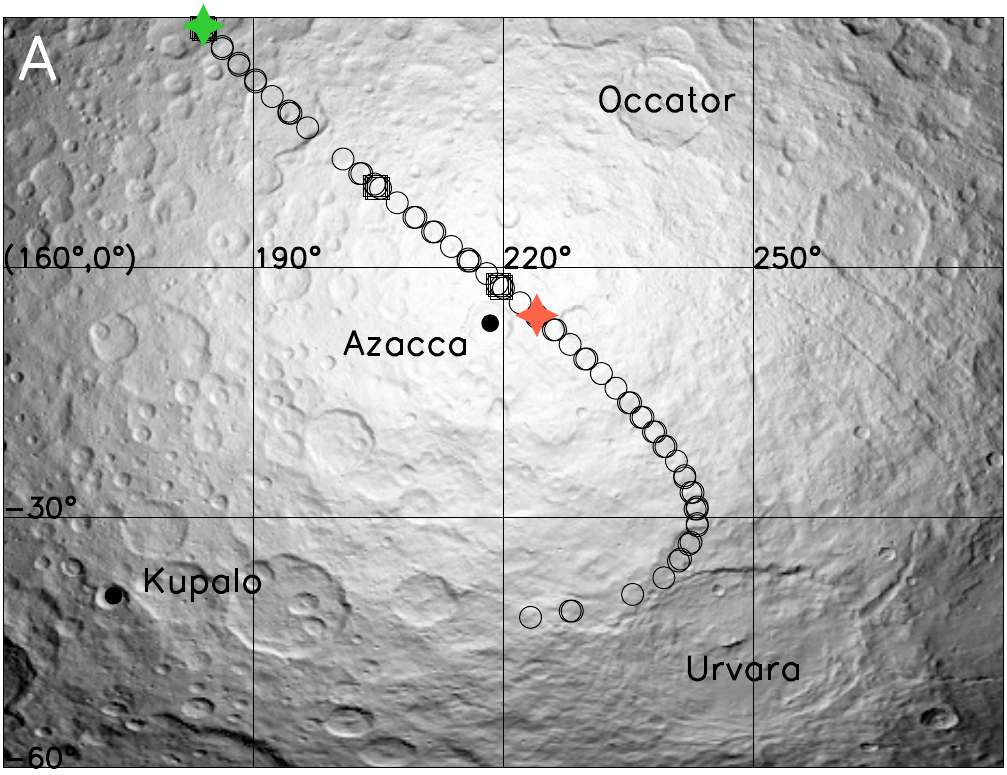}
\includegraphics[width=8cm]{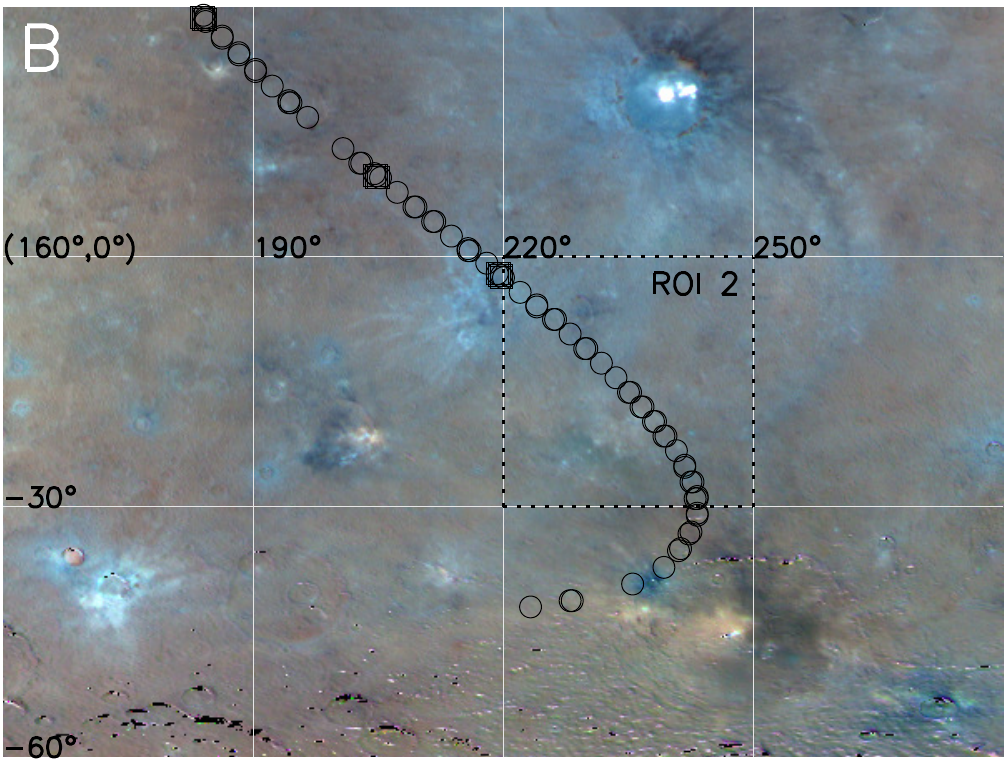}
\caption{Characteristics of ROI~1 (outlined in Fig.~\ref{fig:albedo_map}). \textbf{A}: Shape model with Lambert reflection and illumination conditions from \textit{XMO4} image \textbf{89504}. The green and red stars denote locations selected for detailed study. Craters of interest are indicated. \textbf{B}: Enhanced color composite with filters F5 (965~nm), F2 (555~nm), and F8 (438~nm) in the RGB color channels \citep{S17}. ROI~2 is indicated. We note that the color and brightness of the spots in Occator cannot be displayed correctly here as they are much brighter than their surroundings, exceeding the dynamic range of the figure.}
\label{fig:ROI}
\end{figure*}

ROI~1, which contains the zero-phase path in its entirety, is shown in more detail in Fig.~\ref{fig:ROI}. We find the Juling-Kupalo crater pair at the bottom left and Occator crater at the top right, all fresh craters with characteristic blue ejecta. The large crater at the bottom right, with a reddish interior, is named Urvara. The area close to the path of zero phase is fairly nondescript. Parts of it are slightly bluish, mostly due to the presence of Occator ejecta, but without large excursions in either albedo or color. The one exception is Azacca crater at longitude $218.4^\circ$ and latitude $-6.7^\circ$, also with relatively blue and bright ejecta. This feature plays an important role in the discussion of the OE below. A second region of interest is the area between longitudes $220^\circ$ and $250^\circ$ and latitudes $-30^\circ$ and $0^\circ$. In ROI~2 we find the results of our various photometric analyses to be most reliable, with the least artifacts.

\subsection{Model comparison}
\label{sec:model_comparison}

In this section, we model the phase curve at one particular location on the path of zero-phase angle to show the morphology of the phase curve, including the OE, and to illustrate the availability of data. We compare the performance of the Akimov and Hapke models below. In the following section, we extend our analysis to the entirety of ROI~1 to search for spatial variations of the model parameters.

\subsubsection{Akimov model}

\begin{figure*}
\centering
\includegraphics[width=8cm]{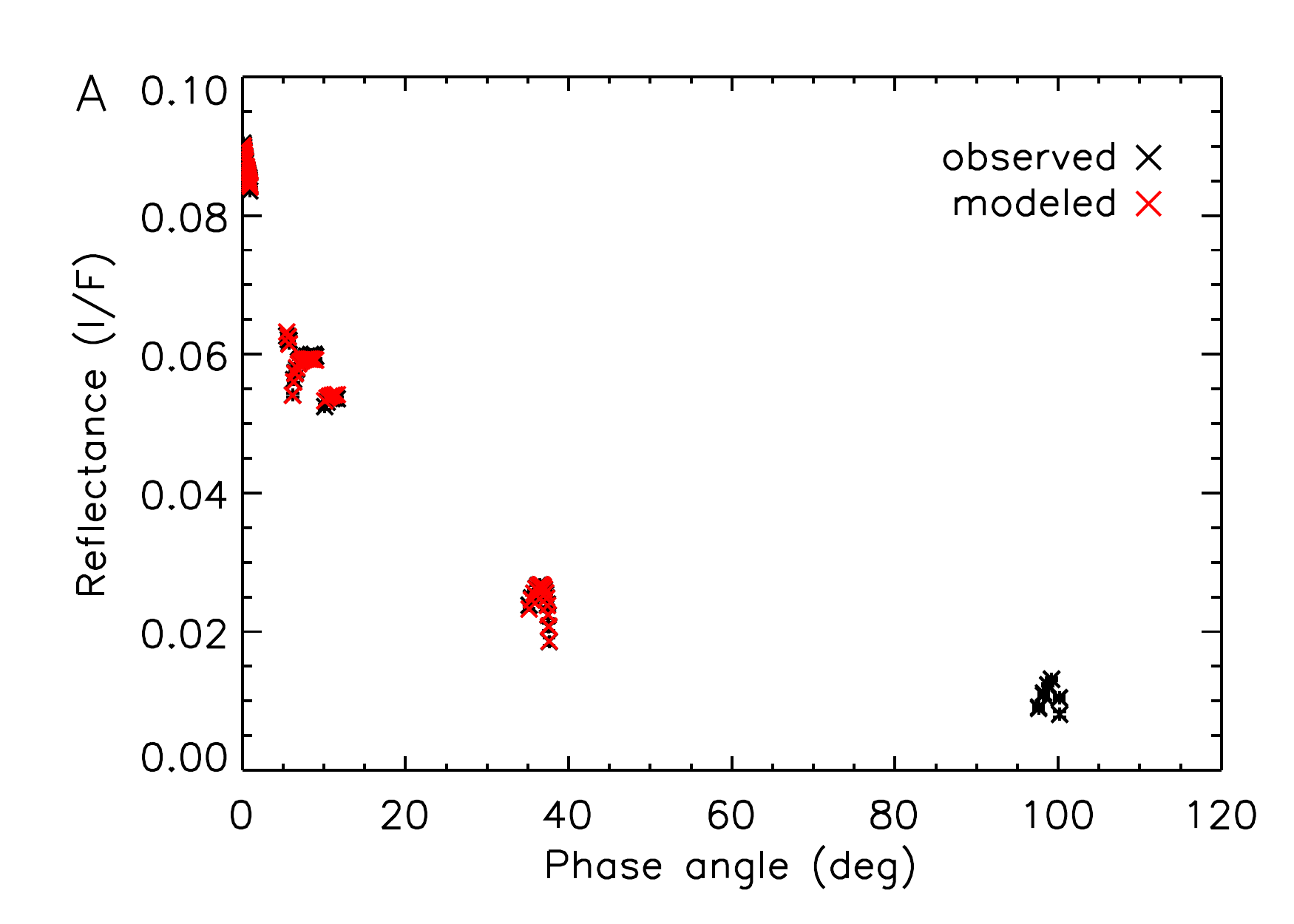}
\includegraphics[width=8cm]{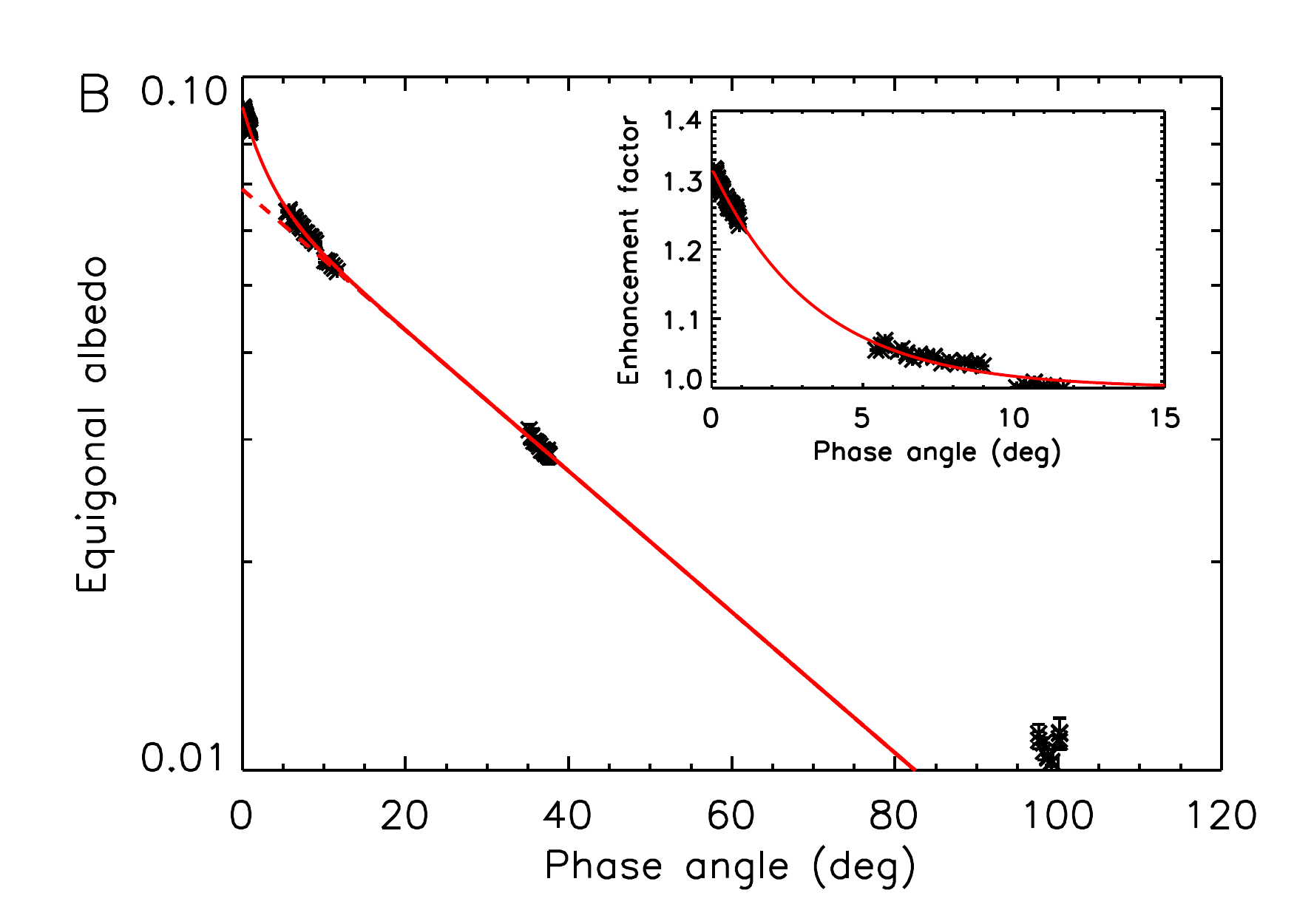}
\includegraphics[width=8cm]{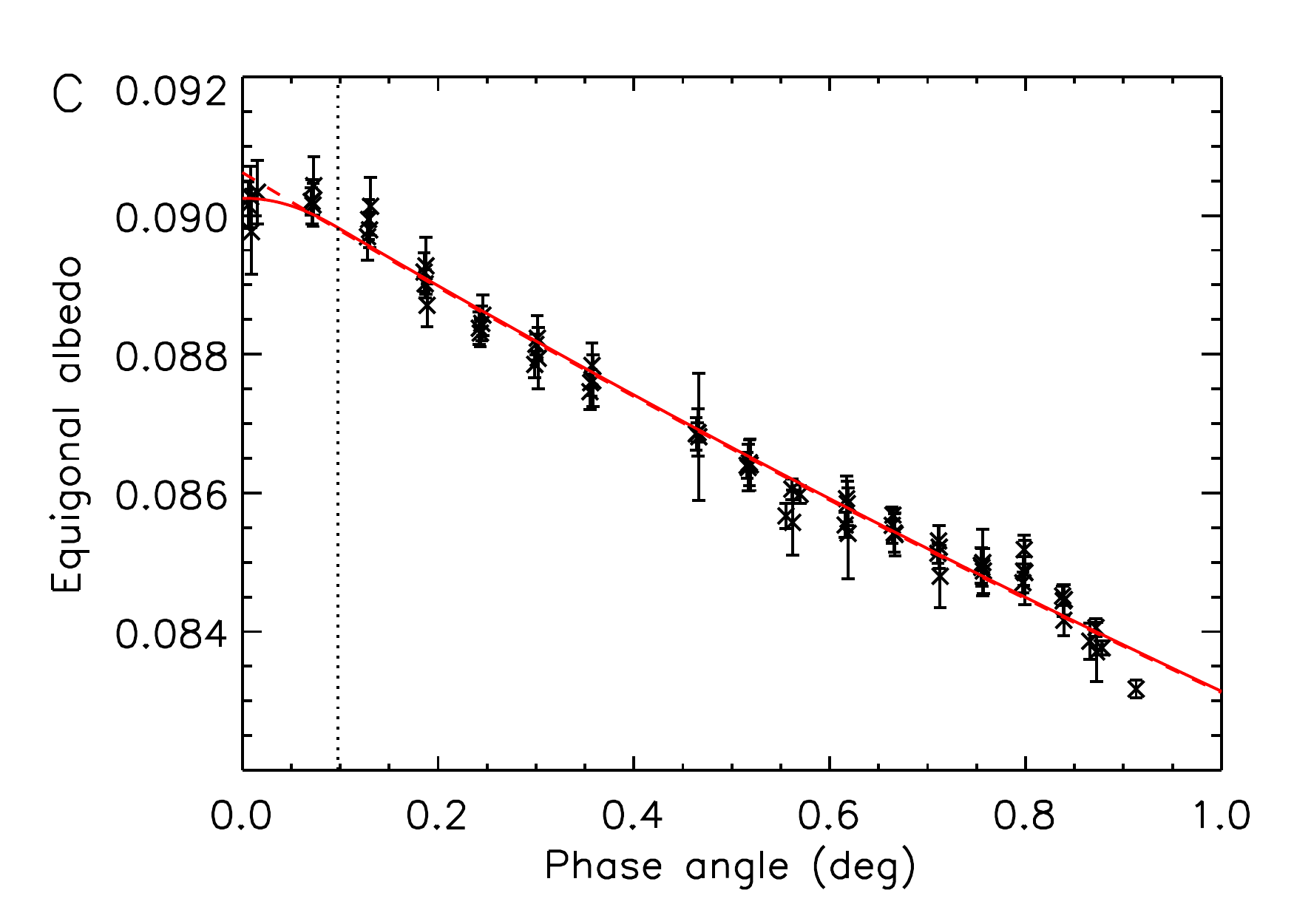}
\includegraphics[width=8cm]{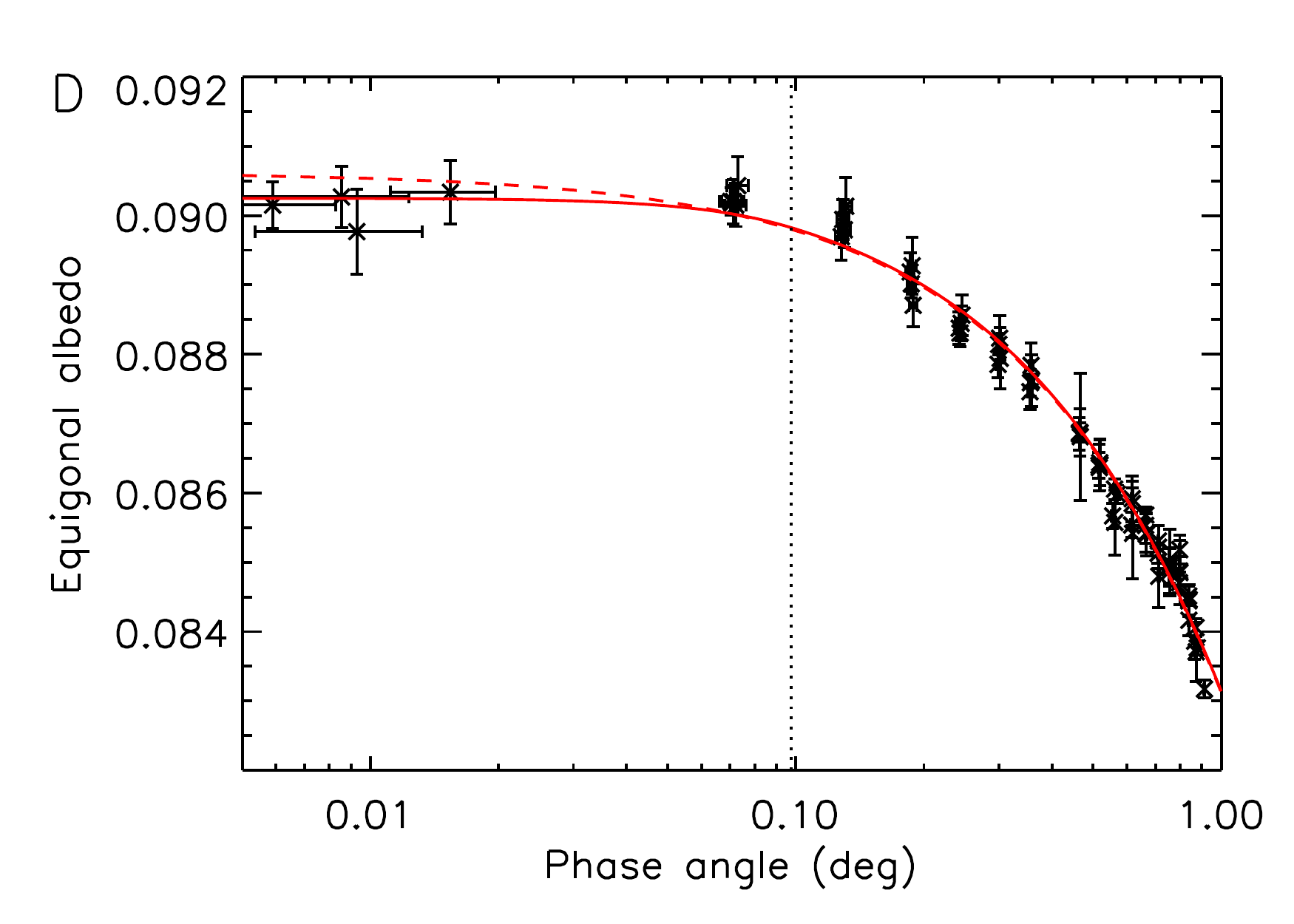}
\caption{Fitting the Akimov model to the average reflectance in a  box of $3 \times 3$ pixels centered on the location of the green star in Fig.~\ref{fig:ROI}A. The fit, with parameters $(A_0, \nu_1, m, \nu_2) = (0.091, 0.023, 0.32, 0.32)$, was limited to data with $0.05^\circ < \alpha < 40^\circ$ and $\iota, \epsilon < 70^\circ$. \textbf{A}: Reflectance. \textbf{B}: Equigonal albedo (logarithmic scale), with the best-fit model from \textbf{A} drawn in red. The inset shows the enhancement factor, or the equigonal albedo divided by the ``linear'' part of the phase curve (dashed red line). \textbf{C}: Zooming in on the phase angle range below $1^\circ$. The solid curve accounts for the finite size of the Sun, whereas the dashed curve does not. The vertical dotted line indicates the angular radius of the Sun at Ceres. \textbf{D}: As in \textbf{C}, but with the phase angle on a logarithmic scale.}
\label{fig:Akimov_phase_curve}
\end{figure*}

We start our analysis by employing the Akimov model (Eqs.~\ref{eq:type_I}-\ref{eq:Akimov_disk_fie}). A known limitation of the double exponential version of this model is its failure to fit observations at high phase angles. \citet{V11} required a third exponential term to accurately model the lunar phase curve over a wide range of phase angles. We restrict the fit to $\alpha < 50^\circ$ and find that two terms suffice. We choose a location at the northern end of the path, indicated by the green star in Fig.~\ref{fig:ROI}A, because of the availability of observations at large phase angle in \textit{RC3} images. Figure~\ref{fig:Akimov_phase_curve}A shows how the reflectance at this location compares to a fit of the Akimov model. The observed reflectance is not a smooth function of phase angle because of variations in incidence and emission angles over the surface, for which we account by dividing out the disk function (Eq.~\ref{eq:Akimov_disk_fie}) to arrive at the equigonal albedo in Fig.~\ref{fig:Akimov_phase_curve}B. We display the equigonal albedo on a logarithmic scale to demonstrate the linear character of the phase curve and the departure from linearity towards zero phase that constitutes the OE \citep{G56}. The inset shows the enhancement factor, or the phase curve divided by the linear part, and we calculate $\zeta(0^\circ) = 1.32$ and \textsc{hwhm}~$= 2.3^\circ$. The Akimov model accurately describes the data at low to moderate phase angles ($\alpha < 40^\circ$). The phase angle coverage below $1^\circ$ is particularly dense, and we zoom in on this region in Fig.~\ref{fig:Akimov_phase_curve}C. The reflectance is well reproduced and reaches a plateau below $0.05^\circ$ phase, likely due to the finite size of the Sun \citep{S99,D12}. To model this phenomenon, we follow \citet{D12} and adopt the empirical limb darkening model from \citet{HM98}:
\begin{equation}
I_\lambda(\hat{r}) = (1 - \hat{r}^2)^{\alpha_\lambda / 2},
\label{eq:limb_darkening}
\end{equation}
with $\hat{r}$ the normalized solar radius. Limb darkening parameter $\alpha_\lambda$ depends on the wavelength, but the F1 filter is broadband. Whereas the CCD detector is most sensitive around 700~nm \citep{Si11}, the Sun is brightest around 500~nm. We therefore adopt $\alpha_{\rm F1} = 0.5$, valid around 570~nm\footnote{\citet{D12} reported using $\beta_\lambda = \alpha_\lambda$ instead of $\alpha_\lambda / 2$ for the exponent in Eq.~\ref{eq:limb_darkening}, which is not consistent with \citet{HM98}.}. The angular size of the Sun at the distance of Ceres during \textit{XMO4} (2.74~AU) is $0.195^\circ$. Figures~\ref{fig:Akimov_phase_curve}C and D show that the finite size of the Sun has a small but significant effect on the phase curve for $\alpha < 0.05^\circ$. We therefore exclude this range from model fits to the data in the remainder of this paper. Coherent backscatter may also cause rounding of OE peak towards zero phase \citep{MD93}, but we do not have the data to distinguish such an effect from rounding due to the finite size of the Sun.

\subsubsection{Hapke model}

\begin{figure*}
\centering
\includegraphics[width=8cm,angle=0]{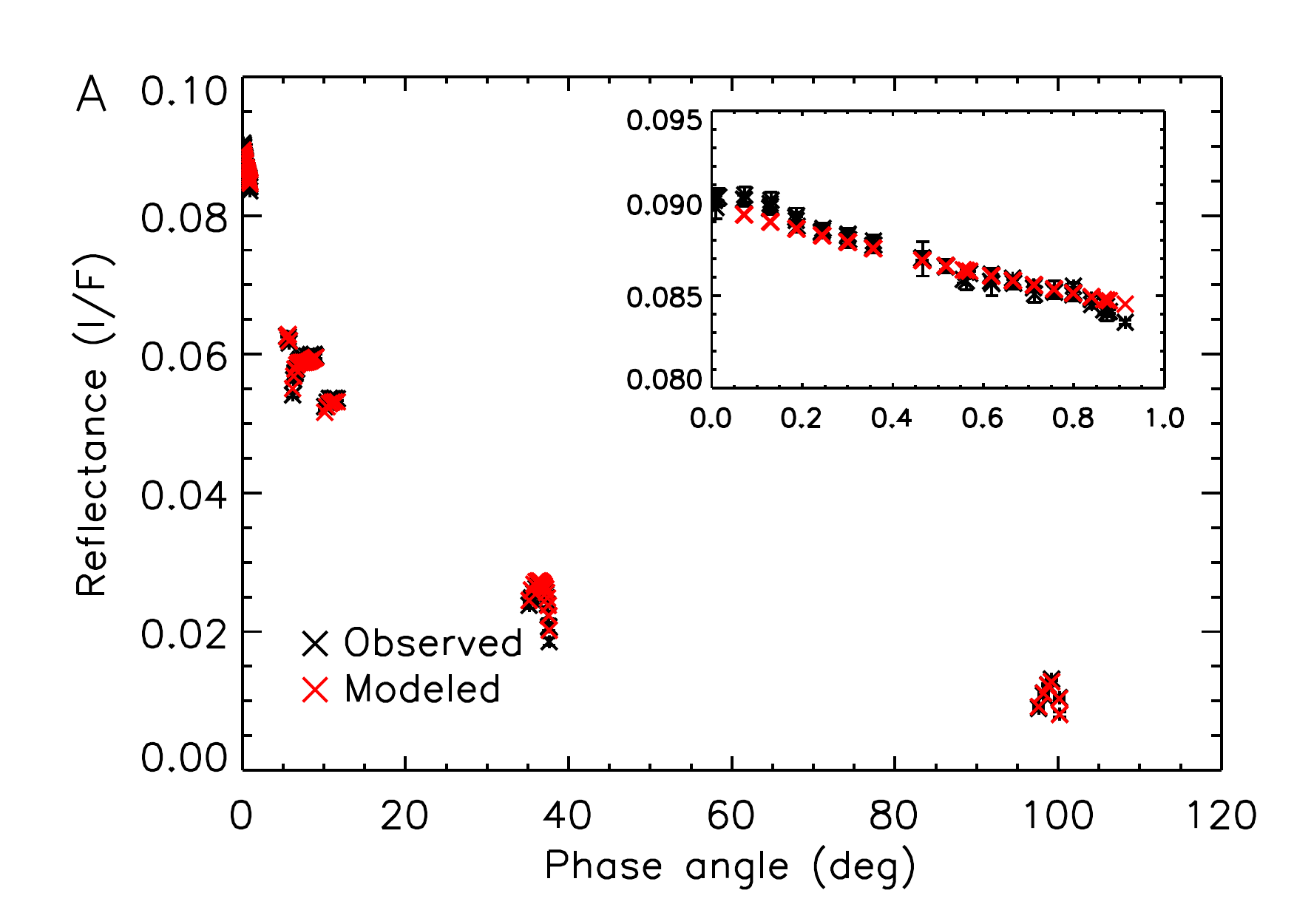}
\includegraphics[width=8cm,angle=0]{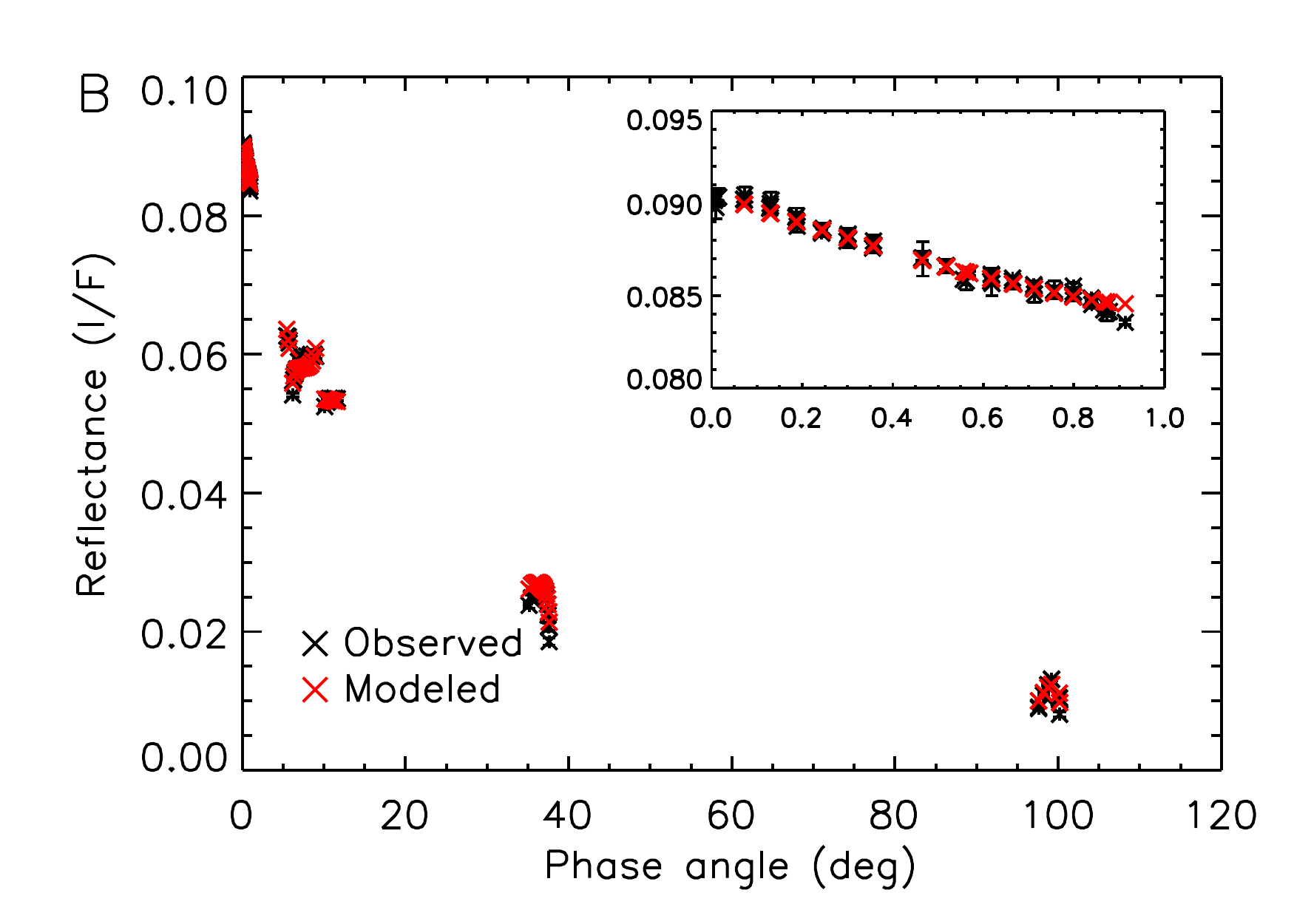}
\includegraphics[width=8cm,angle=0]{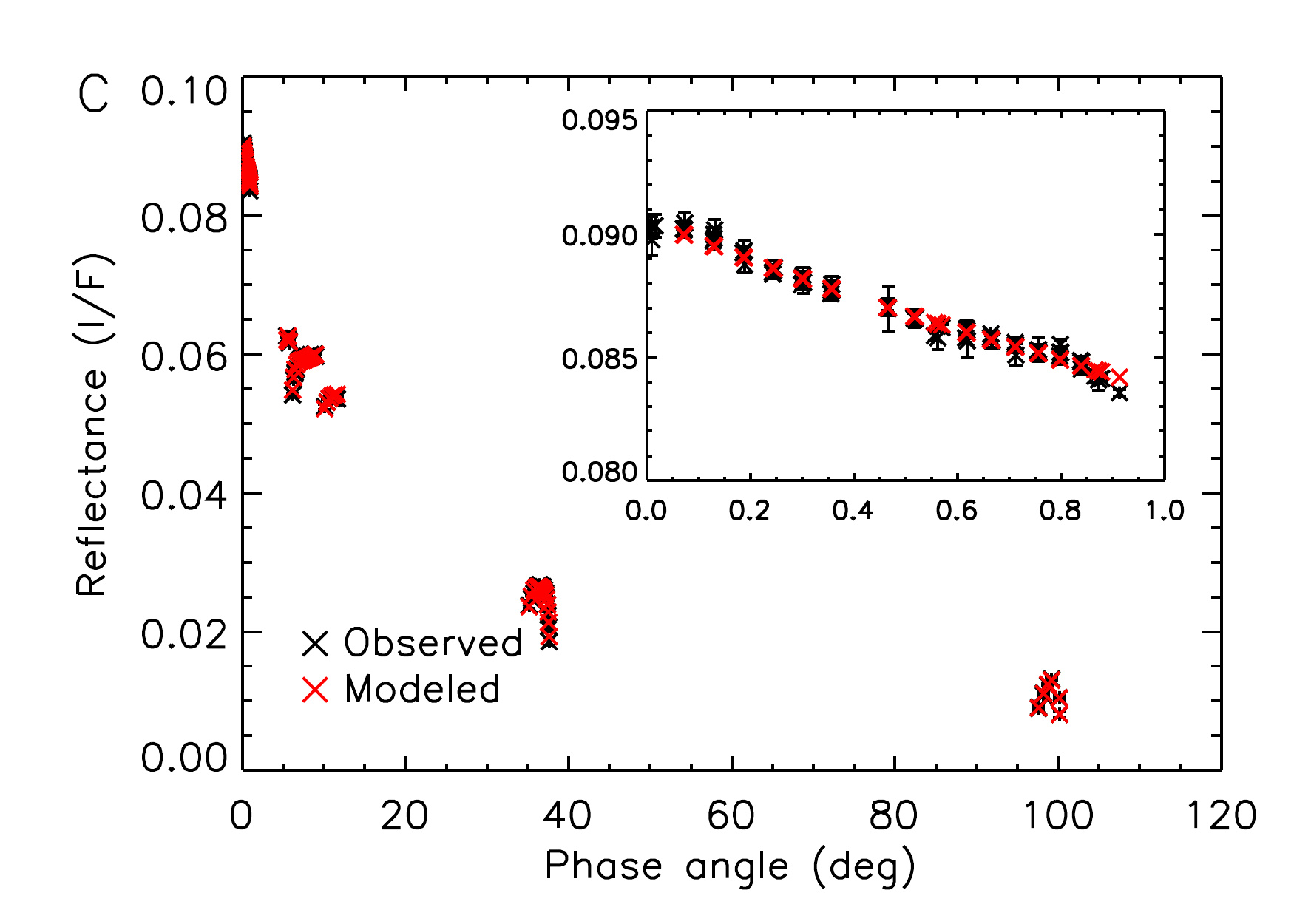}
\includegraphics[width=8cm,angle=0]{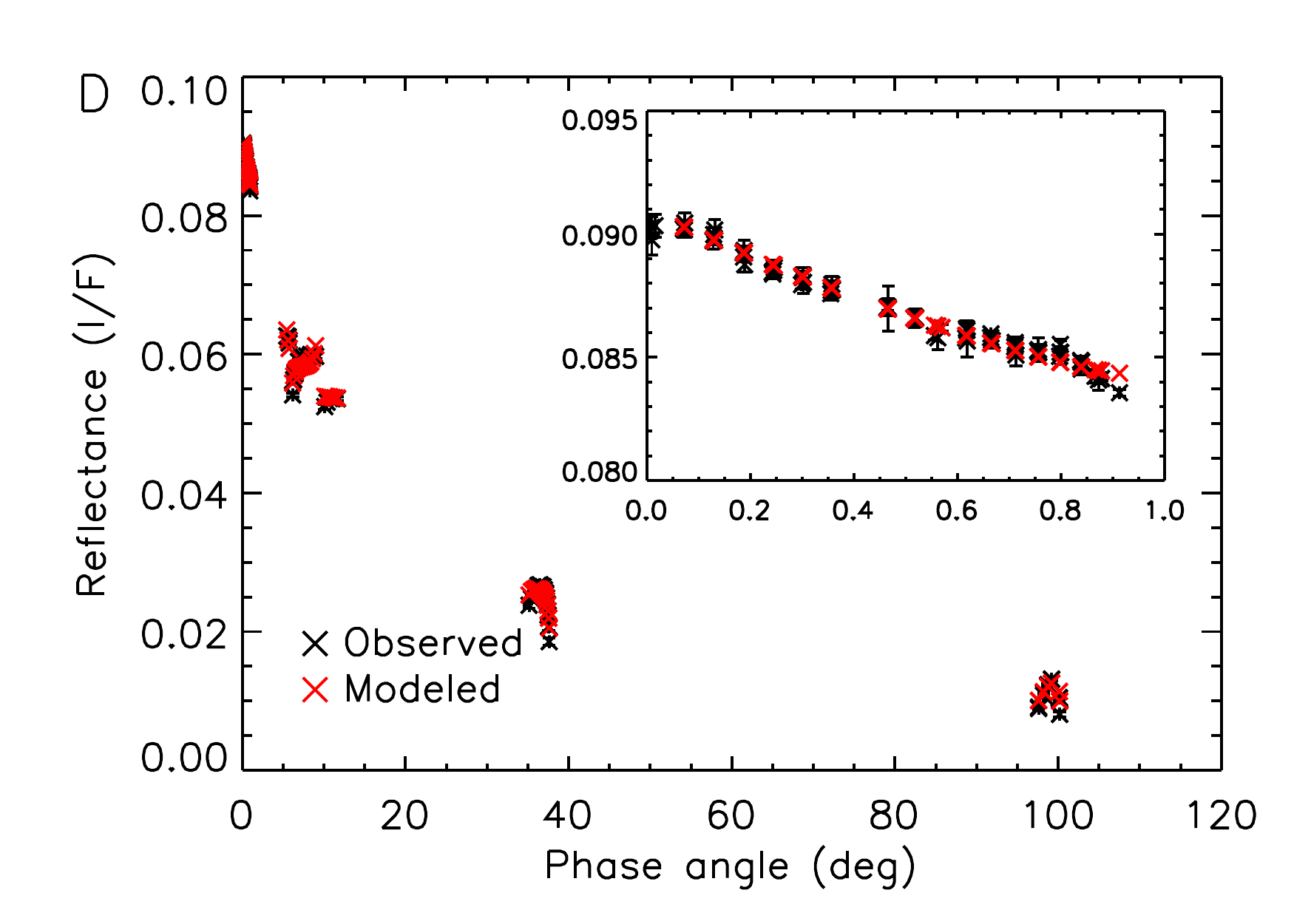}
\caption{Fitting the Hapke model to the average reflectance in a box of $3 \times 3$ pixels  centered on the location of the green star in Fig.~\ref{fig:albedo_map}, with $\alpha > 0.05^\circ$ and $\iota, \epsilon < 70^\circ$. The inset zooms in on the phase angle range below $1^\circ$. These four solutions, with parameters listed in Table~\ref{tab:Hapke_fit_coef}, were identified by the Levenberg-Marquardt optimization algorithm. \textbf{A} and \textbf{B} use the single-term Henyey-Greenstein function, whereas \textbf{C} and \textbf{D} use the double-term Henyey-Greenstein function. The search start value for the photometric roughness was $\bar{\theta}_{\rm start} = 25^\circ$ for \textbf{A} and \textbf{C}, whereas it was $\bar{\theta}_{\rm start} = 1^\circ$ for \textbf{B} and \textbf{D}.}
\label{fig:Hapke_phase_curve}
\end{figure*}

Next, we fit the \citet{H81,H84,H86} model to the same data, where we employ both the single and double term Henyey-Greenstein phase functions. As the double term version offers flexibility to shape the backscattering lobe of the particle phase function, we expect it to fit the data better. As data with phase angles $> 90^\circ$ are available, we can reliably derive the roughness parameter $\bar{\theta}$ \citep{H88}. Figure~\ref{fig:Hapke_phase_curve} shows the best-fit Hapke models, with the corresponding parameters listed in Table~\ref{tab:Hapke_fit_coef}. The best-fitting model overall is the double Henyey-Greenstein version with $\bar{\theta} = 23^\circ$ (model~C in Table~\ref{tab:Hapke_fit_coef} and Fig.~\ref{fig:Hapke_phase_curve}). In our trials, we found that the search algorithm converged to a local rather than the global minimum depending on the start value for the photometric roughness parameter ($\bar{\theta}_{\rm start}$). For $\bar{\theta}_{\rm start} = 0^\circ$, the algorithm would converge to $\bar{\theta} = 1^\circ$, whereas for $\bar{\theta}_{\rm start} = 25^\circ$, the algorithm would converge to values around $\bar{\theta} = 23^\circ$. The change in fit quality due to this difference in $\bar{\theta}$ is partly compensated by the Henyey-Greenstein parameters. Nevertheless, the high-$\bar{\theta}$ models fit the data better than the low-$\bar{\theta}$ models, regardless of whether the single or double-term Henyey-Greenstein version is used. It is tempting to conclude that $\bar{\theta} = 23^\circ$ is the ``correct'' Hapke roughness value for these data. This would be true if the data were 100\% reliable. But apart from known error sources such as photon noise, the data are affected by projection errors, which are very hard to quantify. Therefore, we may expect this kind of ambiguity when extending the Hapke model fit to other parts of the surface. Our analysis concerns primarily the OE parameters, and both $B_{\rm S0}$ and $h_{\rm S}$ show considerable variation in Table~\ref{tab:Hapke_fit_coef}. Moreover, the model that best fits the data in the OE range is actually model~D, which comes only third out of four in the ranking of overall fit quality.

We have shown the Ceres phase curve, including the OE, for one particular location on the surface, and successfully reproduced it with both the Akimov and Hapke models. The reflectance data are densely distributed in regular fashion over the lowest degree of phase angle. At larger phase angles, the data are mostly distributed in widely separated clumps. The distribution of data along phase angle varies over the surface. When we extend the analysis to the entirety of ROI~1, we may expect to see consequences for the model parameters in the form of artifacts \citep{S17}. The resulting ambiguities will be larger for the Hapke model, because it has more free parameters.

\begin{table*}
\centering
\caption{Best-fit parameters of the Hapke models in Fig.~\ref{fig:Hapke_phase_curve}. For cases \textbf{A} and \textbf{B} the single-term Henyey-Greenstein function was used ($c = 1$), for \textbf{C} and \textbf{D} the double-term version. $\bar{\theta}_{\rm start}$ is the photometric roughness start value provided to the optimization algorithm and $\chi^2$ is a measure of the goodness-of-fit (lower is better), with 120 degrees of freedom.}
\begin{tabular}{lllllllll}
\hline\hline
Case & $w$ & $B_{{\rm S}0}$ & $h_{\rm S}$ & $\bar{\theta}$ & $b$ & $c$ & $\bar{\theta}_{\rm start}$ & $\chi^2$ \\
\hline
A & $0.087$ & $3.1$ & $0.081$ & $22^\circ$ & $-0.22$ & $1.00$ & $25^\circ$ & 428 \\
B & $0.091$ & $1.8$ & $0.056$ & $0^\circ$ & $-0.31$ & $1.00$ & $1^\circ$ & 750 \\
C & $0.116$ & $1.6$ & $0.054$ & $23^\circ$ & $0.38$ & $-0.29$ & $25^\circ$ & 165 \\
D & $0.100$ & $1.4$ & $0.046$ & $0^\circ$ & $0.38$ & $-0.62$ & $1^\circ$ & 603 \\
\hline
\end{tabular}
\label{tab:Hapke_fit_coef}
\end{table*}

\subsection{Spatial photometric variations}
\label{sec:spatial_variations}

Our next goal is to widen the photometric analysis to the entire surface in ROI~1 to identify variations in the phase curve, in particular those of the OE amplitude and slope, and correlate these to physical properties of the surface. Again, we distinguish between the Akimov and Hapke models.

\subsubsection{Akimov model}

\begin{figure*}
\centering
\includegraphics[width=8cm]{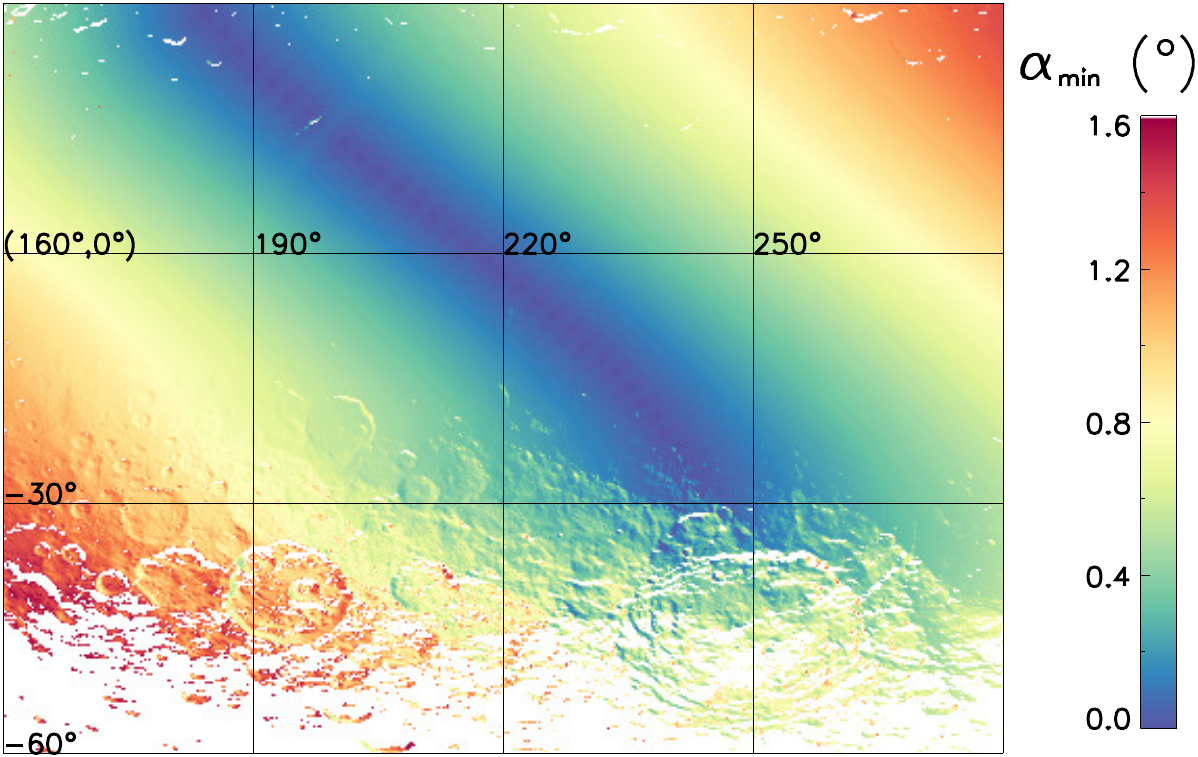}
\includegraphics[width=8cm]{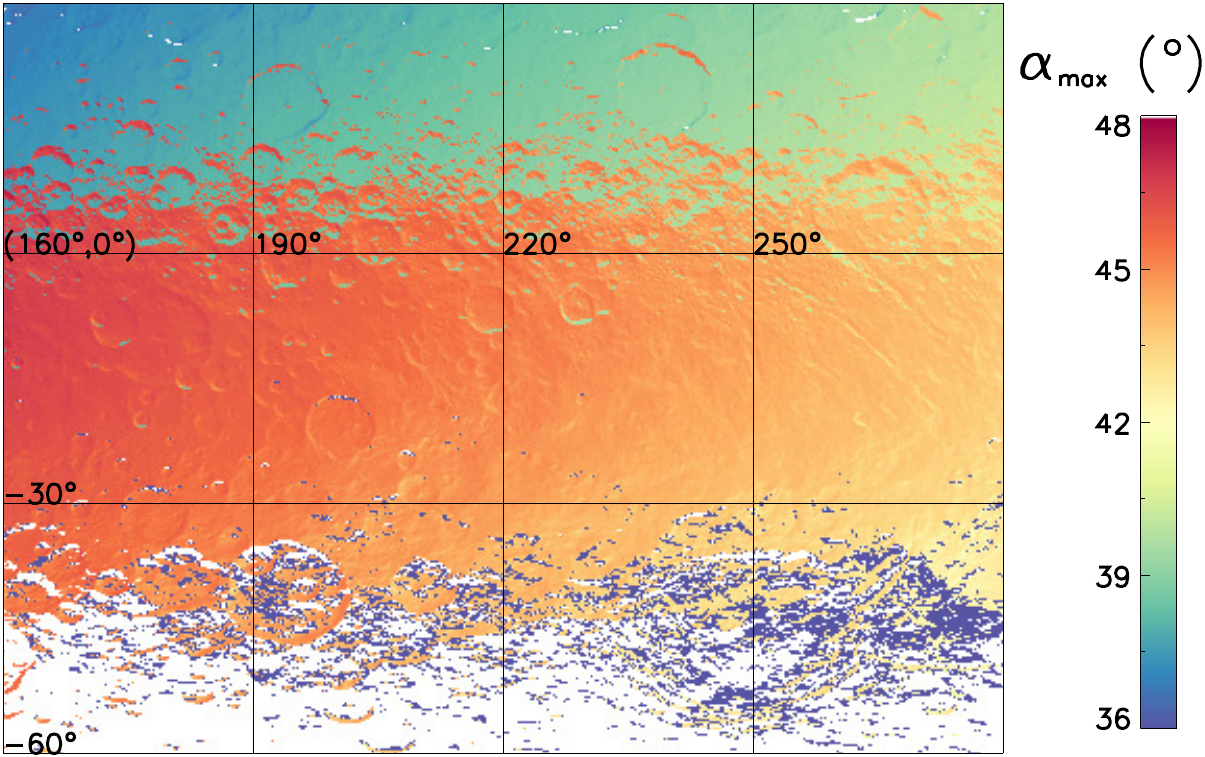}
\caption{The minimum (\textbf{left}) and maximum (\textbf{right}) phase angle of observation in ROI~1 for $\iota, \epsilon < 50^\circ$.}
\label{fig:phase_angle_coverage}
\end{figure*}

\begin{figure*}
\centering
\includegraphics[width=8cm]{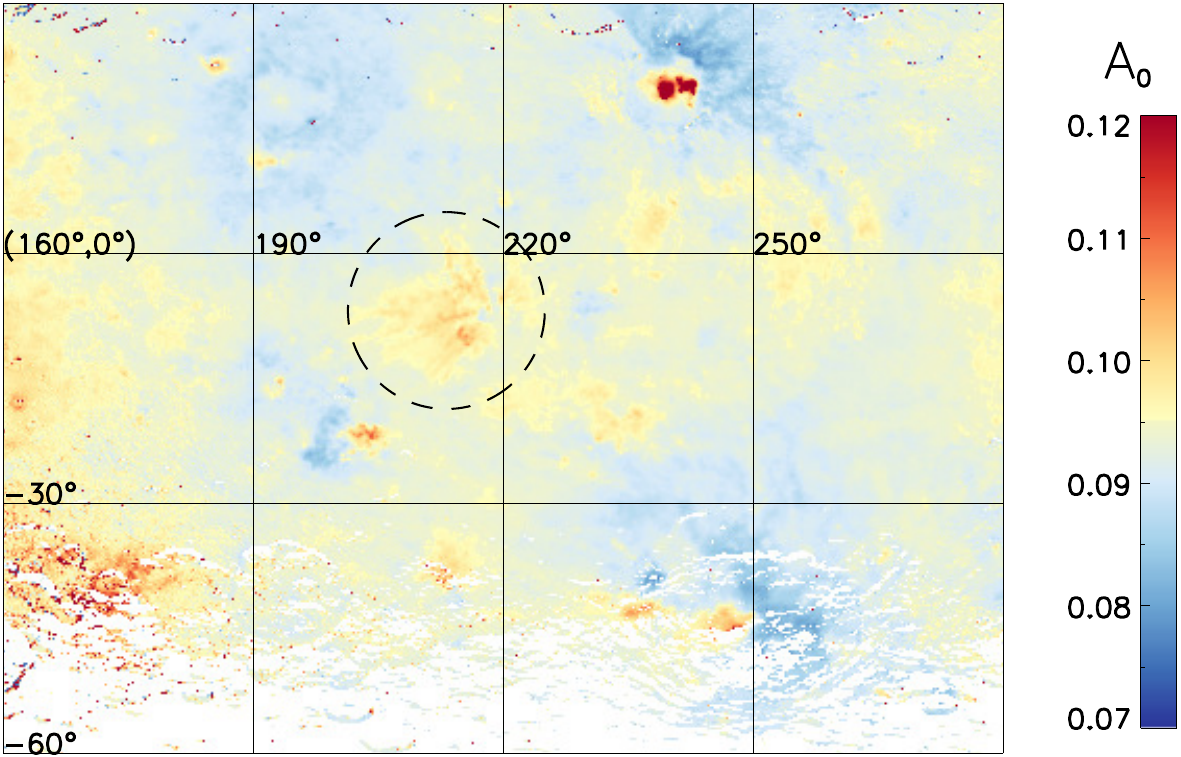}
\includegraphics[width=8cm]{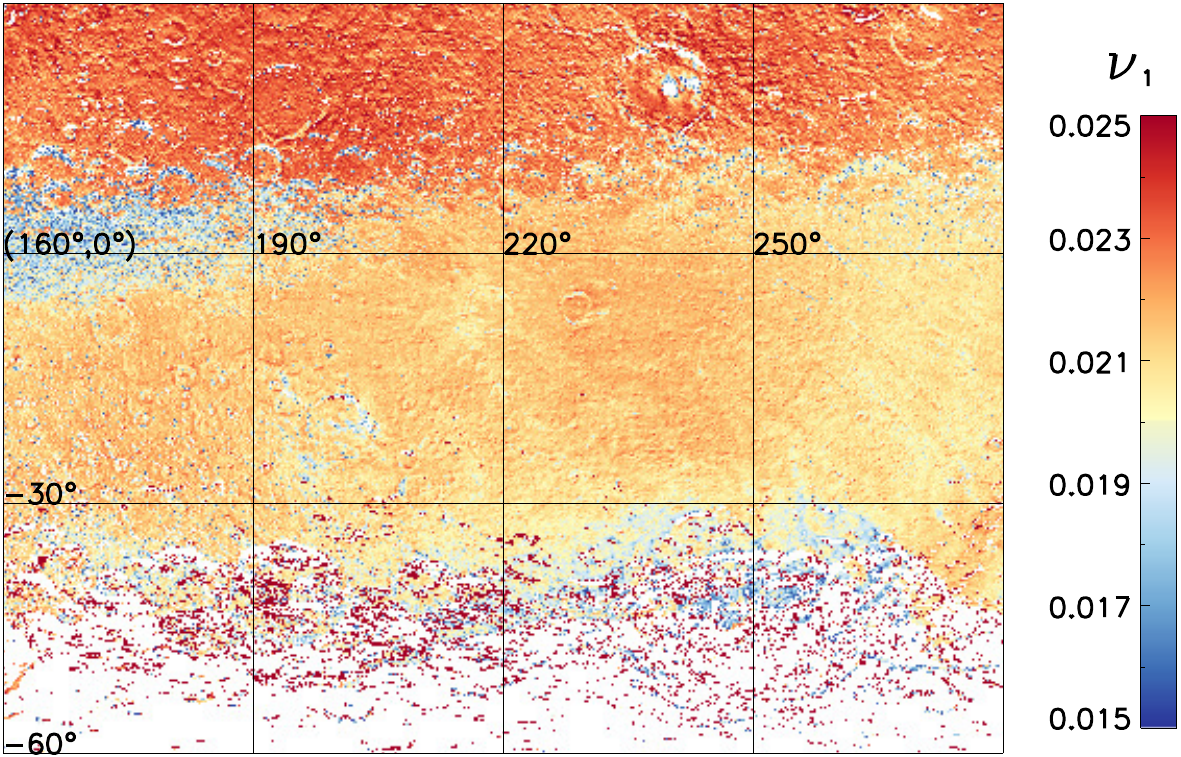}
\includegraphics[width=8cm]{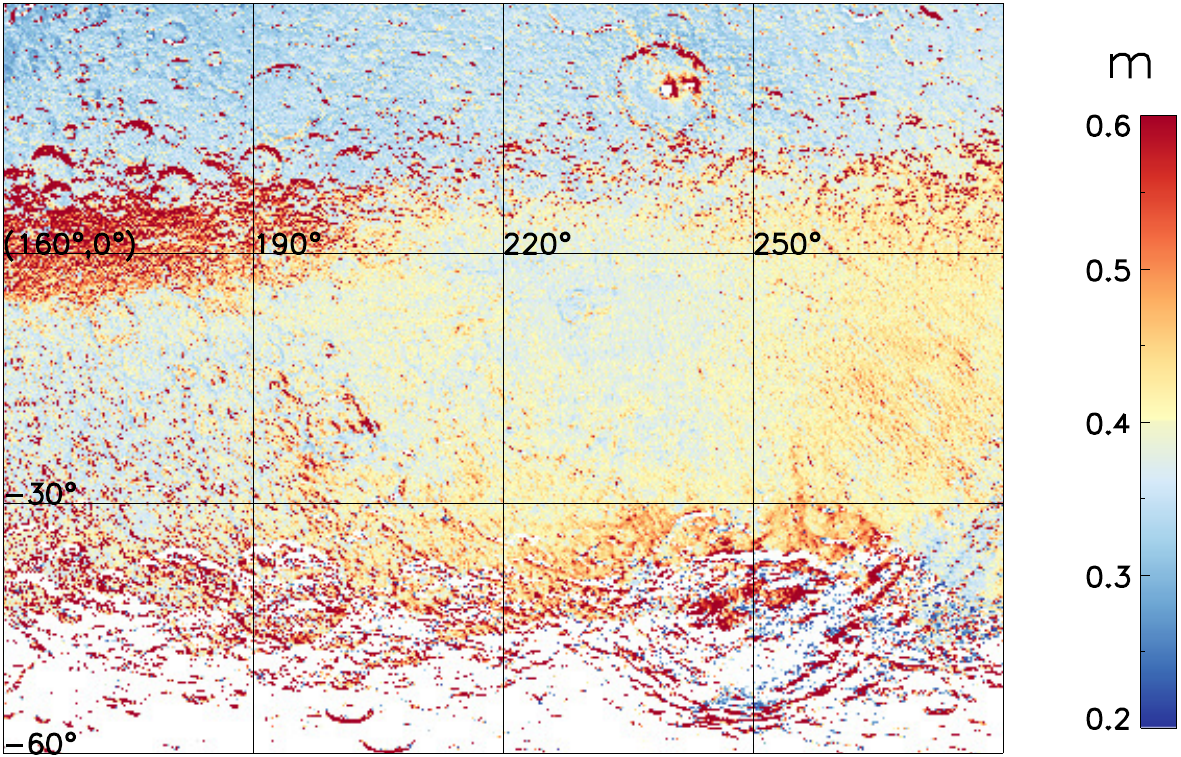}
\includegraphics[width=8cm]{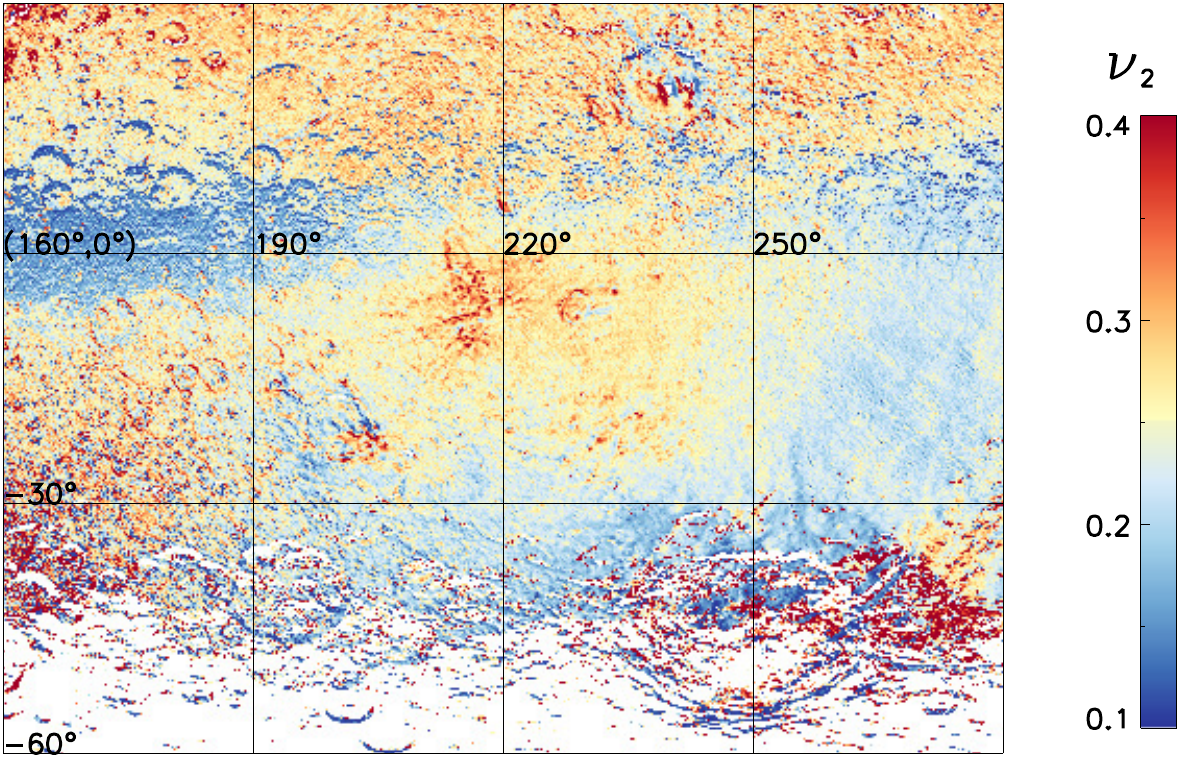}
\caption{Maps of the Akimov model parameters (Eqs.~\ref{eq:equigonal_albedo} and \ref{eq:Akimov_phase_fie}) in ROI~1, with $\iota, \epsilon < 50^\circ$. The Azacca ejecta are encircled in the $A_0$ map.}
\label{fig:Akimov_parameter_maps}
\end{figure*}

First we employ the Akimov model, for which we map the distribution of the $A_0$, $\nu_1$, $\nu_2$, and $m$ parameters. This means that for each projected pixel in the ROI we fit the model to the observations that include that pixel. As the model fails at very large phase angles, we exclude data from \textit{RC3}a and b (see Table~\ref{tab:image_data}) by restricting the maximum phase angle to $48^\circ$. The minimum and maximum phase angles of observation are shown in Fig.~\ref{fig:phase_angle_coverage}. In the map of the minimum phase angle we can clearly trace the path of zero phase diagonally across the center of the ROI. At the bottom of the ROI we find more extreme imaging geometries towards the limb of Ceres. Furthermore, the maximum phase angle shows a discontinuity around $+10^\circ$ latitude. As the model solutions are sensitive to the imaging geometry, we initially restrict the incidence and emission angles to $< 50^\circ$. Figure~\ref{fig:Akimov_parameter_maps} shows maps of the model parameters. The normal albedo ($A_0$) is reliably retrieved with little noise, and agrees nicely with the map in Fig.~\ref{fig:albedo_map}. The maps of the other parameters are much noisier and show obvious correlations with the maximum phase angle. These correlations are so strong that we can only evaluate any spatial variations within the ($-35^\circ, +5^\circ$) latitude band. Subtle variations are associated with the relatively bright ejecta of Azacca crater. The slope of the phase curve ($\nu_1$) is anti-correlated with the normal albedo, consistent with \citet{S17} and the general trend observed for asteroids, which is linked to the role of multiple scattering in the regolith \citep{Lo16}. On the other hand, the OE slope ($\nu_2$) appears to be correlated with the normal albedo, with the Azacca ejecta clearly recognizable in the $\nu_2$ map (but not the $m$ map).

\begin{figure*}
\centering
\includegraphics[width=8cm]{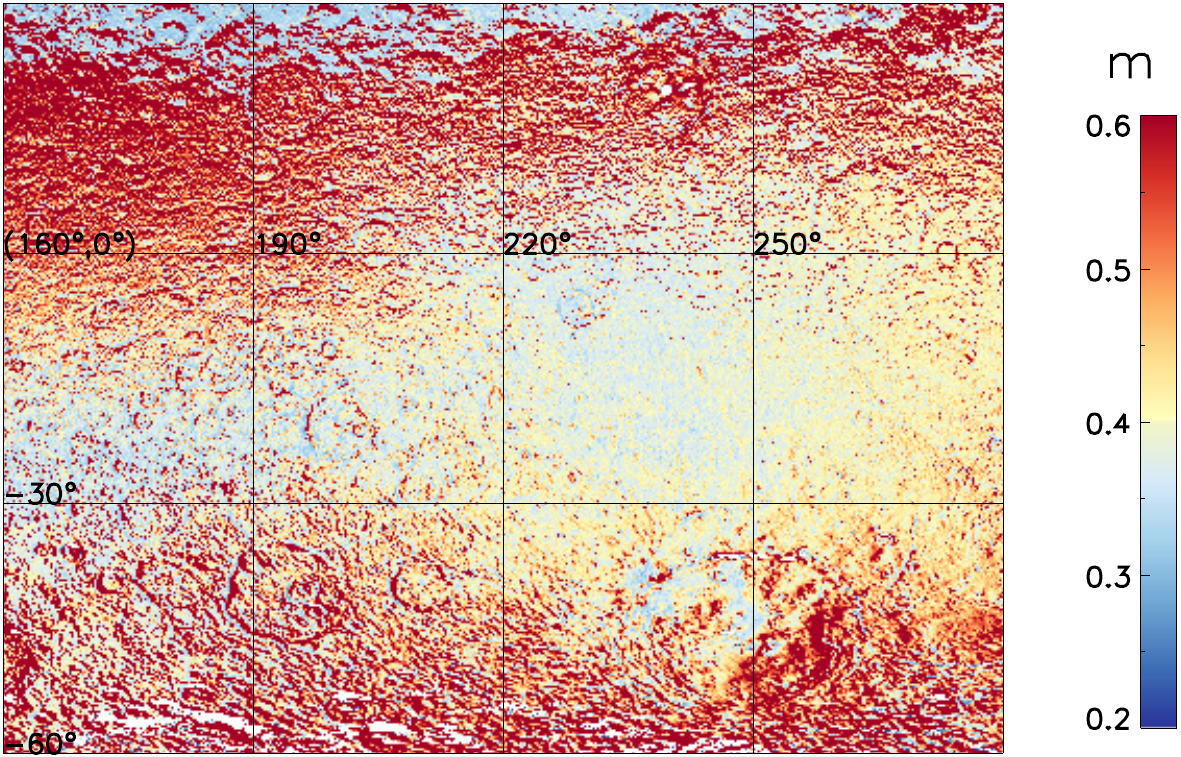}
\includegraphics[width=8cm]{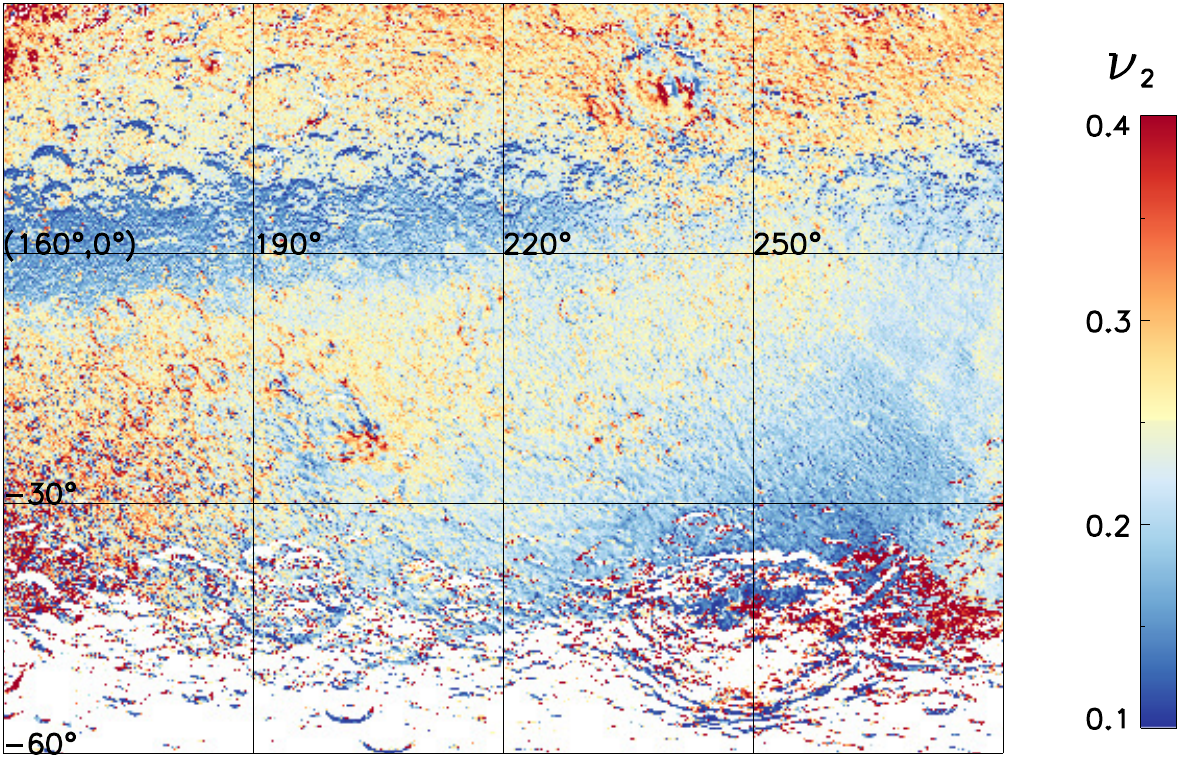}
\caption{Sensitivity analysis of the Akimov model parameter retrieval. \textbf{Left}: Map of $m$ with all phase angles included and $\iota, \epsilon < 70^\circ$. \textbf{Right}: Map of $\nu_2$ with phase angle restricted to $\alpha > 0.5^\circ$ and $\iota, \epsilon < 50^\circ$. See text for details.}
\label{fig:sensitivity}
\end{figure*}

\begin{figure*}
\centering
\includegraphics[width=8cm]{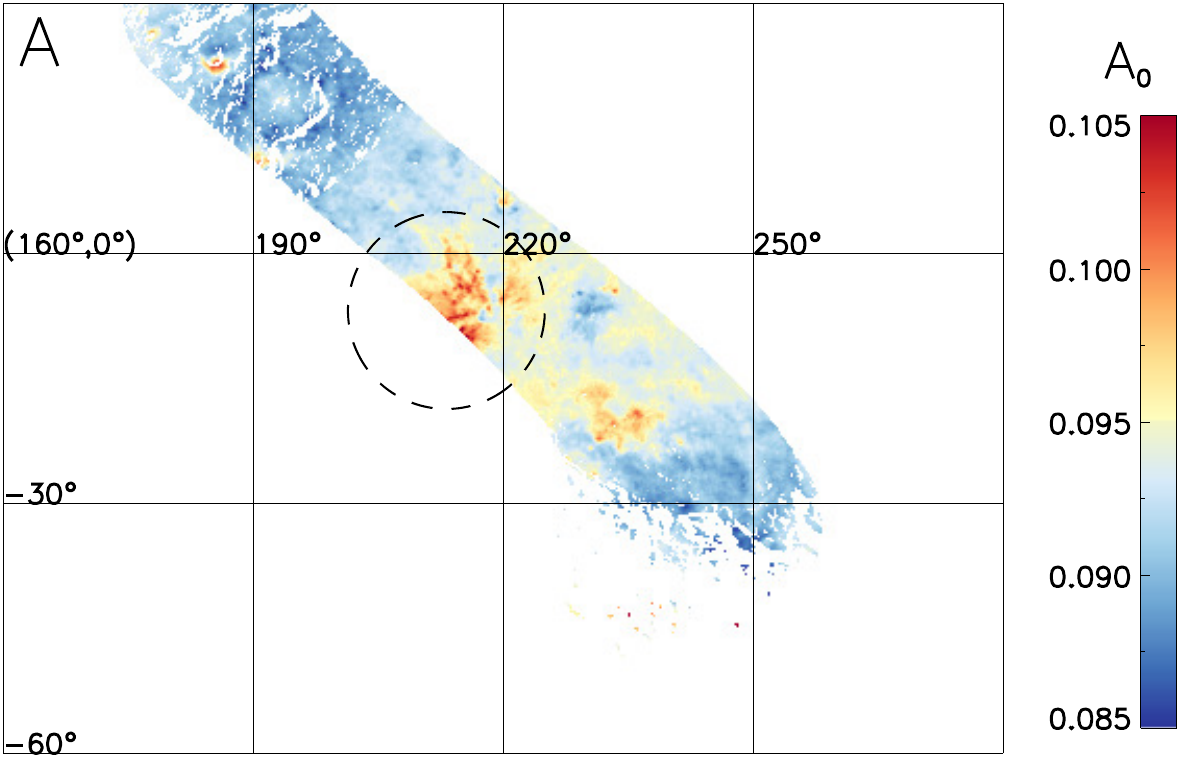}
\includegraphics[width=8cm]{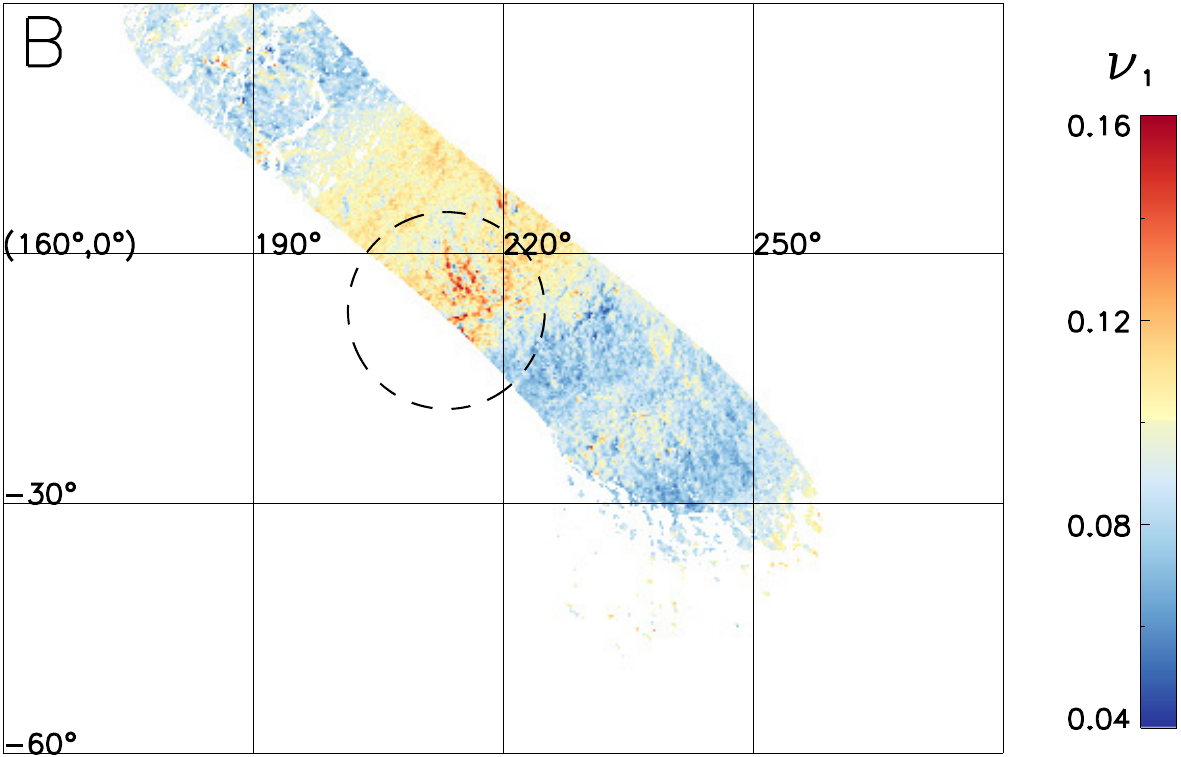}
\includegraphics[width=8cm]{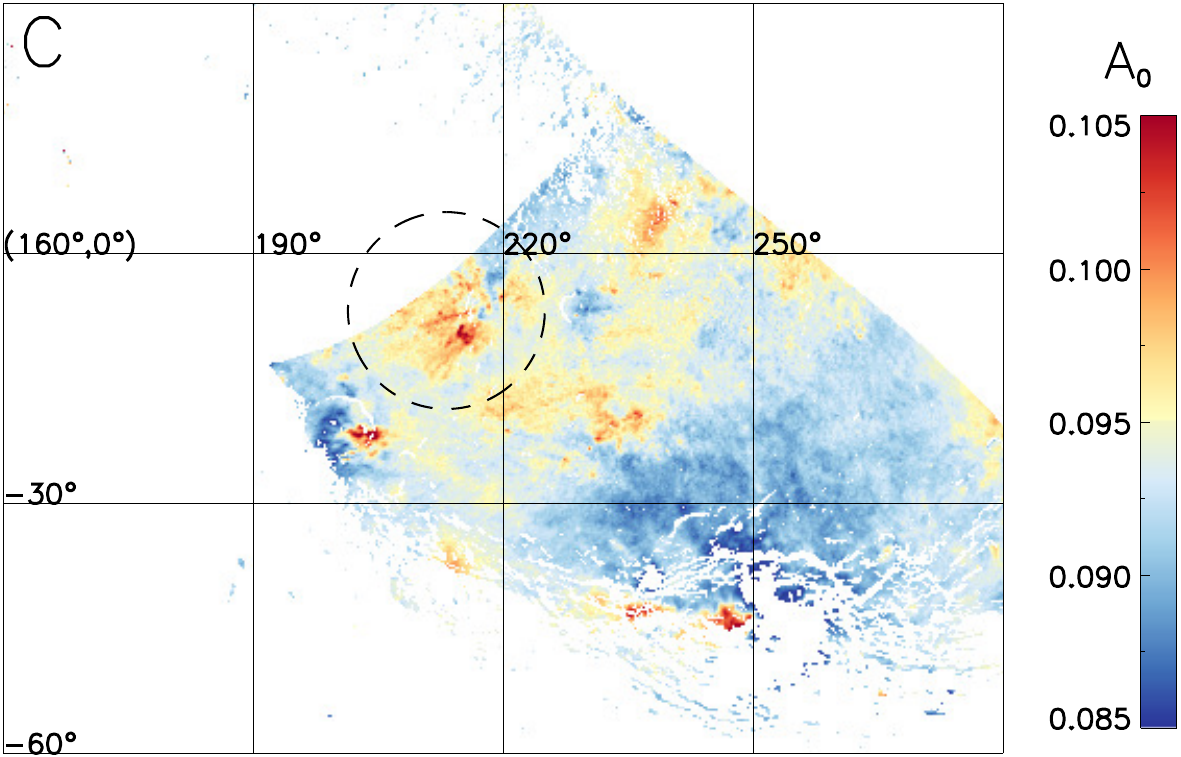}
\includegraphics[width=8cm]{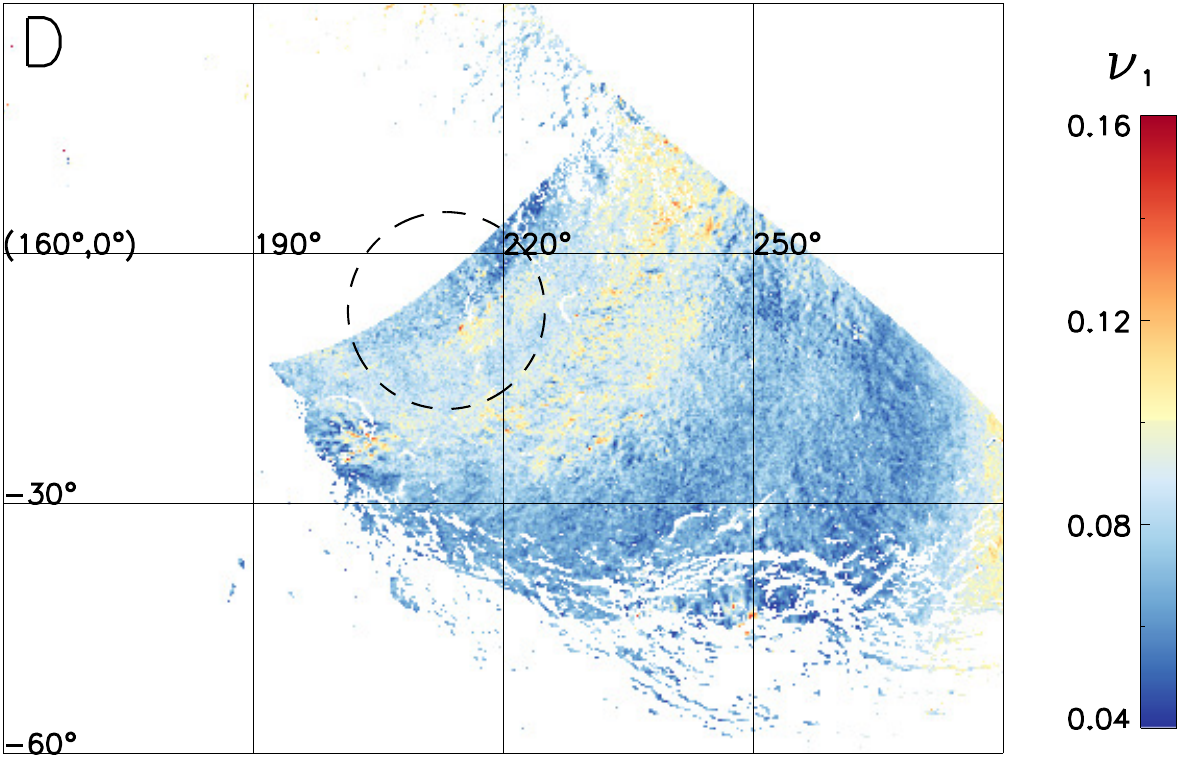}
\caption{Maps of the OE parameters (Eq.~\ref{eq:Akimov_phase_fie} with $m = 0$, and $\nu_1$ the OE slope) for the clear filter, with $\iota, \epsilon < 50^\circ$. The phase angle was restricted to $0.2^\circ < \alpha < 0.6^\circ$ for {\bf A} and {\bf B}, and to $0.6^\circ < \alpha < 1.0^\circ$ for {\bf C} and {\bf D}. The Azacca ejecta are encircled.}
\label{fig:OE_parameters_clear}
\end{figure*}

We investigate the robustness of these findings by changing the maximum incidence and emission angle and minimum phase angle (Fig.~\ref{fig:sensitivity}). Including observations with incidence and emission angles up to $70^\circ$ introduces additional noise in the map of $m$, especially in the western part of the ROI, which justifies our earlier restriction of $\iota, \epsilon < 50^\circ$. The choice of minimum phase angle affects the visibility of the Azacca ejecta in the $\nu_2$ map. When we increase the minimum phase angle to $0.5^\circ$, the Azacca ejecta fade in with the background. In fact, we found the ejecta to \mbox{disappear} gradually when increasing the minimum phase angle in steps from $0.0^\circ$ to $0.6^\circ$. To investigate this disappearance in more detail, we fit single exponential functions (Eq.~\ref{eq:Akimov_phase_fie} with $m = 0$) to the data restricted to two $0.4^\circ$-wide phase angle ranges, and map the parameters. One of these ranges is immediately below $\alpha = 0.6^\circ$ and the other one above. We tried to ensure that these very narrow ranges were fully occupied with data (minimum 3 data points) by requiring the existence of at least one data point with a phase angle both below and above the range. In this way we hope to minimize bias resulting from uneven phase-angle coverage. The maps in Fig.~\ref{fig:OE_parameters_clear} show the results as tracks over the surface, whose width and length are governed by the lower and upper limit of the phase-angle range, respectively. Where they overlap, the normal albedo ($A_0$) tracks (A and C) have a similar appearance for both ranges, with the Azacca ejecta clearly standing out as bright. But whereas the ejecta are clearly seen in the slope ($\nu_1$) map of the lower phase-angle range (B), they are absent from the map associated with the higher phase-angle range (D). This confirms that, below $0.6^\circ$ phase angle, the OE of the Azacca ejecta is steeper, and has a larger amplitude, than that of their surroundings. As the ejecta are visible in both Fig.~\ref{fig:OE_parameters_clear}A and B, it is tempting to conclude that the OE slope variations are correlated with normal albedo. But the slope map shows some distinct differences with the albedo map. We therefore suspect that the Azacca ejecta have physical properties that lead to a steeper OE slope at very low phase.

\subsubsection{Hapke model}

\begin{figure*}
\centering
\includegraphics[width=8cm]{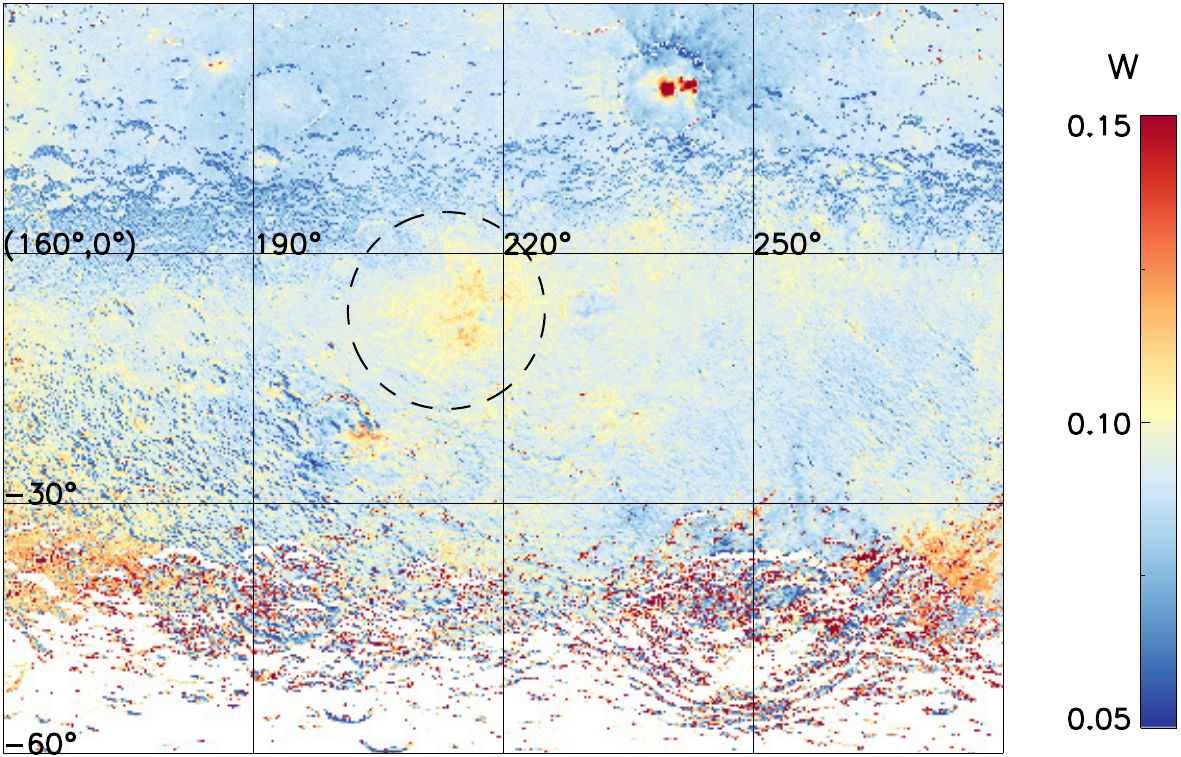}
\includegraphics[width=8cm]{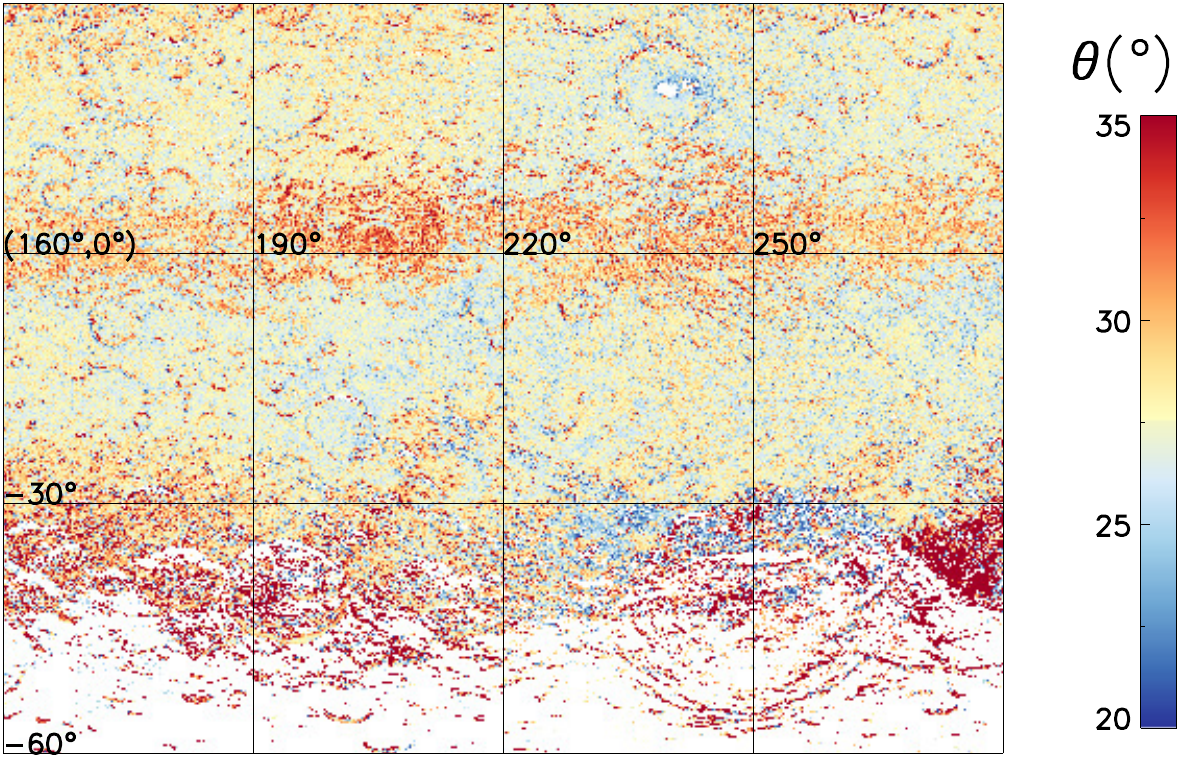}
\includegraphics[width=8cm]{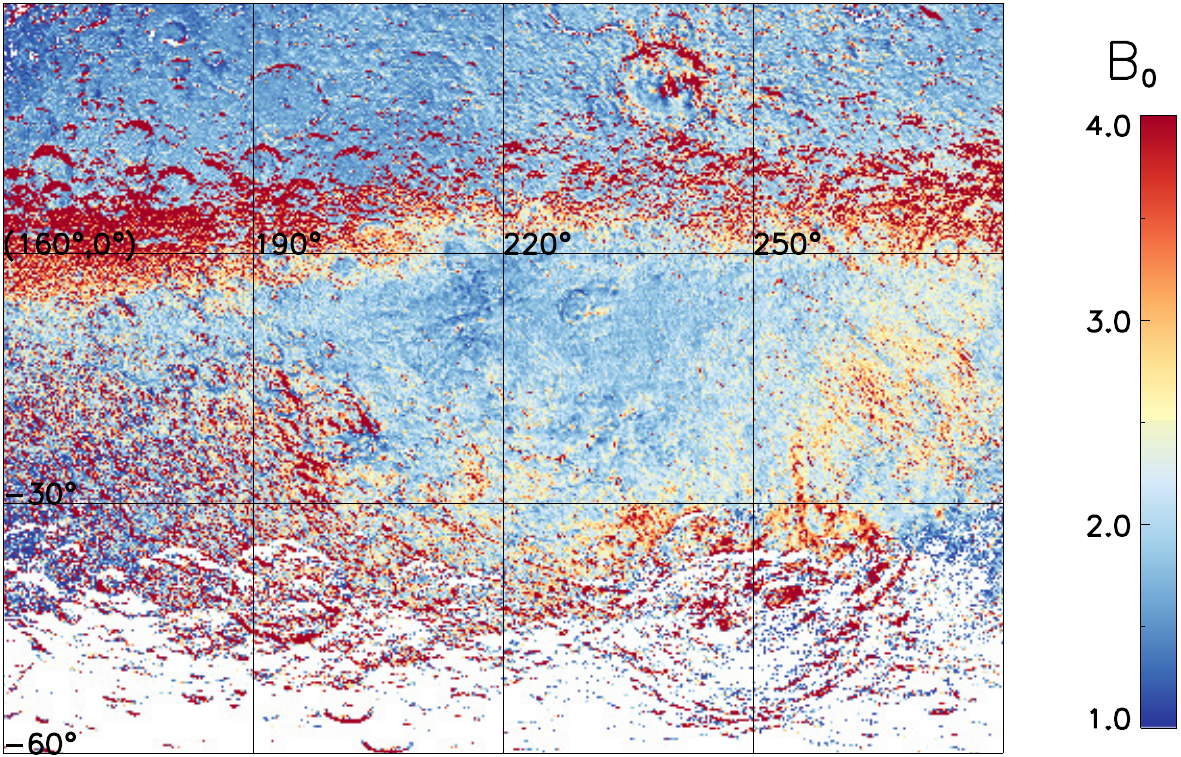}
\includegraphics[width=8cm]{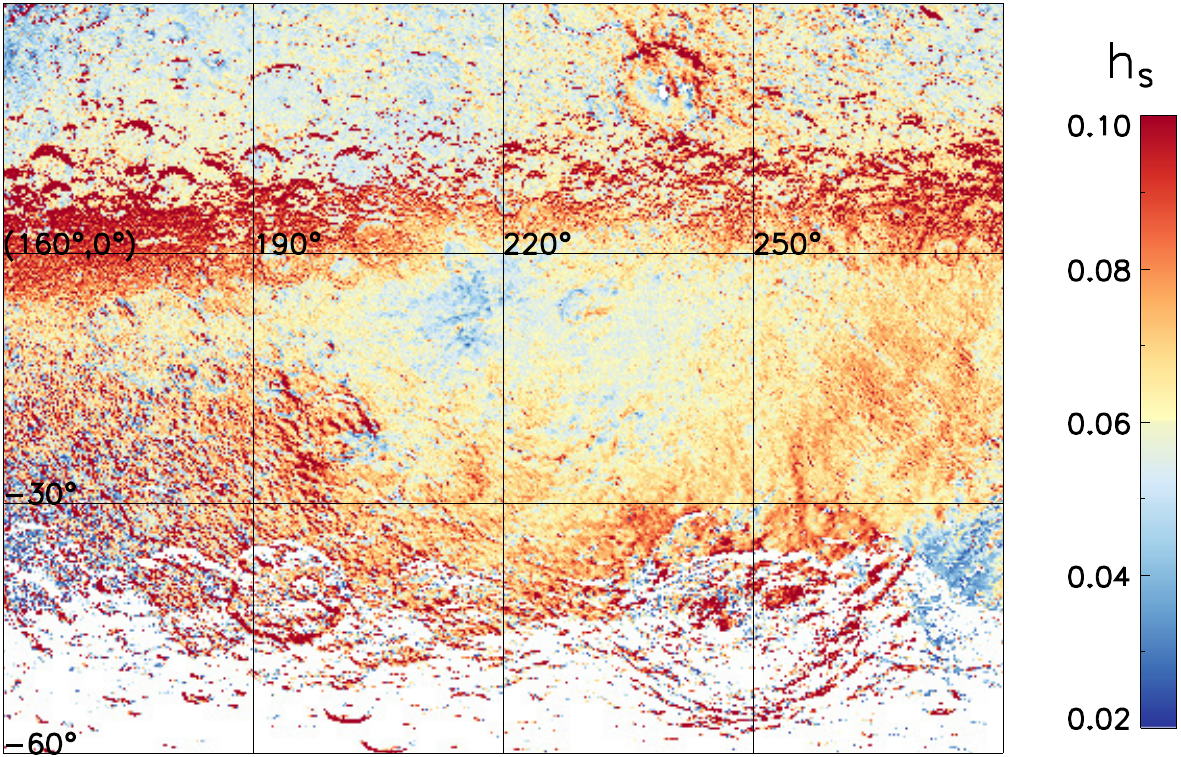}
\includegraphics[width=8cm]{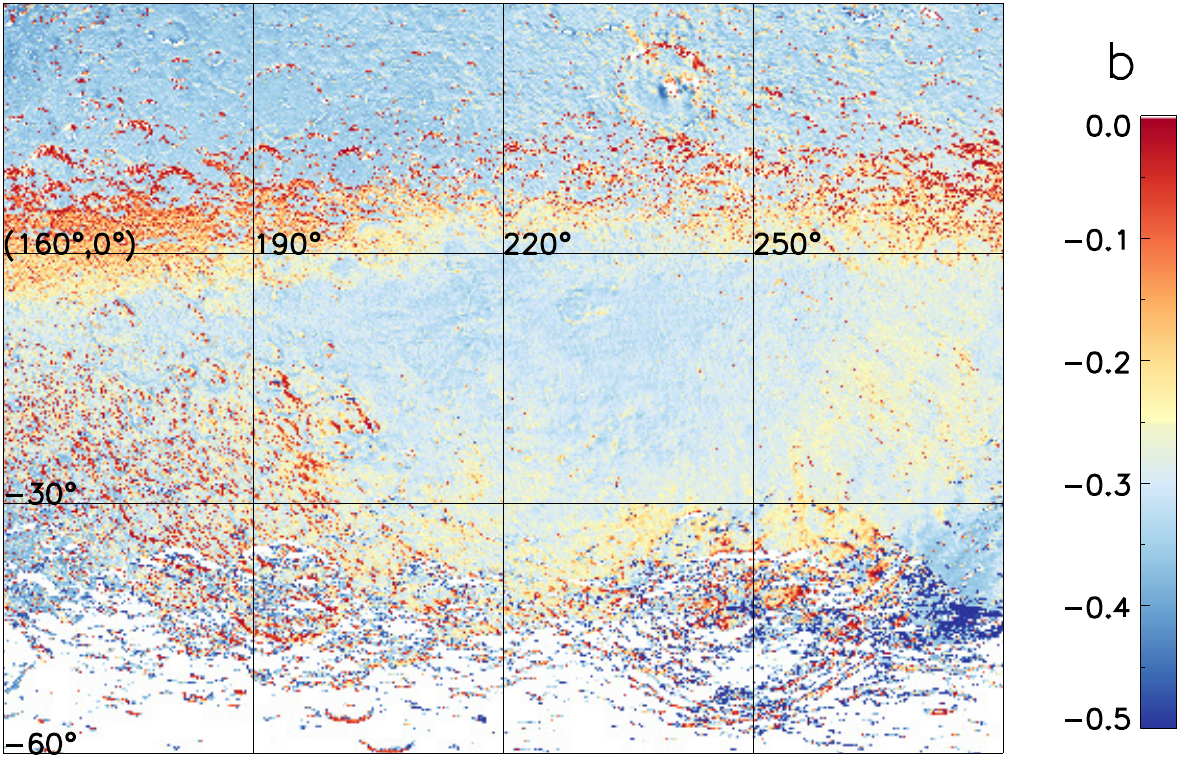}
\caption{Hapke model parameters with the single-term Henyey-Greenstein function and $\iota, \epsilon < 50^\circ$. The minimum and maximum phase angles are as in Fig.~\ref{fig:phase_angle_coverage}. The Azacca ejecta are encircled in the $w$ map.}
\label{fig:Hapke_maps}
\end{figure*}

In the same way, we employ the \citet{H81,H84,H86} model to map its parameters over ROI~1. Even though the Hapke model successfully models the very high phase-angle data, we again restricted the phase angle to roughly $48^\circ$, because we expect that the highly uneven coverage of the very high phase-angle observations (see Table~\ref{tab:image_data}) will systematically affect the parameters. With this restriction we cannot expect that the roughness parameter $\bar{\theta}$ is reliably derived \citep{H88}, but our objective is to study the OE rather than to find the best-fit Hapke parameters for the full phase curve. We trialed both the single and double-term Henyey-Greenstein phase functions, but found that in the double-term version, $w$ and $c$ varied considerably over the surface in a correlated fashion. We therefore proceeded with the single-term Henyey-Greenstein function. Figure~\ref{fig:Hapke_maps} shows the Hapke parameter maps. Again, the parameters exhibit considerable variation over the surface that is associated with unequal phase-angle coverage. South of latitude $-30^\circ$ the parameters are unreliable because of the more extreme illumination and observation geometries. ROI~2 appears to suffer the least artifacts. Here, $\bar{\theta}$ varies between $25^\circ$ and $30^\circ$, a range similar to values derived for global Ceres by \citet{L16} ($20^\circ \pm 3^\circ$), \citet{C17} ($29^\circ \pm 6^\circ$), and \citet{S17} ($22^\circ \pm 2^\circ$). The OE width $h_{\rm S}$ varies between 0.05 and 0.08, consistent with \citet{HV89} ($0.059 \pm 0.006$)\footnote{We quote this value with the caveat that \citet{R15} found Ceres' brightness predicted by the \citet{HV89} model to be off by about 20\% for unknown reasons.}, who modeled the \citet{T83} observations. \citet{R15} found a lower value of $h_{\rm S} = 0.036$, where we note that in both papers $\bar{\theta}$ was fixed at $20^\circ$. The OE amplitude $B_{\rm S0}$ is between 1.5 and 2.0, where \citet{HV89} and \citet{R15} found $1.6 \pm 0.1$ and 2.0, respectively. The Henyey-Greenstein parameter $b$ is around $-0.3$, consistent with \citet{HV89} and \citet{L16}, who derived $-0.40 \pm 0.01$ and $-0.35 \pm 0.05$, respectively. In the $h_{\rm S}$ map we can clearly identify the Azacca ejecta as having smaller values, which correspond to a narrower OE. The single scattering albedo $w$ is higher here, but $B_{\rm S0}$ is lower. As such, here the OE amplitude and width are correlated in the Hapke model, similar to what we found with the Akimov model. When we exclude data with $\alpha < 0.5^\circ$ from the fit, the Azacca ejecta are no longer recognized in the $h_{\rm S}$ map, just as for the $\nu_2$ map in the Akimov model.

In conclusion, we mapped the photometric parameters of the Akimov and Hapke models over ROI~1 to search for spatial variability. Interpretation of the maps is severely hindered by the uneven phase-angle coverage. Nevertheless, we identified variability associated with Azacca crater, whose relatively bright ejecta have a steeper OE slope than their surroundings, but only below $0.5^\circ$ phase. It seems certain physical properties other than brightness are responsible for this phenomenon. We found the Akimov model to offer slightly more flexibility in analyzing the data than the Hapke model, and the interpretation of its parameters more straightforward. The best-fit Hapke model parameter values are consistent with the literature.

\subsection{Opposition effect wavelength dependence}

\begin{figure*}
\centering
\includegraphics[width=8cm]{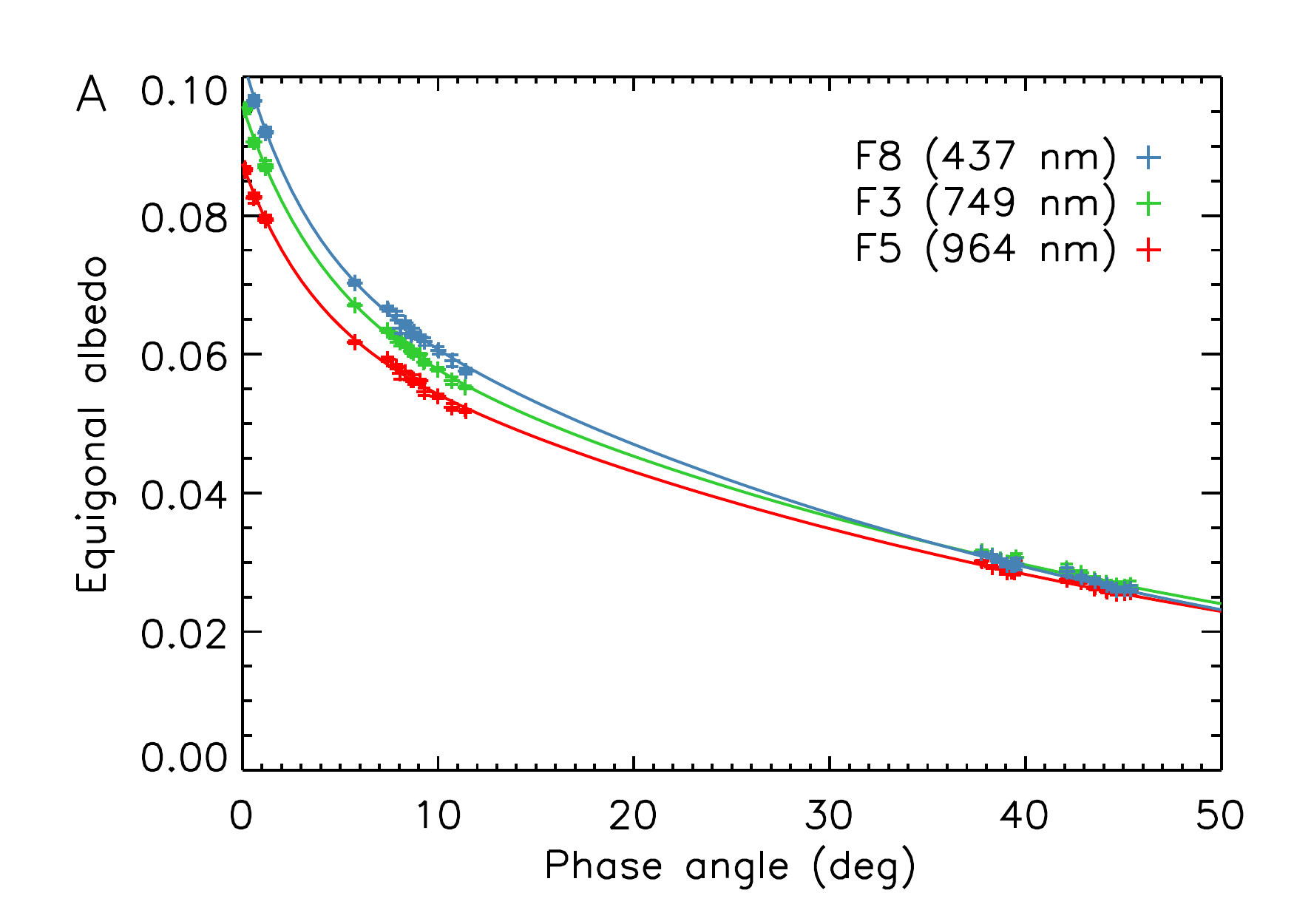}
\includegraphics[width=8cm]{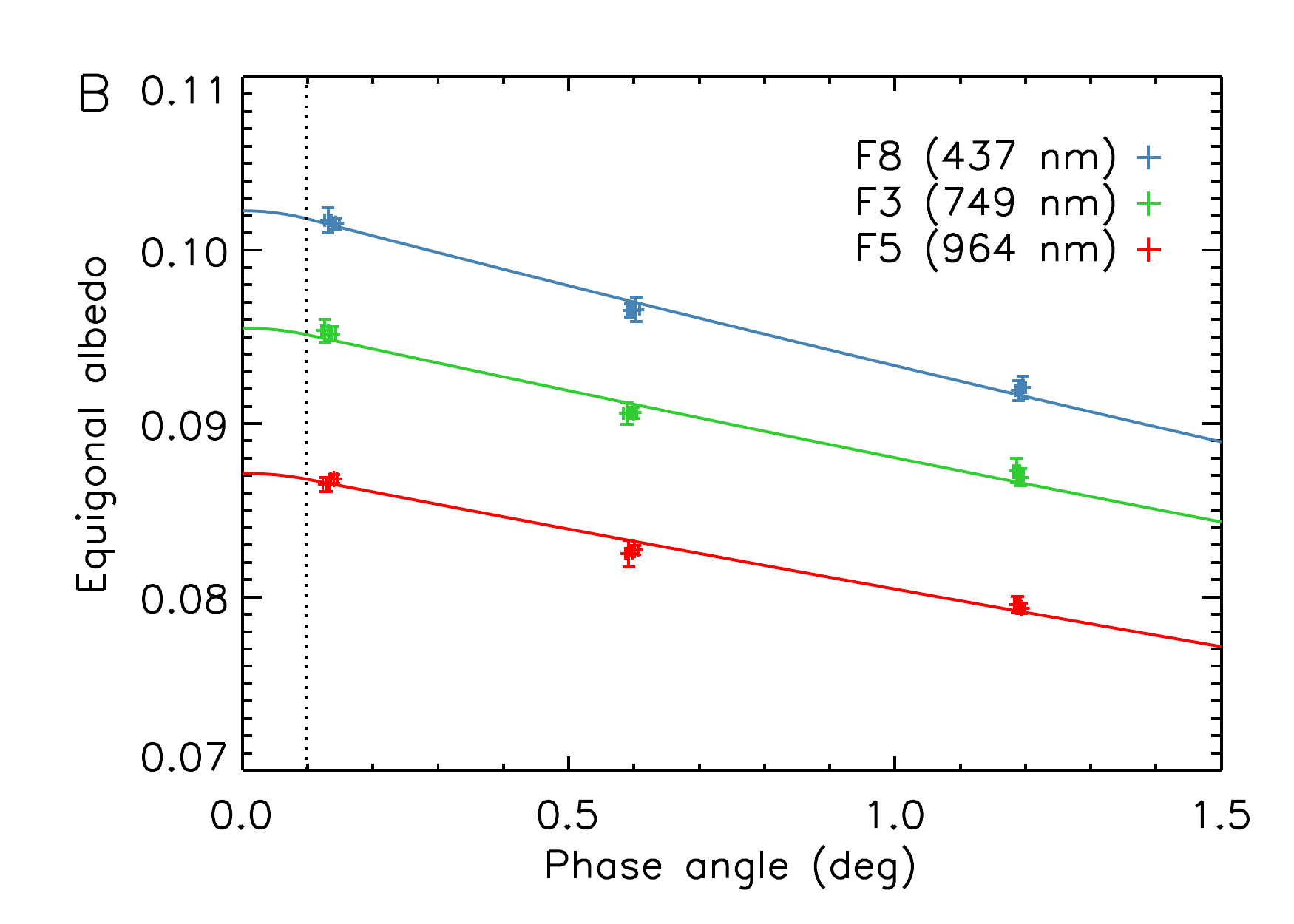}
\caption{Modeling the narrow-band reflectances, calculated as the average of a box of $3 \times 3$ pixels centered on the location of the red star in Fig.~\ref{fig:ROI}A, with $\iota, \epsilon < 50^\circ$. The best-fit model parameters are listed in Table~\ref{tab:Akimov_fit_OE_color}. \textbf{A}: Akimov phase function (Eq.~\ref{eq:Akimov_phase_fie}). \textbf{B}: Exponential phase function (Eq.~\ref{eq:Akimov_phase_fie} with $m = 0$), fitted to $\alpha < 1.5^\circ$. The model curves account for the finite size of the Sun, with the vertical dotted line indicating the angular radius of the Sun at Ceres.}
\label{fig:OE_phase_curve_color}
\end{figure*}

We now turn our attention to the narrow-band filters to search for wavelength-dependent variations of the OE. The analysis as performed for the clear filter in the previous section is not possible for the narrow-band filters, as their surface coverage is much more sparse. Figure~\ref{fig:OE_phase_curve_color} provides an overview of the color data available for a single location, at the red star in Fig.~\ref{fig:ROI}. During \textit{RC3}, narrow-band images were only acquired up to about $50^\circ$ phase angle. The available data for \textit{XMO4} derives from three instances when a set of narrow-band images was acquired, indicated with square symbols in Figs.~\ref{fig:albedo_map} and \ref{fig:ROI}. Each set consists of two images for each of the three narrow-band filters (F3, F5, and F8), or six images in total. The data coverage in the OE phase-angle range in this location is typical for pixels near the path of zero phase over the surface. Looking at Fig.~\ref{fig:OE_phase_curve_color}B, it is clear that we can only study the wavelength dependence of the ``broad OE'' that extends to $10^\circ$-$15^\circ$ phase, and not that of the ``narrow OE'' ($\alpha < 0.5^\circ$) that we uncovered in the previous section. Fits of the Akimov model to the color data over the full phase angle range (Fig.~\ref{fig:OE_phase_curve_color}A) reveal no obvious relation between the OE slope and wavelength at this particular location. Restricting the fit to the smallest phase angles and using a single exponential function again reveals no obvious correlation (Fig.~\ref{fig:OE_phase_curve_color}B).

\begin{figure*}
\centering
\includegraphics[width=8cm]{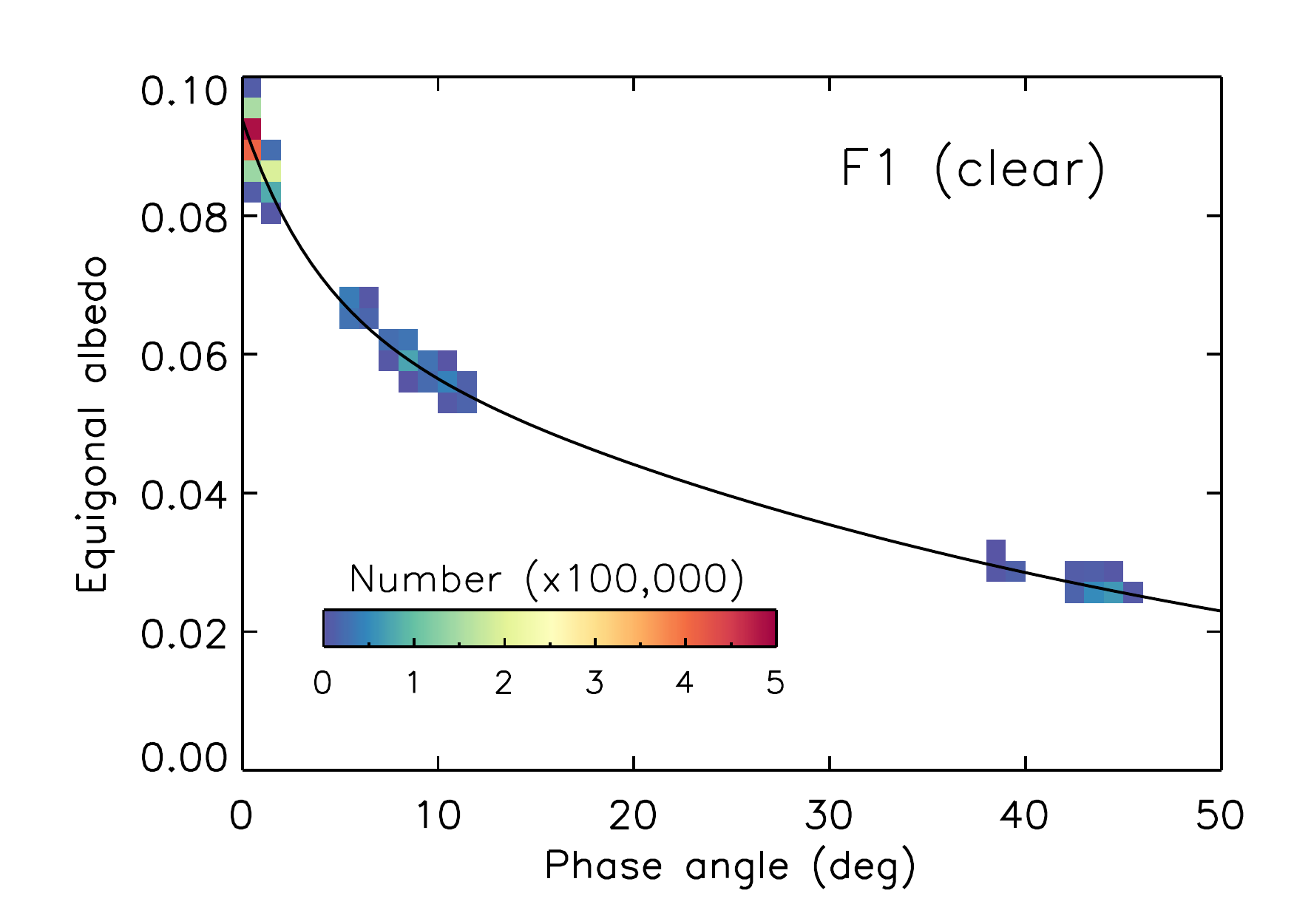}
\includegraphics[width=8cm]{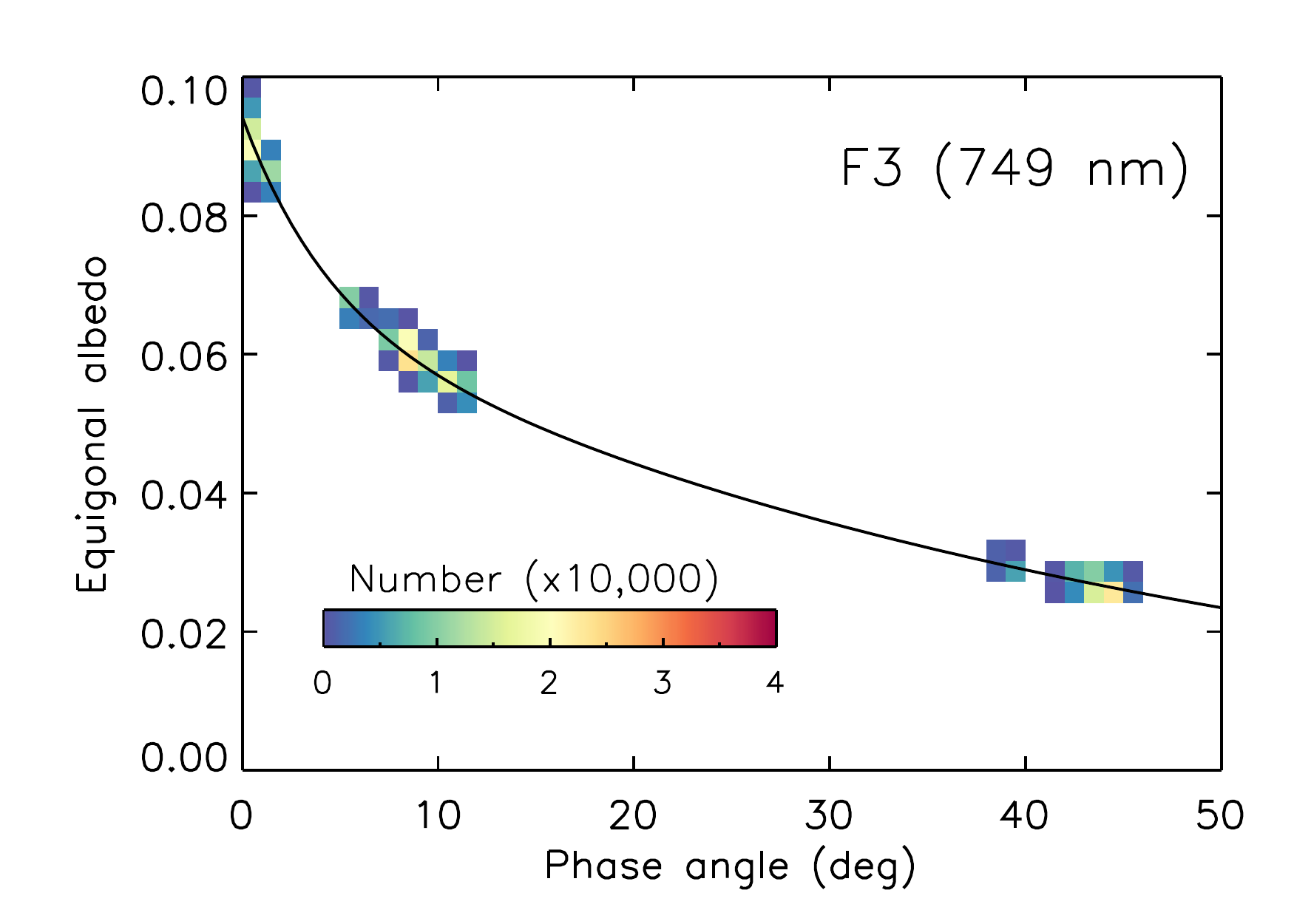}
\includegraphics[width=8cm]{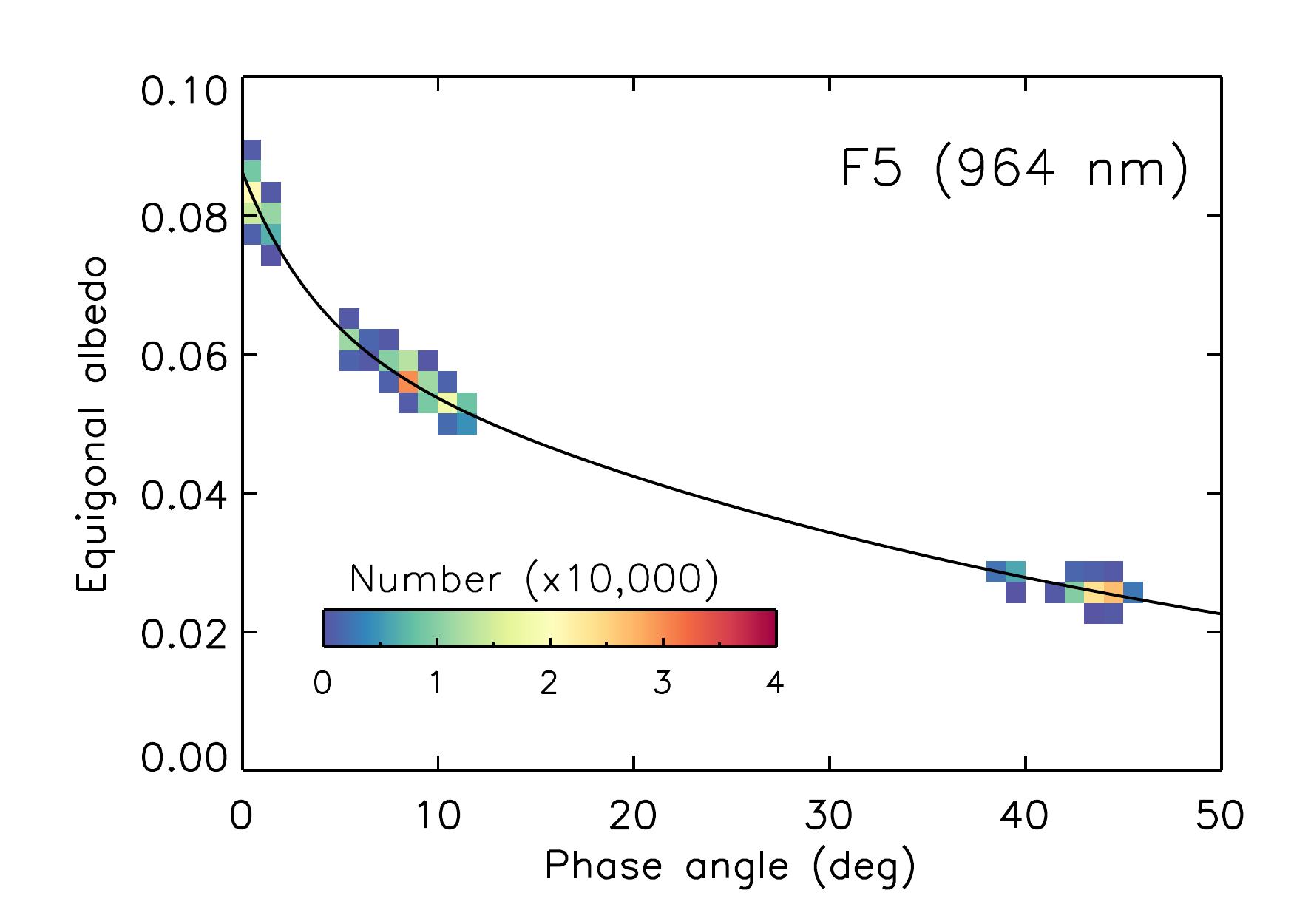}
\includegraphics[width=8cm]{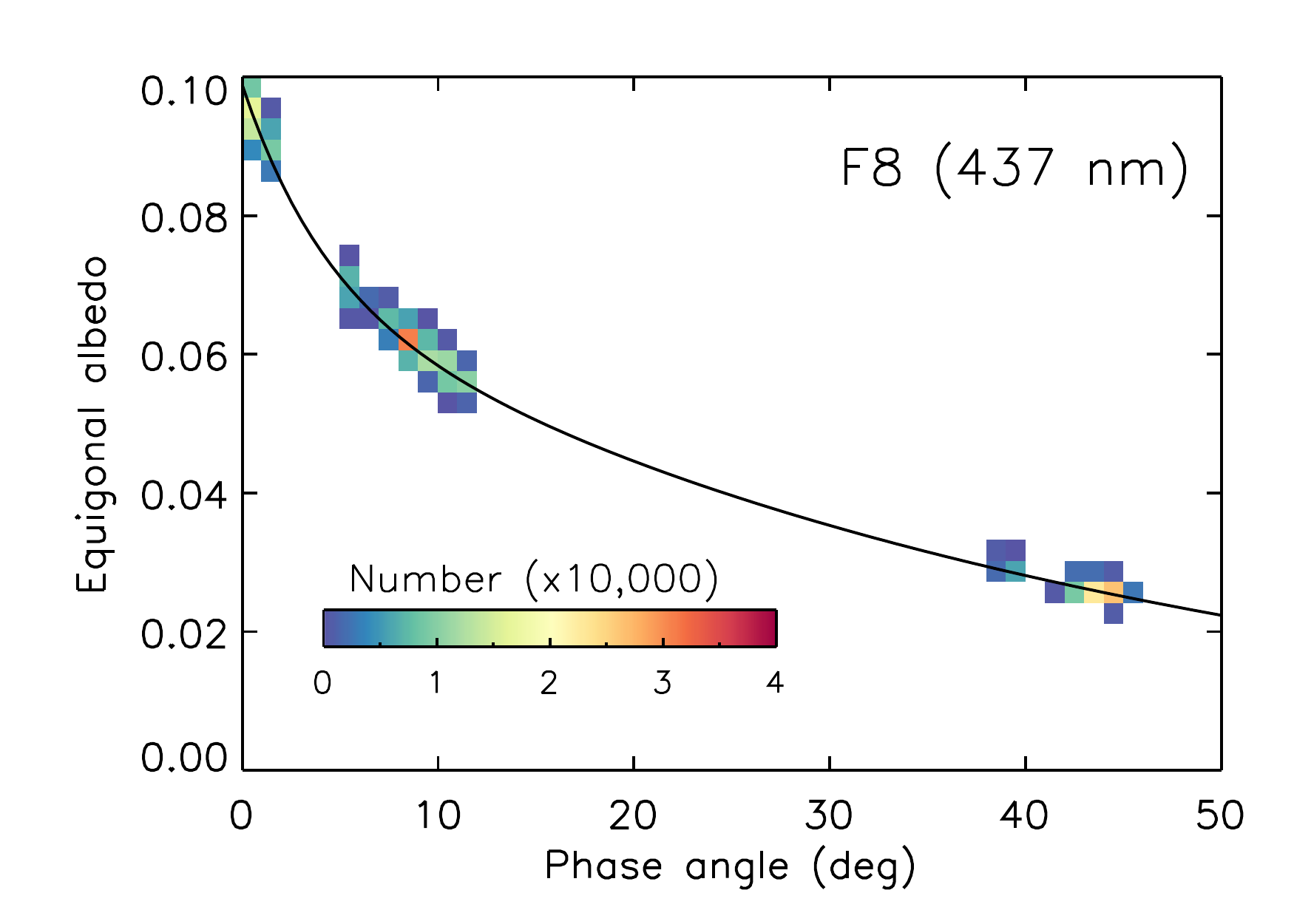}
\caption{Average phase curves of the terrain in ROI~2 (outlined in Fig.~\ref{fig:albedo_map}). The data are displayed as a density in a grid, with the colors reflecting the number of data points in each grid box. The drawn lines are best-fit Akimov phase functions, with the coefficients listed in Table~\ref{tab:Akimov_fit_coef_area}. The fit was restricted to data with $\alpha > 0.05^\circ$ and $\iota, \epsilon < 50^\circ$.}
\label{fig:average_phase_curves}
\end{figure*}

To evaluate the robustness of this finding, we determine the average phase curves for ROI~2. The best-fit Akimov models for the aggregated data are shown in Fig.~\ref{fig:average_phase_curves} for all FC2 filters used during \textit{XMO4}, with parameters in Table~4. The data are shown as a density plot, but the models were fit to the individual data points. ROI~2 was particularly frequently imaged through the clear filter during \textit{XMO4}, which is represented by the high density of data points close to zero phase. For the narrow-band filters, the \textit{XMO4} coverage was more or less similar to that during \textit{RC3}. We find that the normal albedo decreases with wavelength, giving this area a slightly bluish character. The geometric albedo spectrum of Ceres as reconstructed by \citet{R15} is also blue, which the authors attributed to phase reddening. Obvious correlations between the filter wavelength and the OE parameters are absent (Table~\ref{tab:Akimov_fit_coef_area}). Restricting the phase angle to $\alpha < 1^\circ$ and fitting a single exponential phase curve reveals no correlations either (not shown). We also fitted the Hapke model to the aggregated ROI~2 data, and found no systematic wavelength dependence for the OE width parameter ($h_{\rm S}$, Table~\ref{tab:Hapke_ROI2}).

\begin{table}
\centering
\caption{Akimov model fit coefficients associated with the phase curves in Fig.~\ref{fig:OE_phase_curve_color}. $\lambda$ is in nanometers.}
\begin{tabular}{lllllll}
\hline\hline
Figure & Filter & $\lambda_{\rm eff}$ & $A_0$ & $\nu_1$ & $m$ & $\nu_2$ \\
\hline
A & F8 & 437 & 0.103 & 0.024 & 0.37 & 0.30 \\
  & F3 & 749 & 0.096 & 0.021 & 0.39 & 0.25 \\
  & F5 & 964 & 0.087 & 0.021 & 0.34 & 0.29 \\
\hline
B & F8 & 437 & 0.103 & 0.096 & 0.00 & N/A \\
  & F3 & 749 & 0.096 & 0.086 & 0.00 & N/A \\
  & F5 & 964 & 0.087 & 0.084 & 0.00 & N/A \\
\hline
\end{tabular}
\label{tab:Akimov_fit_OE_color}
\end{table}

\begin{table}
\centering
\caption{Akimov model fit coefficients associated with the average phase curves in Fig.~\ref{fig:average_phase_curves} (aggregated ROI~2 data), together with the corresponding enhancement factors and OE \textsc{hwhm}. $\lambda$ is in nanometers.}
\begin{tabular}{llllllll}
\hline\hline
Filter & $\lambda_{\rm eff}$ & $A_0$ & $\nu_1$ & $m$ & $\nu_2$ & $\zeta(0^\circ)$ & \textsc{ hwhm} \\
\hline
F1 & N/A & 0.094 & 0.022 & 0.38 & 0.26 & 1.38 & $2.9^\circ$ \\
F8 & 437 & 0.099 & 0.023 & 0.41 & 0.24 & 1.41 & $3.2^\circ$ \\
F3 & 749 & 0.094 & 0.021 & 0.41 & 0.23 & 1.41 & $3.3^\circ$ \\
F5 & 964 & 0.086 & 0.021 & 0.34 & 0.26 & 1.34 & $2.8^\circ$ \\
\hline
\end{tabular}
\label{tab:Akimov_fit_coef_area}
\end{table}

\begin{table}
\centering
\caption{Best-fit Hapke parameters for the aggregated ROI~2 data, using the single-term Henyey-Greenstein function with $\alpha > 0.05^\circ$ and $\iota, \epsilon < 50^\circ$. $\lambda$ is in nanometers.}
\begin{tabular}{lllllll}
\hline\hline
Filter & $\lambda_{\rm eff}$ & $w$ & $B_{{\rm S}0}$ & $h_{\rm S}$ & $\bar{\theta}$ & $b$ \\
\hline
F1 & N/A & 0.095 & 1.9 & 0.059 & $28^\circ$ & -0.31 \\
F8 & 437 & 0.091 & 2.0 & 0.059 & $30^\circ$ & -0.32 \\
F3 & 749 & 0.095 & 2.0 & 0.063 & $28^\circ$ & -0.29 \\
F5 & 964 & 0.095 & 1.6 & 0.058 & $27^\circ$ & -0.31 \\
\hline
\end{tabular}
\label{tab:Hapke_ROI2}
\end{table}

\begin{figure*}
\centering
\includegraphics[width=8cm]{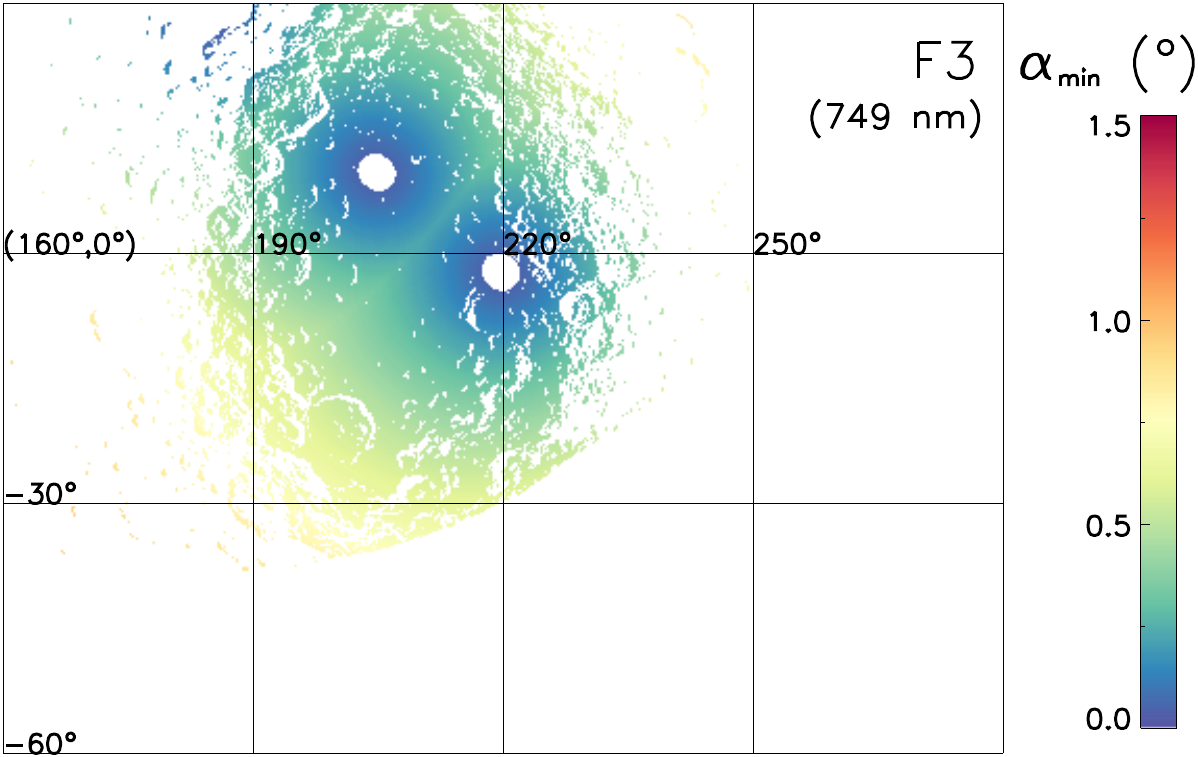}
\includegraphics[width=8cm]{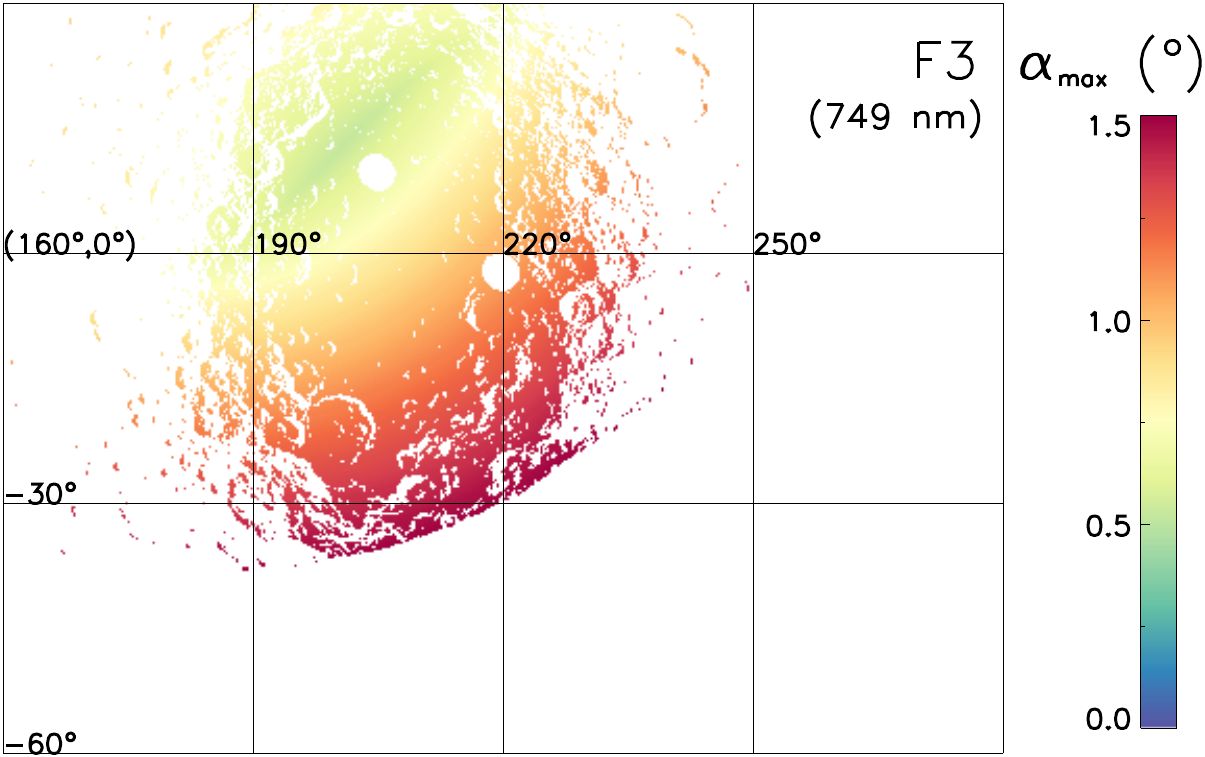}
\includegraphics[width=8cm]{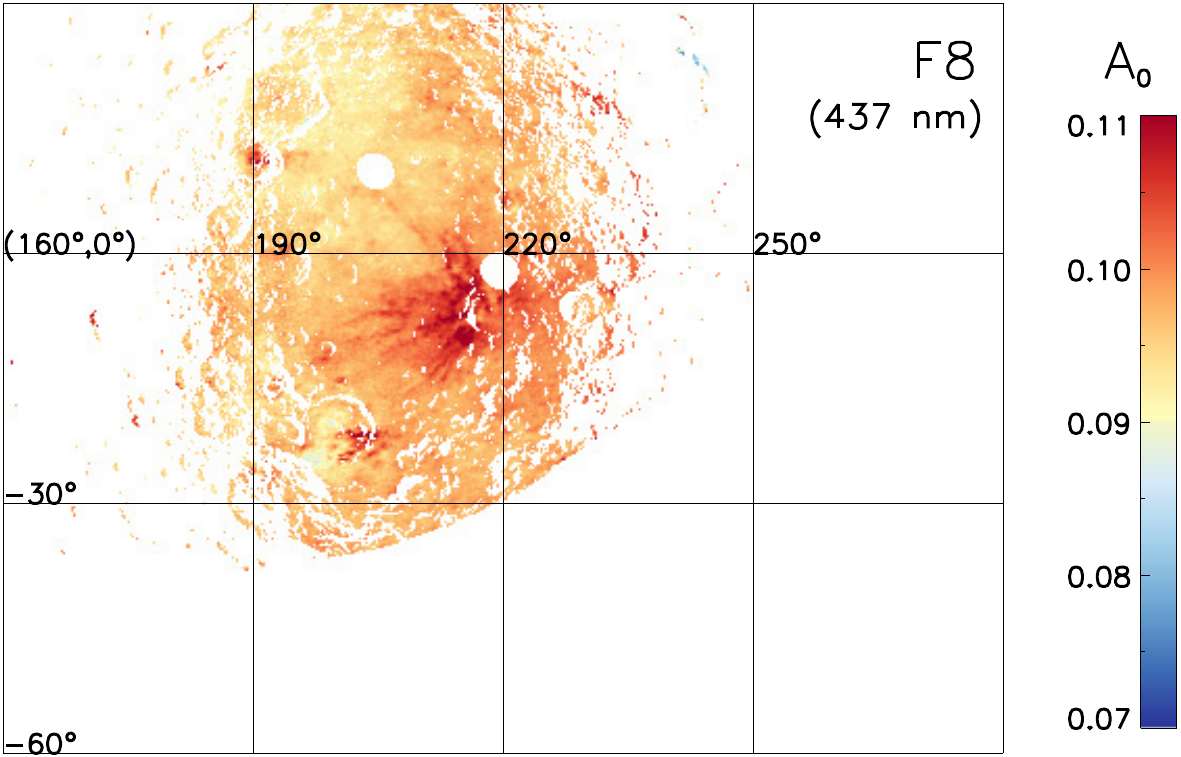}
\includegraphics[width=8cm]{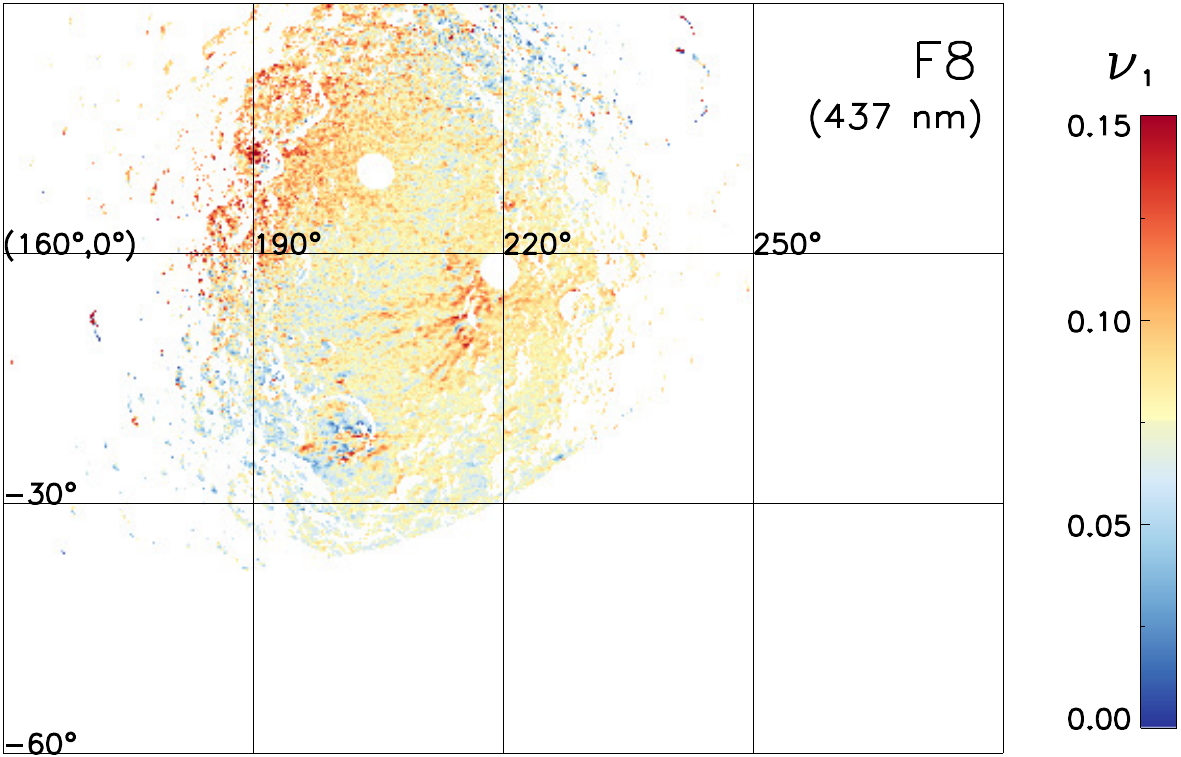}
\includegraphics[width=8cm]{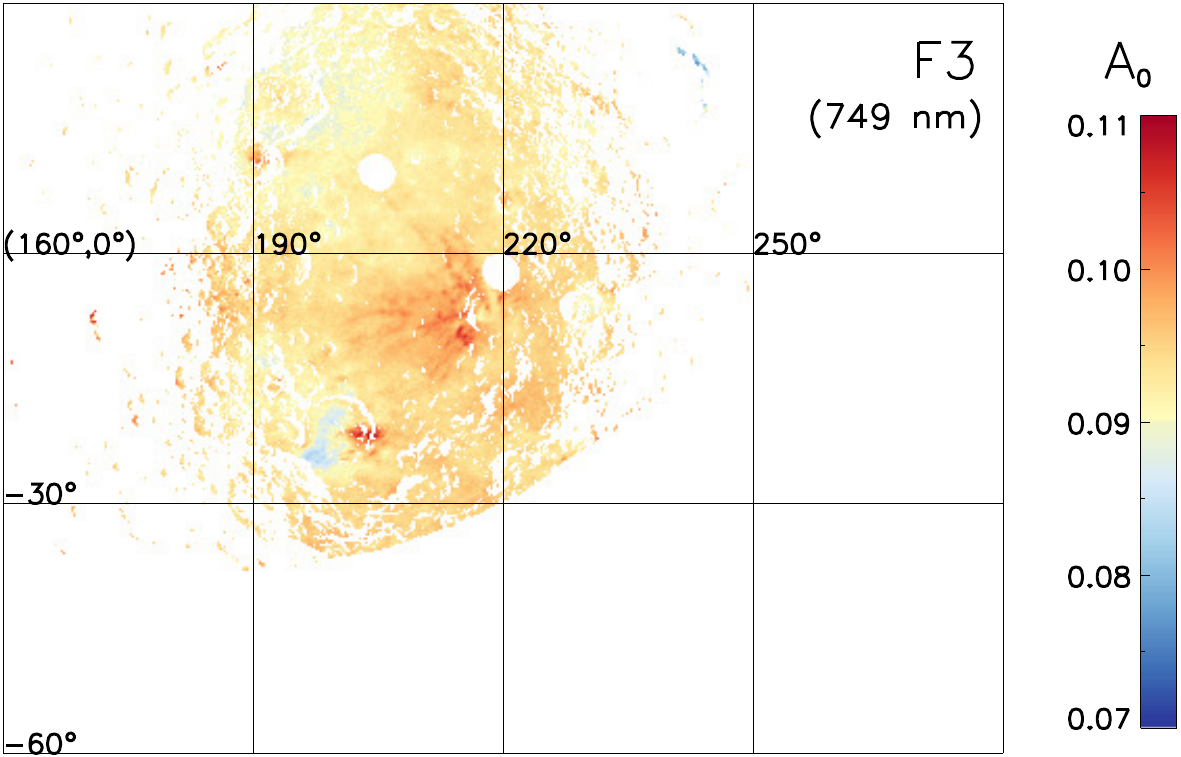}
\includegraphics[width=8cm]{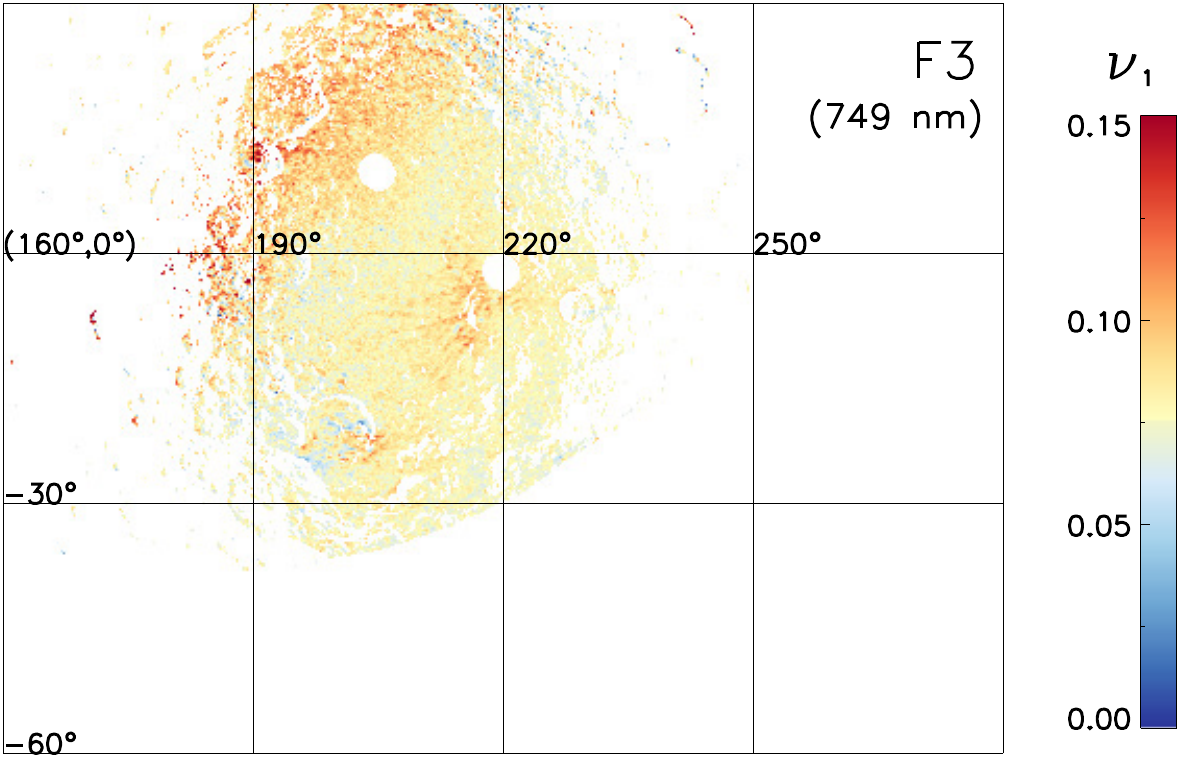}
\includegraphics[width=8cm]{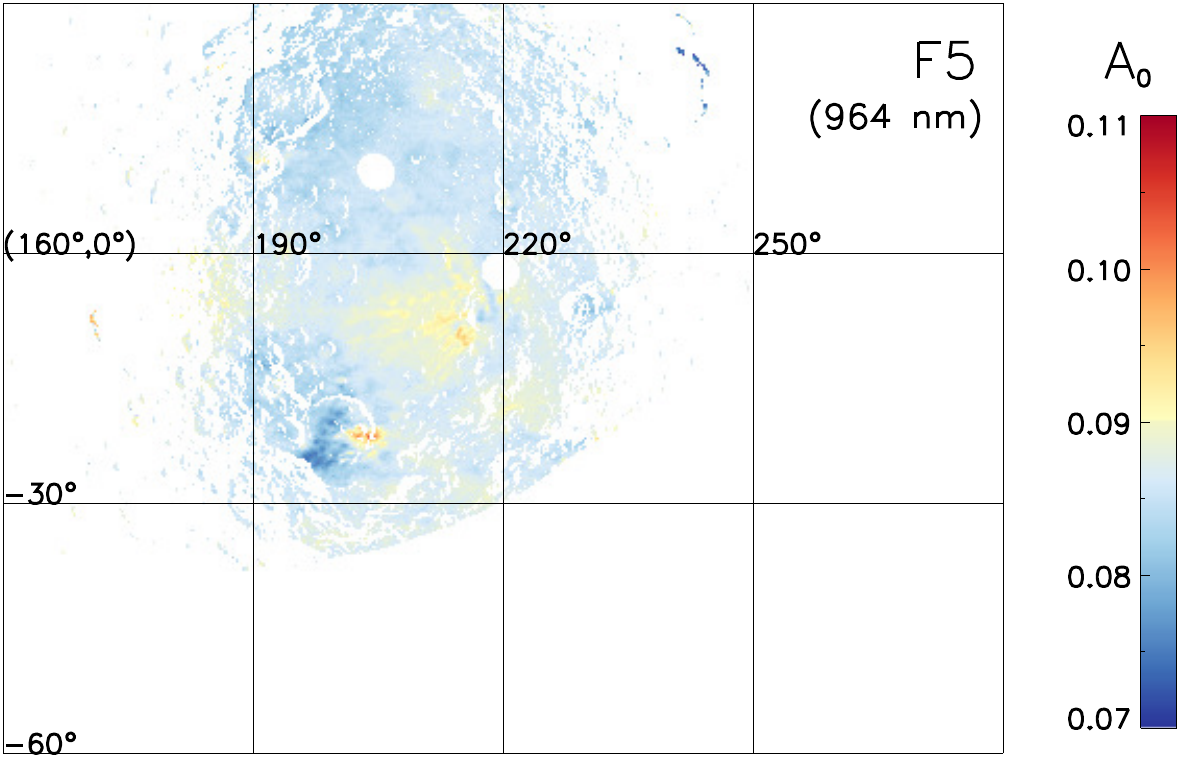}
\includegraphics[width=8cm]{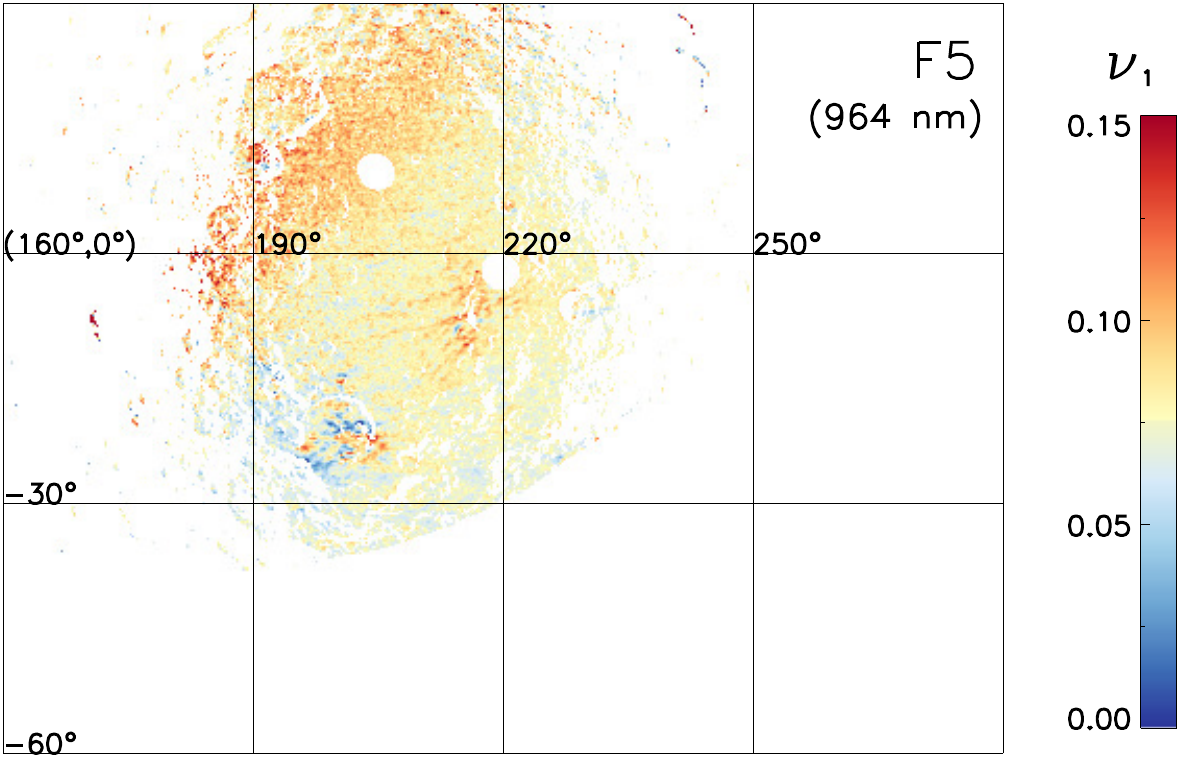}
\caption{Maps of the OE parameters (Eq.~\ref{eq:Akimov_phase_fie} with $m = 0$, and $\nu_1$ the OE slope) for the three narrow-band filters used during \textit{XMO4}. The phase angle was restricted to $0.05^\circ < \alpha < 1.50^\circ$, with $\iota, \epsilon < 50^\circ$. The actual minimum and maximum phase angle of the data, $\alpha_{\rm min}$ and $\alpha_{\rm max}$, are shown for filter F3 (the maps for F5 and F8 are very similar).}
\label{fig:OE_parameters_color}
\end{figure*}

In a final push to uncover a correlation between OE width and wavelength, we map the OE parameters for each of the three narrow-band filters, but only using data at very small phase angles and a single exponential phase function (Eq.~\ref{eq:Akimov_phase_fie} with $m = 0$). As before, the retrieval of the OE parameters may be sensitive to the limits of the phase-angle range for each projected pixel. Figure~\ref{fig:OE_parameters_color} shows these limits ($\alpha_{\rm min}$ and $\alpha_{\rm max}$) together with maps of the normal albedo ($A_0$) and OE slope ($\nu_1$, here playing the role of $\nu_2$). We restrict the phase angle to $0.05^\circ < \alpha < 1.50^\circ$, but in contrast to Fig.~\ref{fig:OE_parameters_clear}, we cannot be sure that each pixel has this range fully occupied with observations. The range covers both the narrow and broad OE, but the data are too sparse to distinguish between the two regimes. There are two ``holes'' in the maps where the phase angle reached zero. A third hole exists but is not visible, as we only consider data with $\iota, \epsilon < 50^\circ$. The $A_0$ maps appear unaffected by the uneven phase-angle distribution. This is not the case for the $\nu_1$ maps, in which the decrease of $\alpha_{\rm max}$ towards the northwest is accompanied by an increase of $\nu_1$. Restricting the phase-angle range to a narrower $0.05^\circ < \alpha < 1.00^\circ$ decreases the coverage but otherwise has only a minor effect on the appearance of the maps. The uneven distribution of $\alpha_{\rm min}$ appears inconsequential for both $A_0$ and $\nu_1$. The ejecta of Azacca crater are recognizable in all maps, meaning they are brighter and have a narrower OE than their surroundings, consistent with our findings for the clear filter. The three $A_0$ maps appear different because of the decrease of the average $A_0$ with wavelength. The $\nu_1$ maps appear similar, indicating that the OE slope is not correlated with wavelength. We can quantify this as follows. If we consider only projected pixels with the phase-angle range restricted to $0.2 < \alpha_{\rm min} < 0.4$ and $0.7 < \alpha_{\rm max} < 0.9$ ($n = 3861$), then the average $\nu_1$ at wavelength 437~nm is $0.0818 \pm 0.0015$, at 749~nm it is $0.0820 \pm 0.0011$ (749~nm), and at 964~nm it is $0.0819 \pm 0.0010$.

In conclusion, we do not find evidence for a wavelength-dependent OE width. However, due to the limited availability of narrow-band data, this conclusion applies to the regular, ``broad'' OE, and not to the ``narrow'' OE ($\alpha < 0.6^\circ$) that we uncovered in the previous section for the ejecta of the Azacca crater.

\subsection{Visible geometric albedo}

Judging from the global albedo map in Fig.~\ref{fig:albedo_map}, ROI~2 represents a fairly typical part of the surface. We estimate Ceres' visible geometric albedo ($p_{\rm V}$) from the average phase curve derived for this area (Fig.~\ref{fig:average_phase_curves}). The geometric albedo equals the normal albedo ($A_0$) if the disk function is constant at zero phase, which is approximately true for Ceres \citep{S17}. And because Ceres' reflectance spectrum is very flat in the visible \citep{R15,L16}, we can use the clear filter as a proxy for the V-band ($\lambda_{\rm eff} = 551$~nm). We therefore determine the visible geometric albedo as $p_{\rm V} = 0.094 \pm 0.005$, where we assume an error margin of 5\%. This value agrees well with the $p_{\rm V} = 0.10 \pm 0.01$ derived by \citet{T89} from IRAS observations. \citet{T02} revised their estimate to $p_{\rm V} = 0.113 \pm 0.005$, which seems a little on the high side. Our estimate is also consistent with the geometric albedo of $0.099 \pm 0.003$ reported by \citet{R15}, who observed Ceres at phase angles as low as $0.85^\circ$, and the geometric albedo of 0.094 of \citet{ST06}, who derived the diameter of Ceres from stellar occultations and accounted for the OE. \citet{C17} determined $p_{\rm V} = 0.094 \pm 0.007$ by adopting the Hapke OE parameters from \citet{H97}, as their data do not go below $7^\circ$ phase angle.

\subsection{Cerealia Facula}

While Cerealia Facula was not located on the path of zero phase, it was imaged at phase angles smaller than $1^\circ$. The facula was barely resolved in \textit{XMO4} images, measuring only 2-3 pixels across. We analyze the clear filter images in combination with the \textit{RC3} images, which have a spatial resolution that is  1.5
times higher (Table~\ref{tab:image_data}). The area was observed up to phase angle $5.1^\circ$ during \textit{XMO4} and down to $8.1^\circ$ during \textit{RC3}. We compare our results with those of \citet{S17}, who analyzed the photometric properties of Cerealia Facula using only \textit{RC3} images and found its phase curve to be unusually steep. \citet{L18} obtained similar results using data acquired by Dawn's visual and infrared mapping spectrometer (VIR), suggesting the steepness may be due to high surface roughness. We found that the projection of Cerealia Facula in \textit{XMO4} images was not always very accurate, and we therefore calculated the average reflectance in a relatively large box of $5 \times 5$ projected pixels centered on Cerealia Facula. To further minimize the influence of projection inaccuracies we restricted the photometric angles to $\iota, \epsilon < 70^\circ$, where \citet{S17} allowed $\iota, \epsilon < 85^\circ$. The standard deviation associated with the average reflectance was calculated by accounting for photon noise only. Of course, the true uncertainty is larger because of projection errors. We fit two photometric models to the data: (1) The combination of the Akimov phase function and Lambert/Lommel-Seeliger disk function, and (2) the Hapke model.

\begin{figure*}
\centering
\includegraphics[width=8cm]{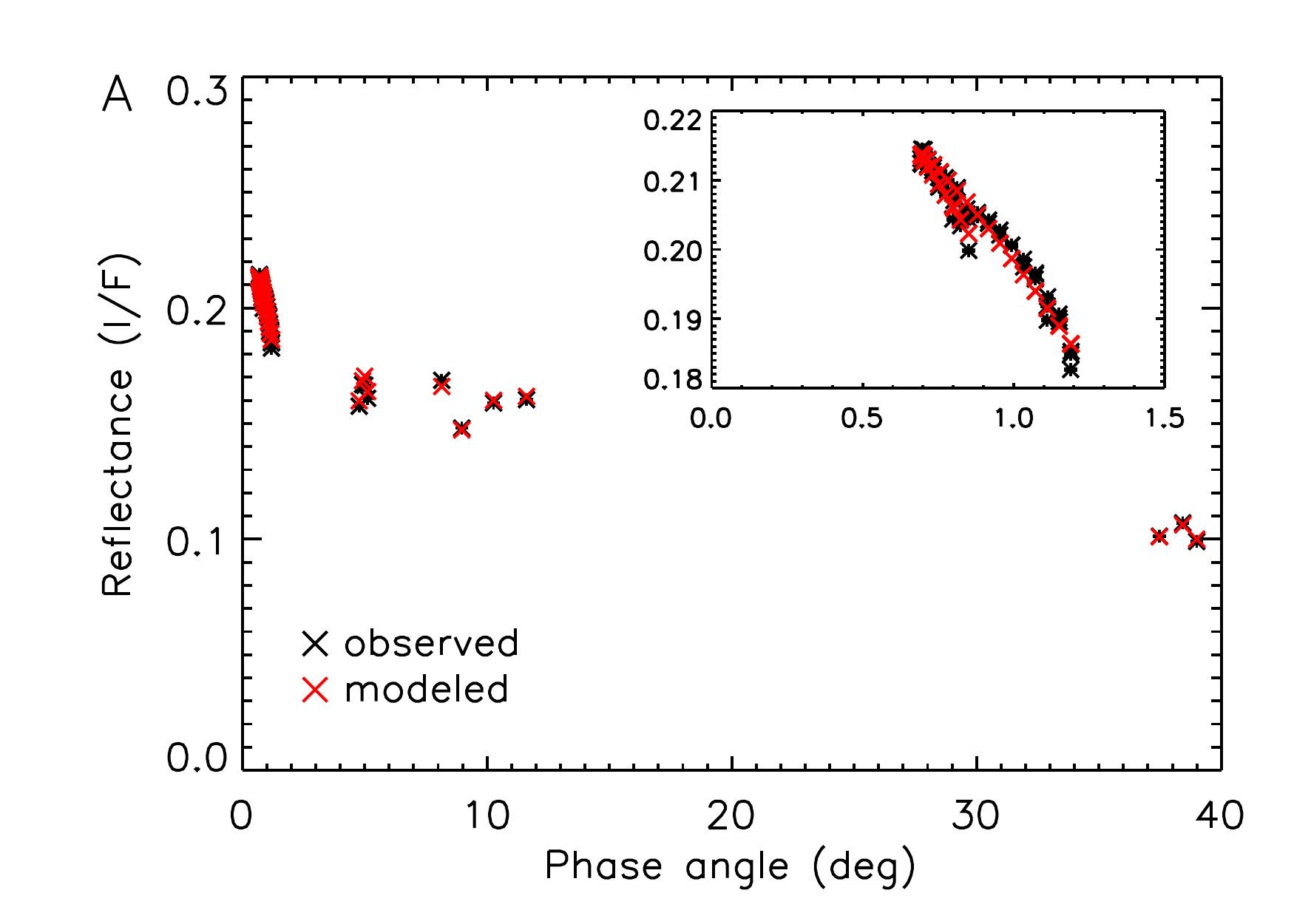}
\includegraphics[width=8cm]{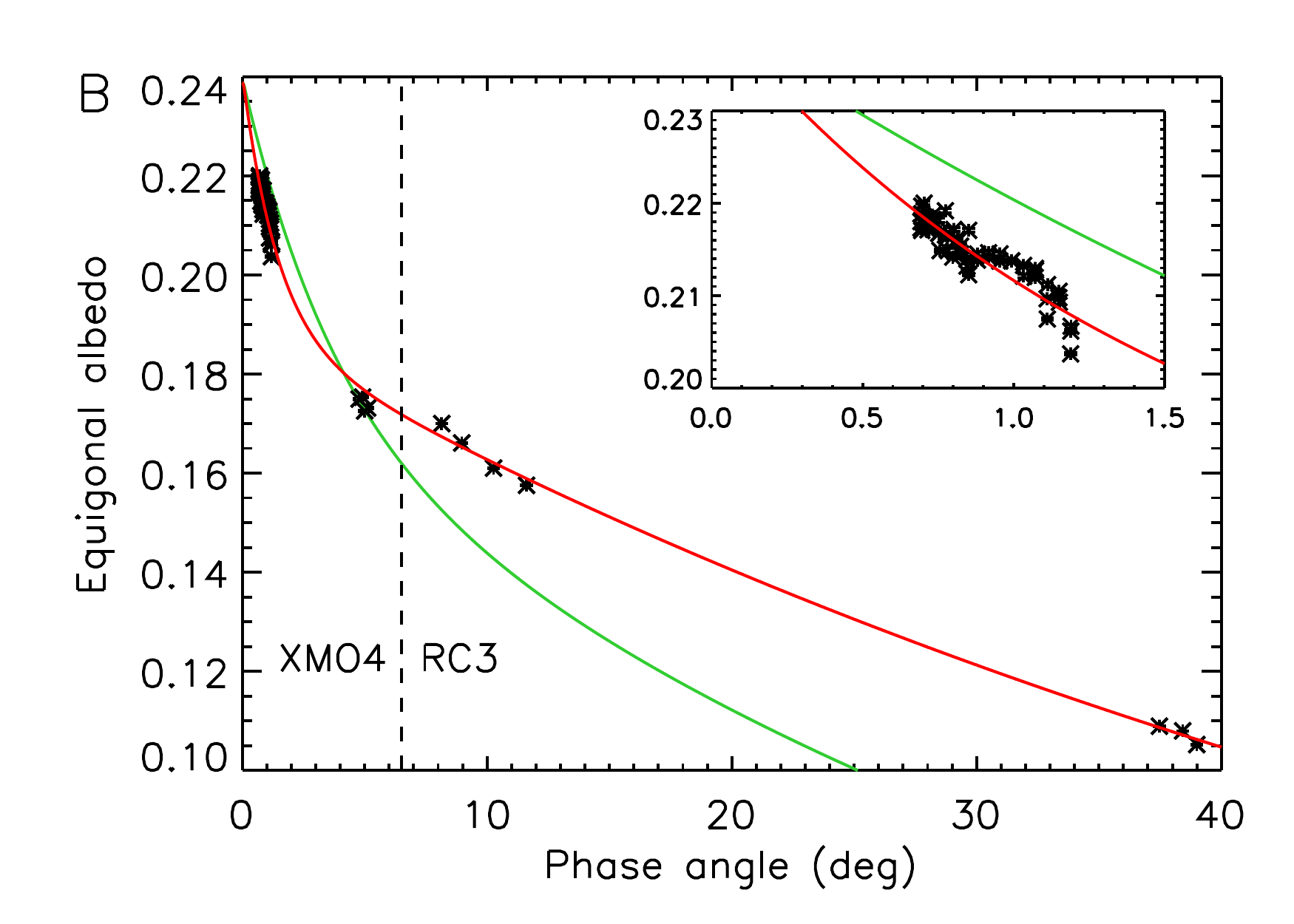}
\includegraphics[width=8cm]{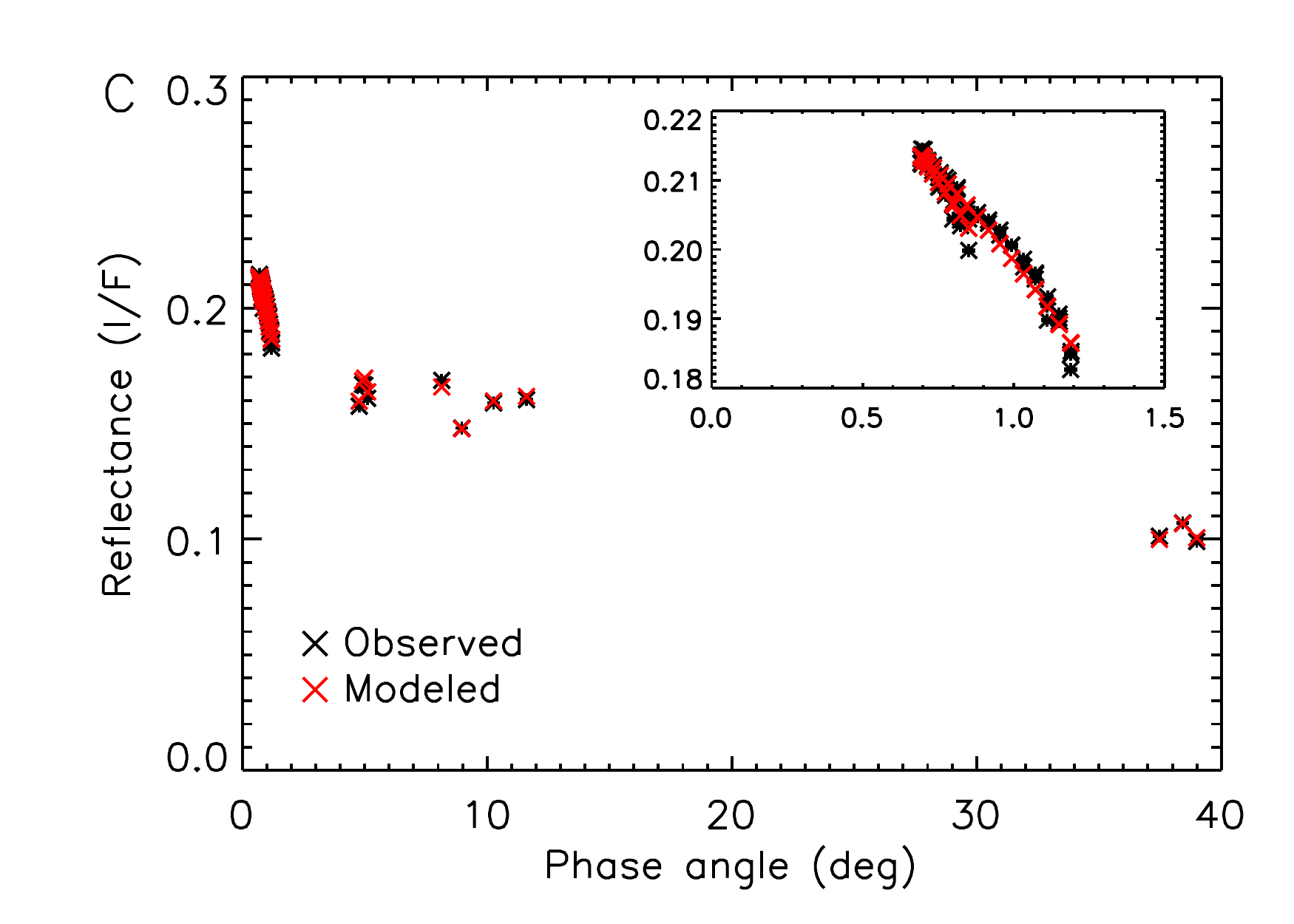}
\includegraphics[width=8cm]{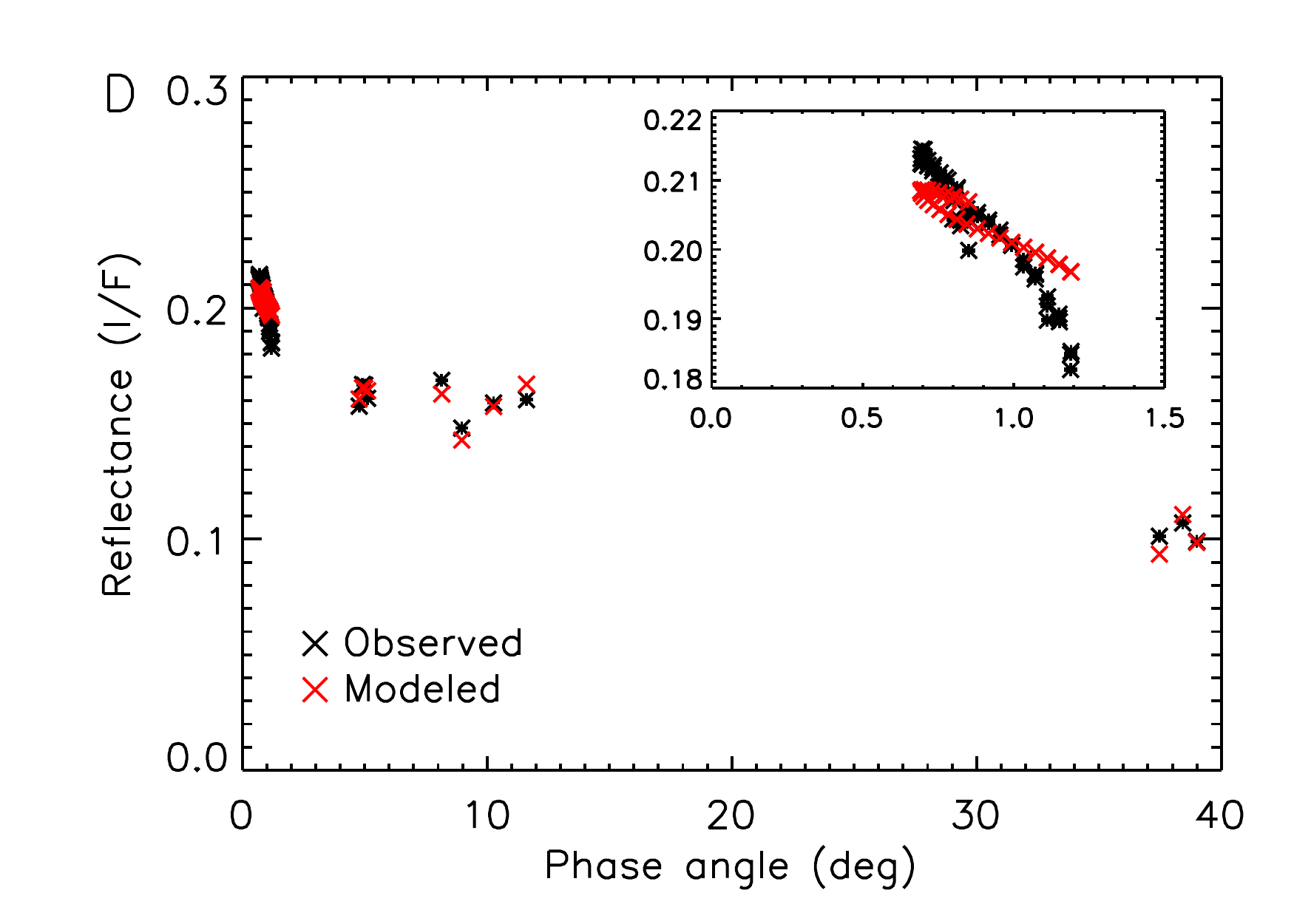}
\caption{Fitting photometric models to the average reflectance in a $5 \times 5$ pixel sized box centered on Cerealia Facula in \textit{XMO4} and \textit{RC3} clear filter images, with $\iota, \epsilon < 70^\circ$ ($n = 69$). The insets zooms in on the smallest phase angles. \textbf{A}: Akimov phase function with Lambert/Lommel-Seeliger disk function. \textbf{B}: As (A), showing the equigonal albedo. The boundary between the two data sets (\textit{XMO4} and \textit{RC3}) is indicated. The best-fit model is drawn in red, while the best-fit model for Ceres average (Fig.~\ref{fig:average_phase_curves}), scaled to match the former at $\alpha = 0^\circ$, is drawn in green. \textbf{C}: Best-fit Hapke model, which has $\bar{\theta} = 0^\circ$. \textbf{D}: Best-fit Hapke model for $\bar{\theta} = 59^\circ$ (fixed).}
\label{fig:Cerealia_Facula}
\end{figure*}

(1) The phase and disk function parameters were fit simultaneously. The best-fit model is shown in Figs.~\ref{fig:Cerealia_Facula}A and B, and has parameters $A_0 = 0.24$, $\nu_1 = 0.015$, $m = 0.28$, and $\nu_2 = 0.70$, with $c_{\rm L} = 0.75 - 0.0015 \alpha$ (with $\alpha$ in degrees). The disk function has a lower contribution of the Lambert term at all phase angles than found by \citet{S17}, who used \textit{RC3} images only ($c_{\rm L} = 0.23 - 0.0017 \alpha$). This is expected, given that the 25 projected \textit{XMO4} pixels contain more of the surrounding terrain than the 6 projected \textit{RC3} pixels used by \citeauthor{S17}. Let us compare the best-fit Akimov model parameters with those of ``Ceres average'', defined by the F1 curve in Fig.~\ref{fig:average_phase_curves}. Obviously, the normal albedo is much higher than average ($A_0 = 0.24$ vs.\ 0.094). It is not as high as the normal albedo of 0.6 estimated by \citeauthor{S17}, because we averaged the reflectance over an area much larger than Cerealia Facula itself. The phase curve is shallower than average ($\nu_1 = 0.015$ vs.\ 0.022). This is also expected, as multiple scattering is more prevalent in a bright surface. The OE characteristics are clearly different from average: The OE amplitude is only slightly lower ($m = 0.28$ vs.\ 0.38), but the OE peak is substantially narrower ($\nu_2 = 0.70$ vs.\ 0.27). The different quality of the OE can be seen most clearly in Fig.~\ref{fig:Cerealia_Facula}B, showing the equigonal albedo. However, this figure also reveals how the \textit{XMO4} and \textit{RC3} data sets poorly match. The apparent discontinuity around $7^\circ$ phase angle, most likely due to the difference in spatial resolution (Table~\ref{tab:image_data}), suggests that the modeling results presented here cannot be considered fully reliable. Still, this issue would mostly affect the phase curve parameters. The OE characterization for Cerealia Facula is relatively reliable, as it rests almost exclusively on \textit{XMO4} data.

(2) The best-fit Hapke model is shown in Fig.~\ref{fig:Cerealia_Facula}C, and has virtually the same fit quality as the model in Fig.~\ref{fig:Cerealia_Facula}A. The parameters consistently converged to $w = 0.62$, $B_{{\rm S}0} = 0.96$, $h_{\rm S} = 0.0069$, $\bar{\theta} = 0.0^\circ$, $b = 0.54$, and $c = 0.68$, regardless of the starting values fed to the fitting algorithm. Adopting $\iota, \epsilon < 85^\circ$ as in \citet{S17} led to a worse model fit, a higher $B_{{\rm S}0}$, a lower $h_{\rm S}$, but similar values for the other parameters with again $\bar{\theta} = 0^\circ$. Clearly, this result is inconsistent with the $\bar{\theta} = 59^\circ$ determined from only \textit{RC3} data by \citeauthor{S17}, who marked their value as unusually high for a planetary surface. When we fixed $\bar{\theta}$ at $59^\circ$, we obtained the model fit in Fig.~\ref{fig:Cerealia_Facula}D, with $w = 0.65$, $B_{{\rm S}0} = 10$ (limit), $h_{\rm S} = 0.00019$, $b = 0.15$, and $c = -1.0$, which fails to fit the data in the OE range (inset). When we subsequently removed the limit on the OE amplitude parameter, the model converged to $B_{{\rm S}0} = 821$, without noticeably improving the fit quality. We then tried to fit the \textit{XMO4} alone with $\bar{\theta} = 59^\circ$, but found that we could only achieve reasonable fits for $\bar{\theta} < 30^\circ$. Therefore, we conclude that the OE observations are not consistent with an unusually high value for the roughness parameter. At the same time, our results confirm that the spatial resolution of the \textit{XMO4} images is too low compared to that of the \textit{RC3} images for the two data sets to be successfully combined. The present situation is somewhat puzzling. We trust the high photometric roughness derived from the higher-resolution \textit{RC3} data\footnote{\citet{S17} minimized the consequences of projection errors by inspecting each projected image and removing those with large errors from their sample.}, which was confirmed by \citet{L18}. Perhaps it is the consequence of large-scale topography inside Cerealia Facula, which has a dome at its center \citep{SB16}. At the very minimum, the OE observations do not reinforce the notion that the high roughness value found by \citet{S17} has physical significance for the regolith properties of Cerealia Facula.

Unfortunately, an analysis of the OE wavelength dependence for Cerealia Facula using only data at small phase angles (as in Fig.~\ref{fig:OE_phase_curve_color}B) is not possible, as there are only two groups of data points with the restriction of $\iota, \epsilon < 70^\circ$, too few for a meaningful exponential model fit. We also found that the modeling of the \textit{XMO4} color data for Cerealia Facula is very sensitive to the disk function, while at the same time the $c_{\rm L}$ disk function parameter is ill constrained due to the low spatial resolution and scarcity of data.

\section{Ceres in context}

Earlier we estimated the enhancement factor and {\sc hwhm} of the average Ceres OE as 1.38 and $2.8^\circ$, respectively (Table~\ref{tab:Akimov_fit_coef_area}). How does the Ceres OE compare to that of other asteroids? \citet{C17} suggest that C-type asteroids are the closest match to Ceres' spectrophotometric properties, so we may expect their OE to be similar. Let us consider the clear filter OE in Fig.~\ref{fig:average_phase_curves} to be representative for Ceres in the visible. Published reviews of asteroid OE parameters generally consider the integrated brightness in magnitudes, whereas we consider the (resolved) surface reflectance. The integrated phase curve is steeper than the resolved one, because it includes the effect of the diminishing size of the illuminated part of the asteroid disk with increasing phase angle. A simple geometric correction for the decreasing illuminated fraction of the disk with phase angle is the factor $(1 + \cos \alpha) / 2$ \citep{C15,Lo16}. Because this factor is close to unity in the OE range (e.g., it is 0.998 for $\alpha = 5^\circ$), we ignore differences between resolved and integrated phase curves.

\begin{figure*}
\centering
\includegraphics[width=8cm]{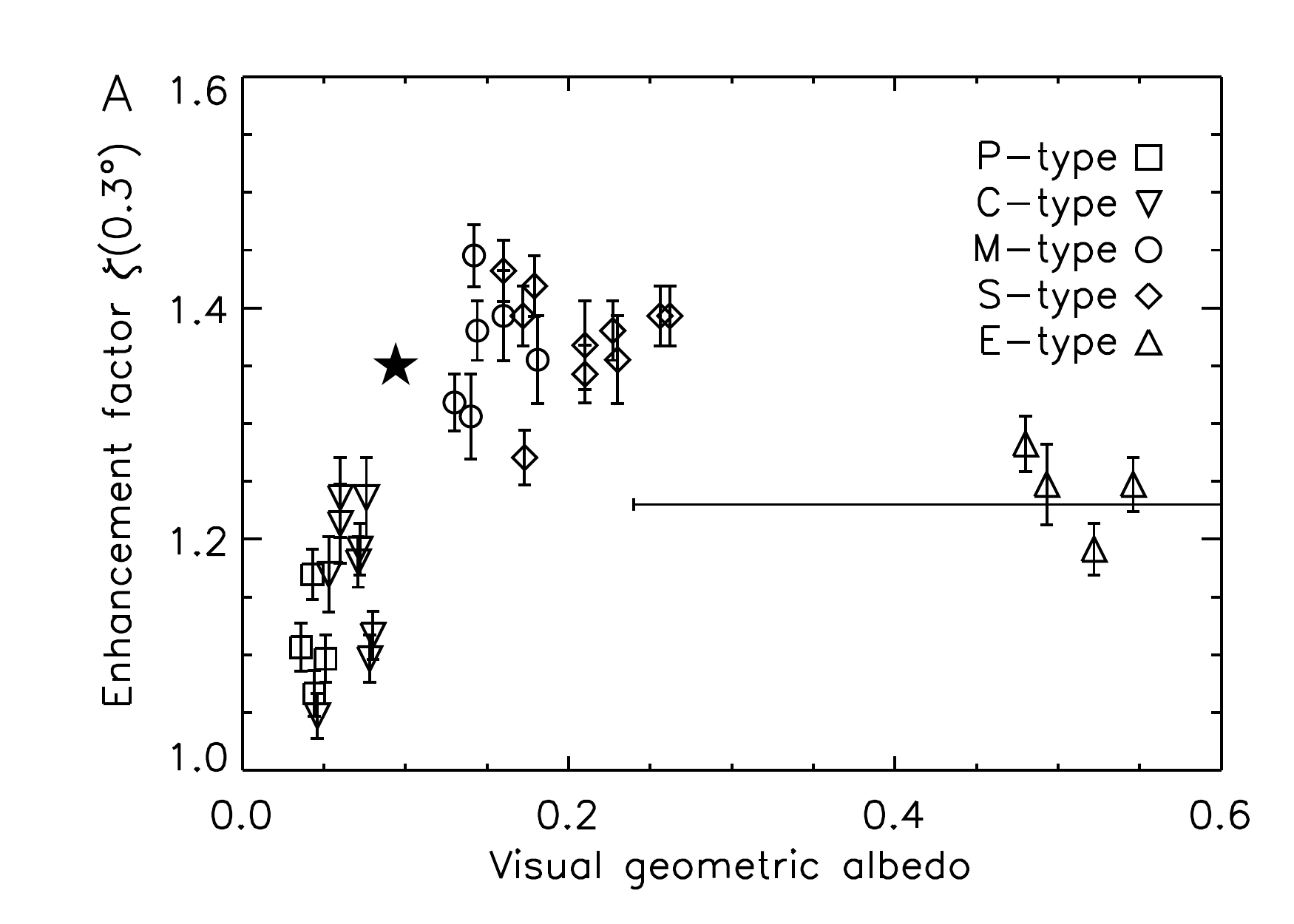}
\includegraphics[width=8cm]{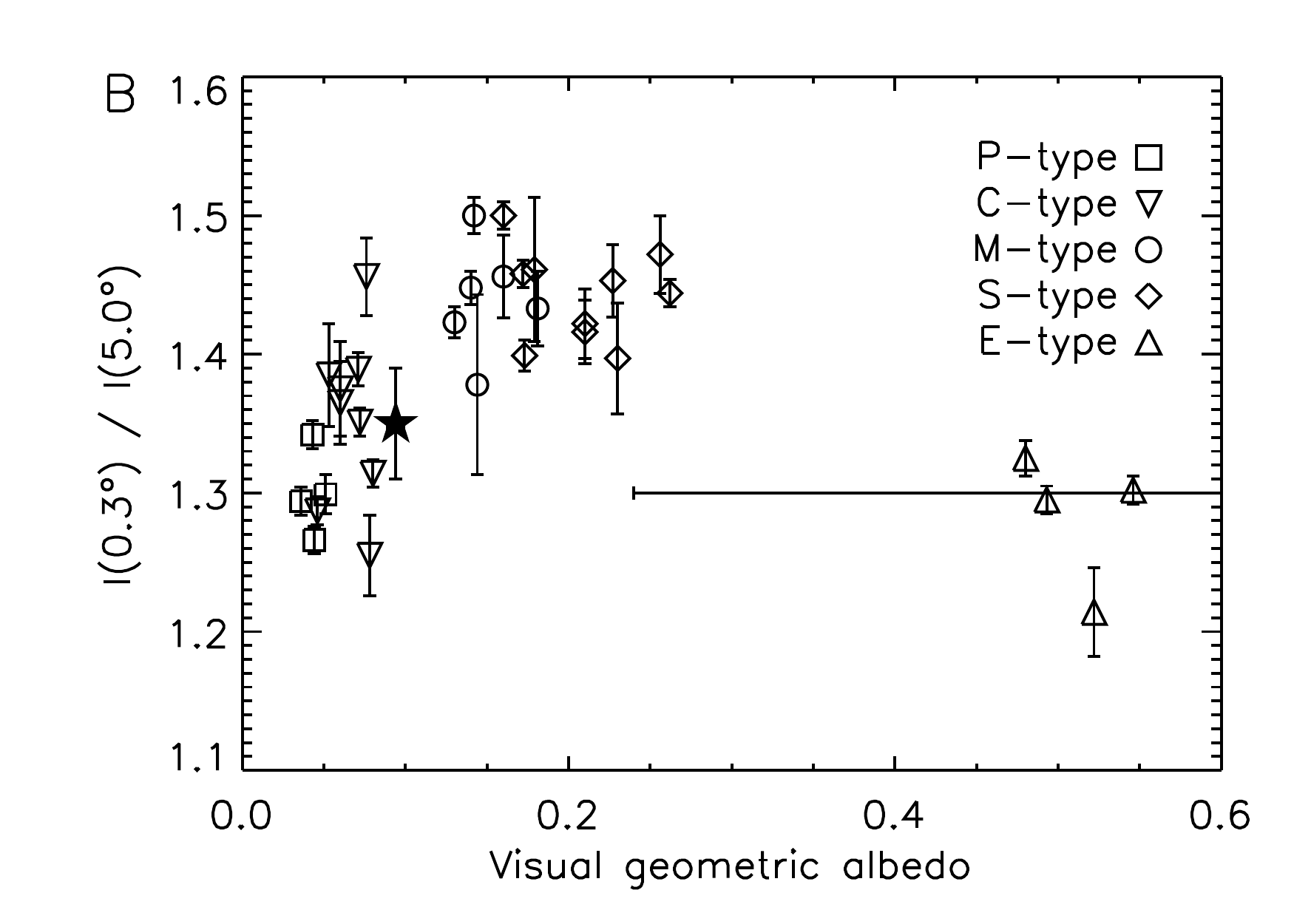}
\caption{The Ceres OE compared to that of the asteroids in \citet{BS00,B03,S16}. (\textbf{A}) The enhancement factor $\zeta(0.3^\circ)$, (\textbf{B}) the ratio of the intensities at $0.3^\circ$ and $5.0^\circ$. The Ceres values are those for the clear filter in Table~\ref{tab:Akimov_fit_coef_area} (black star); the formal error of the enhancement factor is on the order of the plot symbol. The Cerealia Facula values are located somewhere on the horizontal line. To simplify the plot we included the F- and G types into the C type.}
\label{fig:comparison_BS00}
\end{figure*}

First we compare the Ceres OE parameters with those of the asteroids in \citet{BS00,B03,S16}. The authors provided two observables: the OE amplitude at $0.3^\circ$ phase angle, in magnitudes, and the ratio of intensity at $0.3^\circ$ and $5^\circ$ phase angle. The former refers to the brightness at $0.3^\circ$ phase angle relative to the extrapolated brightness of the linear part of the phase curve (on the magnitude scale) at the same phase angle. We converted this quantity to the enhancement factor $\zeta(0.3^\circ)$. The intensity ratio is that of the observed integrated intensity, on a linear scale, at two different phase angles. We calculated both quantities for the average Ceres surface. Figure~\ref{fig:comparison_BS00}A shows that the enhancement factor has a maximum for asteroids of intermediate albedo: the M- and S  types. The low albedo C-type asteroids have the lowest factor, whereas the high albedo E-type asteroids have a factor intermediate to that of the C- and M/S types. \citet{BS00} explain the peak at intermediate albedos as resulting from the balance between shadow hiding (SH) and coherent backscatter (CB). The former dominates at low albedos, whereas the latter dominates at high albedos. For asteroids of intermediate albedo the two mechanisms may contribute approximately equally, creating a local maximum. However, the contribution of CB as reconstructed by \citeauthor{BS00} is the same for asteroids of both intermediate and high albedo, which is somewhat counterintuitive given that the CB amplitude should increase with increasing albedo \citep{H93}. But, as the authors note, variations in regolith particle size from one asteroid type to another may also affect the OE amplitude. The figure shows that Ceres' enhancement factor is typical for that of M/S types, being a little higher than that of the C types. The intensity ratio in Fig.~\ref{fig:comparison_BS00}B shows a similar maximum as $\zeta(0.3^\circ)$. Here, the ratio for Ceres fits right in between the C- and M/S types, as expected for an asteroid with a geometric albedo of 0.09.

\begin{figure*}
\centering
\includegraphics[width=8cm]{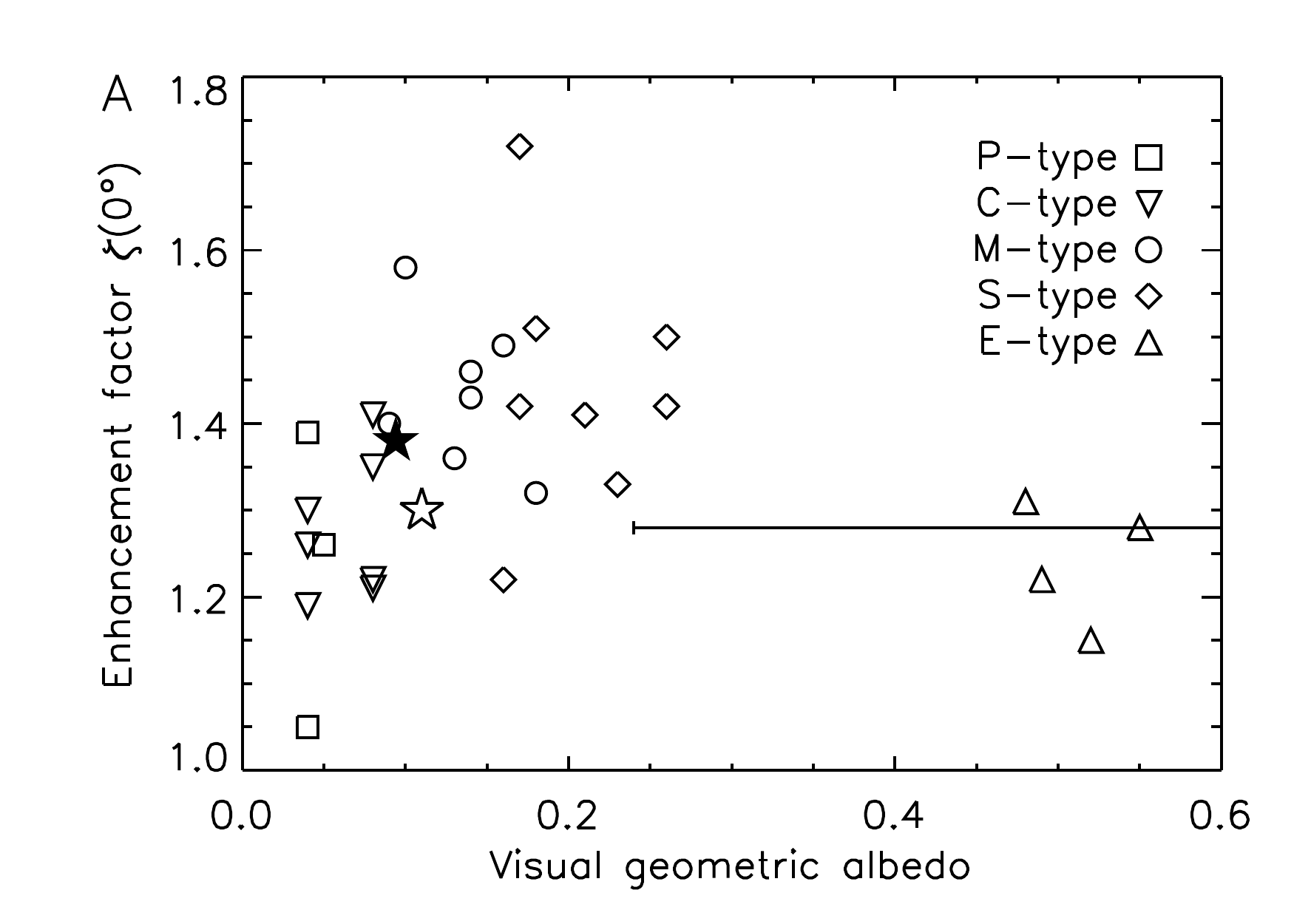}
\includegraphics[width=8cm]{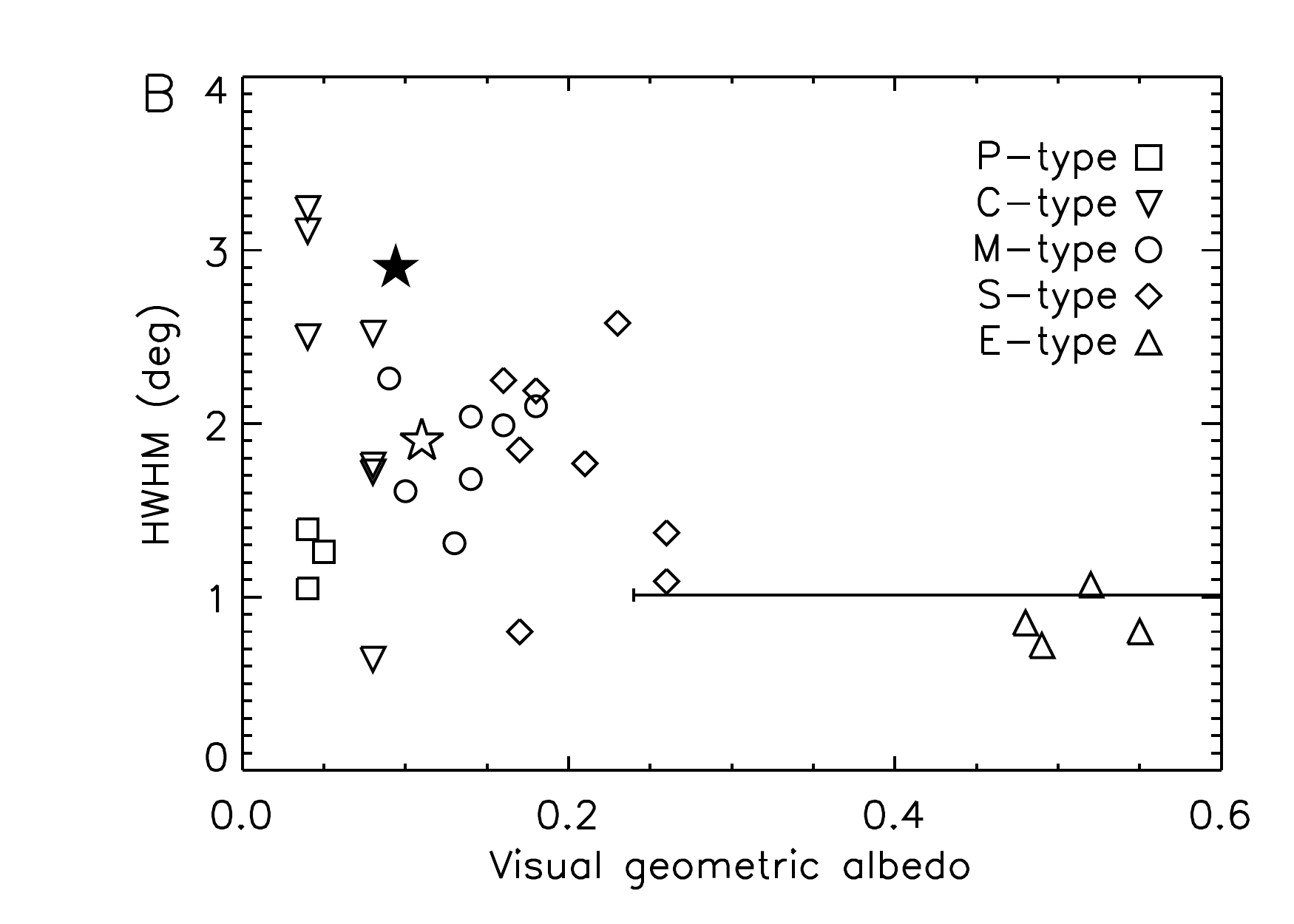}
\caption{The Ceres OE compared to that of the asteroids in \citet{R02}. (\textbf{A}) The enhancement factor $\zeta(0^\circ)$, (\textbf{B}) the \textsc{hwhm}. The Ceres values are those for the clear filter in Table~\ref{tab:Akimov_fit_coef_area} (black star) and from \citeauthor{R02} (white star). The Cerealia Facula values are located somewhere on the horizontal line. To simplify the plot we included the G type into the C type.}
\label{fig:comparison_R02}
\end{figure*}

\citet{R02} evaluated how the \textsc{hwhm} and enhancement factor $\zeta(0^\circ)$ depend on the geometric albedo for asteroids of various types, partly using the same data as the previous authors. The enhancement factor in Fig.~\ref{fig:comparison_R02}A shows a similar maximum as seen earlier. The \textsc{hwhm} in Fig.~\ref{fig:comparison_R02}B appears to be anti-correlated with albedo, although there is a lot of scatter in the data at low albedo. The location of the Ceres parameters in the two figures is consistent with its status as a C type asteroid, although its \textsc{hwhm} is at the high end of the C-type range. The data points in the figure lack error bars\footnote{\citet{R02} provide the following estimated error ranges: 0.01-0.08 for $\zeta(0^\circ)$ and $0.1^\circ$-$0.7^\circ$ for the \textsc{hwhm}.}. Ceres was also included in the \citet{R02} survey, with data from \citet{T83}, and its ``old'' location is shown as a separate data point in Fig.~\ref{fig:comparison_R02}. The distance to the new location can be considered as an indication of the uncertainty of the data in the figure. \citet{L18} found that the slope of the Ceres phase curve is close to that of other C-type asteroids. We conclude that the Ceres OE is also typical for an asteroid of its geometric albedo. We can also evaluate the enhancement factor and \textsc{hwhm} over the resolved surface using the Akimov model parameter maps in Fig.~\ref{fig:Akimov_parameter_maps}, concentrating on the area around Azacca crater where the maps are most reliable. As the enhancement factor is calculated as $\zeta(0^\circ) = 1 + m$ (Eq.~\ref{eq:Akimov_zeta}), the map of the enhancement factor is essentially the map of $m$. We do not recognize the bright Azacca ejecta in the map of $m$, which implies that the enhancement factor has no clear trend with increasing albedo, where we had expected a slightly positive trend. The reason may be that the OE of the Azacca ejecta is not governed by albedo but by other physical properties, as we suggested in Sect.~\ref{sec:spatial_variations}. The \textsc{hwhm} is calculated according to Eq.~\ref{eq:Akimov_HWHM}, and is about $0.5^\circ$-$1.0^\circ$ lower for the Azacca ejecta compared to its darker surroundings. As such, the \textsc{hwhm} shows a clear negative trend with increasing albedo, in line with Fig.~\ref{fig:comparison_R02}B.

The figures also show the enhancement factors, full width at half maximum ({\sc fwhm)}, and intensity ratio for Cerealia Facula, which we calculated from the best-fit Akimov model in Fig.~\ref{fig:Cerealia_Facula}. Determining the corresponding geometric albedo is not straightforward. The Cerealia Facula phase curve was derived for an area that includes both the facula and its surroundings. However, the shape of the curve will primarily be governed by the facula because it is so much brighter. Therefore, the appropriate geometric albedo is found somewhere in the range shown, for which the lower limit is the normal albedo from the Akimov model ($A_0 = 0.24$), and the upper limit is the geometric albedo estimate from \citet{S17} ($p_{\rm V} = 0.6$). The upper part of this range overlaps the cluster of E-type asteroid data points. We conclude that the Cerealia Facula OE compares well to that of the E-type asteroids, which have a similar albedo. We infer from the comparison with other asteroids that the OE characteristics of the different Ceres terrains are primarily determined by the local albedo, rather than any other physical properties.

\section{Discussion}

\begin{figure}
\resizebox{\hsize}{!}{\includegraphics{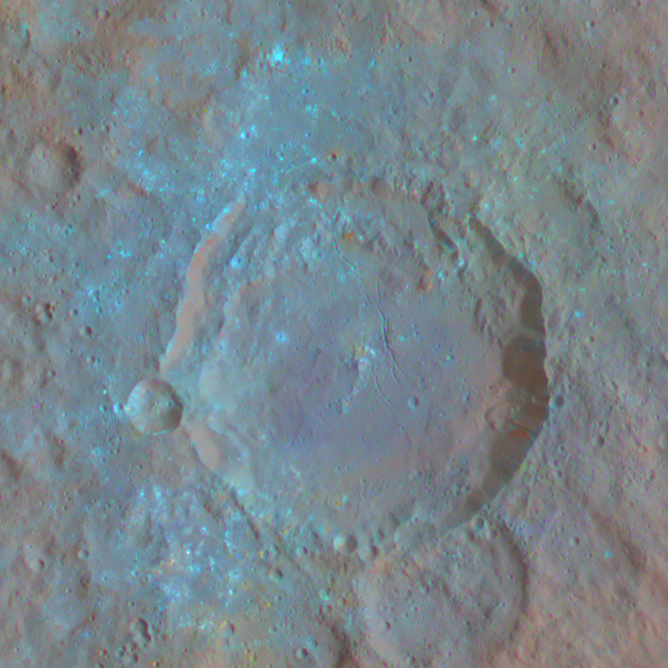}}
\caption{Azacca crater in enhanced color (red: image {\bf 45867}, 965~nm; green: image {\bf 45870}, 555~nm; blue: image {\bf 45864}, 438~nm).}
\label{fig:Azacca}
\end{figure}

We have characterized the OE over a strip on the surface of Ceres that is the path of zero phase as observed by the Dawn cameras. Using different photometric models, we attempted to uncover correlations between the photometric parameters to assess the relative contribution of shadow hiding (SH) and coherent backscatter (CB) to the OE. Coherent backscatter is thought to dominate the OE at very small phase angles \citep{M92,MD93,H97,H98}. We find that the OE has very uniform characteristics over the investigated area, and has an enhancement factor of 1.4 and a {\sc fwhm} of $3^\circ$. If we consider this average, ``broad OE'' to be representative for the Ceres OE, its characteristics are typical for an asteroid of low to moderate albedo. Azacca crater (Fig.~\ref{fig:Azacca}) forms a notable exception. The OE of its relatively bright and blue ejecta is enhanced in a very narrow phase angle range ($< 0.5^\circ$). For asteroid Itokawa, \citet{LI18} found a narrow ($< 1.4^\circ$) enhancement of the OE that correlates with the normal albedo over the surface, which they attributed to CB. Their finding is similar to ours, but the Ceres ``narrow OE'' is even narrower than that of Itokawa and the associated albedos are lower. The narrow Ceres OE may be highly localized. Due to the absence of a clear correlation of the OE slope with normal albedo, we suspect that the Azacca ejecta have physical properties that lead to the enhancement, possibly related to their emplacement. The blue color on Ceres is an indication of youth \citep{SK16,S17}, therefore the emplacement must have been relatively recent. We note that the narrow OE on both Ceres and Itokawa is in the phase angle range of the polarization OE (a narrow, negative polarization spike) found for E-type asteroids, that \citet{R09} suggested to be related to CB. Although the albedo of the Azacca ejecta is much lower than that of a typical E-type asteroid surface, we speculate that the narrow OE enhancement is also due to CB. According to theory, the width of the {\sc cboe} should be substantially dependent on wavelength \citep{M92,H02}. We observed no systematic wavelength-dependent variations of the width of the regular, broad OE ({\sc fwhm} $= 3^\circ$) over the 0.44-0.96~\textmu m range, which is expected if it is exclusively due to SH. Unfortunately, we do not have sufficient data to establish whether or not the width of Azacca's narrow OE is dependent on wavelength.

Let us look closer at the {\sc cboe} width prediction. \citet{M92} gives the {\sc cboe} width (in radians) for optically thick media as:
\begin{equation}
{\rm HWHM}_{\rm CB} = \frac{\epsilon \lambda}{2 \pi \Lambda_{\rm tr}},
\label{eq:HWHM_CB}
\end{equation}
with $\epsilon$ being a constant, $\lambda$ the wavelength of light, and $\Lambda_{\rm tr}$ the transport mean free path. The latter can be regarded as the average distance a photon travels before scattering changes its direction by a large angle \citep{H12}, and depends on the regolith properties as:
\begin{equation}
\Lambda_{\rm tr}^{-1} = n \langle \sigma \rangle Q_{\rm S} (1 - \langle \cos g \rangle),
\label{eq:l_tr}
\end{equation}
where $n$ is the number of particles per unit volume (assumed to be well-separated), $\langle \sigma \rangle$ the mean particle geometric cross section, $Q_{\rm S}$ the particle scattering efficiency, and $g$ the scattering angle. If the Ceres narrow OE is indeed due to CB, then we can roughly estimate the {\sc hwhm} as $0.5^\circ$ and find a transport mean free path of about 7~\textmu m (with $\epsilon = 0.5$ and $\lambda = 732$~nm, the effective wavelength of the clear filter). We note that if the {\sc hwhm} is smaller, $\Lambda_{\rm tr}$ is larger. The evaluation of Eq.~\ref{eq:HWHM_CB} is not straightforward. Therefore, \citet{M92} provided model curves of the \textsc{hwhm} as a function of particle size, material refractive index ($n$), and filling factor (fraction of the volume occupied by the particles), assuming a size distribution of spherical particles. The \textsc{hwhm} in all model curves reaches a maximum for particles with a diameter near the wavelength; it approaches zero for particles much smaller or larger than the wavelength. The \textsc{hwhm} increases with filling factor. For refractive indices characteristic for silicates ($n = 1.45$-1.60), the \textsc{hwhm} is around $2^\circ$ for the largest filling factor modeled. On the basis of his model curves, \citet{M92} concluded that the \textsc{hwhm} should be substantially dependent on wavelength for silicate surfaces. Expanding on this work, \citet{MD93} predicted that CB contributes substantially to the OE of E-type asteroids, and should lead to wavelength-dependent changes of the angular width.

The \citet{M92} prediction has driven a search for wavelength-dependent changes of the OE of rocky solar system bodies over the last few decades. \citet{R09} observed E-type asteroid 44~Nysa and uncovered a narrow brightness OE accompanied by a similarly narrow polarization OE, which they attributed to CB. The authors found evidence for wavelength-dependent changes of the OE width in the visible range, but their results were inconclusive due to limited data availability. For S-type asteroid 433~Eros, \citet{C02} identified a decrease of the OE width parameter in the \citet{H81,H84,H86} model over the 0.8-1.5~\textmu m wavelength range, which they considered to be consistent with CB. For S-type asteroid Itokawa, \citet{K08} derived minor, irregular variations in the OE angular width parameter in the \citet{H02} model over the 0.85-2.1~\textmu m wavelength range, which they considered to be suggestive of CB. However, \citet{LI18} reported no wavelength-dependent variations. \citet{Sp12} did not report such variations either for asteroid \v{S}teins, but may not have searched for them. \citet{H16} did not find evidence for such variations for asteroid Lutetia over the 0.38-0.63~\textmu m wavelength range. Several studies failed to uncover a wavelength dependence for the OE width for the Moon \citep{B96,S99,H12}. However, \citet{K13} found the slope of the lunar OE to increase with wavelength over the 0.5-3.0~\textmu m range for most terrains under study, which they attributed to CB.

The results of the various studies are not always easy to compare because of the use of different OE definitions and the diversity of analytical methods. Nevertheless, it appears that variations of the OE \textsc{hwhm} with wavelength are rarely, if ever, found for planetary surfaces. Failure to find clear wavelength-dependent changes of the lunar OE width has evoked different responses in the literature. \citet{S99} argued that the change of wavelength in Eq.~\ref{eq:HWHM_CB} is compensated for by a similar change in the transport mean free path, a structural property of the regolith they termed ``quasifractal''. On the other hand, the failure led \citet{H12} to question our understanding of the CBOE.  \citet{TM10}, however, argued that we fully understand the physics of CB, but that simulating light scattering by a planetary surface is an extremely complicated effort, as it requires taking into account a complex convolution of contributions from morphologically different surface types with varying albedos. They wrote: ``...the only way to establish the CB nature of a laboratory or remote-sensing observation is to compare the measurement results with the results of theoretical computations''. Their argument implies that the \citet{M92} and \citet{MD93} predictions were too simplistic, and that the mere absence or presence of wavelength-dependent variations of the OE width is not diagnostic for CB.

It is not obvious that CB can play a role on surfaces as dark as that of Ceres, and it seems that there are no observational photometric criteria that are unequivocally diagnostic for CB. The \citet{M92} prediction of a wavelength-dependent OE width may be physically correct, but is not necessarily valid for a complex planetary regolith. That said, the very challenging nature of the Dawn observations prevented us from even investigating the wavelength dependence of the narrow OE of the Azacca ejecta, which is our most promising candidate for CB. Perhaps our best chance of demonstrating the role of CB in the Ceres OE lies in comparing the Dawn observations with physically based simulations as described by \citet{TM09}, although these models are not easy to handle in practice. In addition, OE observations made by the on-board VIR spectrometer may help to further constrain wavelength-dependent behavior and extend the analysis to larger wavelengths.

\begin{acknowledgements}
The authors wish to thank the editor and an anonymous referee for their helpful suggestions.
\end{acknowledgements}

\bibliography{Ceres_opposition}

\end{document}